\documentclass[twocolumn]{aastex62}
\usepackage{amsmath}
\usepackage{amstext}
\usepackage{gensymb}
\usepackage{apjfonts}
\usepackage{bm}
\usepackage{multirow}
\usepackage{hyperref}
\hypersetup{
            pdftitle={Sub-percent Photometry: Faint DA White Dwarf Spectrophotometric Standards for Astrophysical Observatories},
            pdfauthor={Narayan et al.}
            bookmarks=true,
            bookmarksnumbered=true,
            colorlinks=true,
            anchorcolor=blue,
            citecolor=blue,
            linkcolor=blue,
           }
\usepackage[figure,figure*]{hypcap}
\usepackage[graphicx]{realboxes}
\graphicspath{{figures/}}

\newcommand*\mean[1]{\bar{#1}}

\shorttitle{Sub-percent Spectrophotometric Standards}
\shortauthors{Narayan, Matheson, Saha et al.}

\defcitealias{Bergeron1992}{B92}
\defcitealias{Fitzpatrick99}{F99}
\defcitealias{Bohlin2014}{B14}
\defcitealias{Narayan16}{N16}
\defcitealias{Hermes17} {H17}
\defcitealias{Toonen17}{T17}
\defcitealias{Calamida18}{C19}
\defcitealias{Supercal}{S15}

\begin{document}

\title{Sub-percent Photometry: Faint DA White Dwarf Spectrophotometric Standards for Astrophysical Observatories}

\author[0000-0001-6022-0484]{Gautham Narayan}
\altaffiliation{Lasker Fellow}
\email{gnarayan@stsci.edu}
\affiliation{Space Telescope Science Institute, 3700 San Martin Drive, Baltimore, MD 21218}
\affiliation{National Optical Astronomy Observatory, 950 North Cherry Avenue, Tucson, AZ 85719}

\author[0000-0001-6685-0479]{Thomas Matheson}
\affiliation{National Optical Astronomy Observatory, 950 North Cherry Avenue, Tucson, AZ 85719}

\author[0000-0002-6839-4881]{Abhijit Saha}
\affiliation{National Optical Astronomy Observatory, 950 North Cherry Avenue, Tucson, AZ 85719}

\author{Tim Axelrod}
\affiliation{The University of Arizona, Steward Observatory, 933 North Cherry Avenue, Tucson, AZ 85719}

\author{Annalisa Calamida}
\affiliation{Space Telescope Science Institute, 3700 San Martin Drive, Baltimore, MD 21218}

\author{Edward Olszewski}
\affiliation{The University of Arizona, Steward Observatory, 933 North Cherry Avenue, Tucson, AZ 85719}

\author{Jenna Claver}
\affiliation{National Optical Astronomy Observatory, 950 North Cherry Avenue, Tucson, AZ 85719}

\author[0000-0001-9846-4417]{Kaisey S. Mandel}
\affiliation{Institute of Astronomy and Kavli Institute for Cosmology, Madingley Road, Cambridge, CB3 0HA, UK}
\affiliation{Statistical Laboratory, DPMMS, University of Cambridge, Wilberforce Road, Cambridge, CB3 0WB, UK}

\author[0000-0001-9806-0551]{Ralph C. Bohlin}
\affiliation{Space Telescope Science Institute, 3700 San Martin Drive, Baltimore, MD 21218}

\author{Jay B. Holberg}
\affiliation{The University of Arizona, Lunar and Planetary Laboratory, 1629 East University Boulevard, Tucson, AZ 85721}

\author[0000-0003-2823-360X]{Susana Deustua}
\affiliation{Space Telescope Science Institute, 3700 San Martin Drive, Baltimore, MD 21218}

\author[0000-0002-4410-5387]{Armin Rest}
\affiliation{Space Telescope Science Institute, 3700 San Martin Drive, Baltimore, MD 21218}
\affiliation{Department of Physics and Astronomy, Johns Hopkins University, Baltimore, MD 21218, USA}

\author[0000-0003-0347-1724]{Christopher W. Stubbs}
\affiliation{Harvard University, Department of Physics, 17 Oxford Street, Cambridge, MA 02138}
\affiliation{Harvard-Smithsonian Center for Astrophysics, 60 Garden Street, Cambridge, MA 02138}

\author{Clare E. Shanahan}
\affiliation{Space Telescope Science Institute, 3700 San Martin Drive, Baltimore, MD 21218}

\author{Amali L. Vaz}
\affiliation{The University of Arizona, Steward Observatory, 933 North Cherry Avenue, Tucson, AZ 85719}

\author{Alfredo Zenteno}
\affiliation{Cerro Tololo Inter-American Observatory, Casilla 603, La Serena, Chile}

\author[0000-0002-1652-420X]{Giovanni Strampelli}
\affiliation{Space Telescope Science Institute, 3700 San Martin Drive, Baltimore, MD 21218}

\author{Ivan Hubeny}
\affiliation{The University of Arizona, Steward Observatory, 933 North Cherry Avenue, Tucson, AZ 85719}

\author{Sean Points}
\affiliation{Cerro Tololo Inter-American Observatory, Casilla 603, La Serena, Chile}

\author{Elena Sabbi}
\affiliation{Space Telescope Science Institute, 3700 San Martin Drive, Baltimore, MD 21218}

\author[0000-0001-6529-8416]{John Mackenty}
\affiliation{Space Telescope Science Institute, 3700 San Martin Drive, Baltimore, MD 21218}


\begin{abstract}
We have established a network of 19 faint (16.5~mag $< V < $19~mag) northern and equatorial DA white dwarfs as spectrophotometric standards for present and future wide-field observatories. Our analysis infers SED models for the stars that are tied to the three CALSPEC primary standards. Our SED models are consistent with panchromatic Hubble Space Telescope (\emph{HST}) photometry to better than 1\%. The excellent agreement between observations and models validates the use of non-local-thermodynamic-equilibrium (NLTE) DA white dwarf atmospheres extinguished by interstellar dust as accurate spectrophotometric references. Our standards are accessible from both hemispheres and suitable for ground and space-based observatories covering the ultraviolet to the near infrared. The high-precision of these faint sources make our network of standards ideally suited for any experiment that has very stringent requirements on flux calibration, such as studies of dark energy using the Large Synoptic Survey Telescope (LSST) and the Wide-Field Infrared Survey Telescope \emph{(WFIRST}).
\end{abstract}
 



\section{Introduction}\label{sec:Intro}

The volume of astronomical data has grown by an order of magnitude from the first-generation of wide-field surveys such as the Sloan Digital Sky Survey\footnote{\url{http://www.sdss.org}} (SDSS) and the Two Micron All-Sky Survey\footnote{\url{https://www.ipac.caltech.edu/2mass/}} (2MASS) to the second-generation of surveys including Pan-STARRS\footnote{\url{http://pan-starrs.ifa.hawaii.edu/public/}}, the Dark Energy Survey\footnote{\url{http://www.darkenergysurvey.org/}} (DES), and \textit{Gaia}\footnote{\url{http://sci.esa.int/gaia/}}. Together, these surveys have produced terabytes of images and catalogs, measuring the sky from the UV to the NIR. While each individual project reports internal relative photometric accuracies of 1--2\%, all these surveys must be placed on a single \emph{consistent} flux scale in order to inter-compare across wavelength and redshift.

The most precise existing network of spectrophotometric standard stars for flux calibration, the 93 extrasolar CALSPEC\footnote{\url{http://www.stsci.edu/hst/observatory/crds/calspec.html}} standards~\citep[][hereafter \citetalias{Bohlin2014}]{Bohlin14, Bohlin2014}, have mean \textit{V}$ = $11.1~mag, which is close to or brighter than the saturation limit of many of these surveys. While the faint limit of the CALSPEC stars is \textit{V}$ \sim16$~mag, most of the objects near this limit are extremely red stars with complex SEDs that are difficult to model. Consequently, surveys have had to resort to establishing networks of secondary or tertiary standards together with synthetic color transformations to tie their natural systems to a common flux scale, such as that defined by the broadband measurements of \citet[][]{Landolt92}. Calibration errors are compounded as the degree of separation from the primary standards increase, and these systematic errors in the photometric calibration propagate into the analyses of several high-precision experiments. The measurement of the equation of state $w$ of dark energy from multi-survey compilations of Type Ia supernovae (SNIa) is particularly afflicted by such systematic errors in the photometric calibration~\citep{Scolnic14}. Each of the surveys observe SNIa with different telescopes, instruments and passbands at different sites, and target different mean redshifts. While there are thousands of well-measured SNIa, the constraints on $w$ are dominated by systematic uncertainties. 

With the advent of the third-generation of wide-field surveys, including the Zwicky Transient Facility\footnote{\url{http://ztf.caltech.edu}} (ZTF) and the Large Synoptic Survey Telescope\footnote{\url{http://www.lsst.org}} (LSST), the volume of observational data will increase by an additional order of magnitude, exacerbating the impact of calibration errors and the mismatch between the standards suitable for calibration, and what is practically observable by ground-based facilities. To reduce these systematic errors, and obtain less biased measurements of source fluxes, we need more precise and more accurate photometric calibration. \citet{Stubbs16} review efforts to establish precise astronomical flux calibration, and divides these efforts into two categories: i) approaches where the metrology standard is the spectral energy distribution an emissive source that can be determined from fundamental physics (e.g., a blackbody spectrum) ii) approaches where the metrology standard is based on a detector with known quantum efficiency (e.g. NIST-calibrated Si photodiodes). We employ the first approach in this work. 

We have been observing a new all-sky network of faint (16.5~mag $ < V < $19~mag) pure-hydrogen (DA) white dwarf (WD) stars with three \emph{HST} programs (GO-12967 and GO-13711 for northern and equatorial standards, and GO-15113 for southern standards, P.I.~Abhijit Saha), and we have obtained multi-band above-the-atmosphere photometry to complement ground-based spectroscopy and temporal monitoring. The goal of this work is to use these observations to infer calibrated synthetic SED models, which can then be used as transfer standards by any facility.

Our team presented a proof-of-concept end-to-end analysis in \citet[][hereafter \citetalias{Narayan16}]{Narayan16} examining the feasibility of modeling the \emph{HST} observations for 4 of these twenty three stars in our cycle 20 program. \citetalias{Narayan16} assembled a pipeline that largely used existing software packages to process and analyze the data, and demonstrated that synthetic DA white dwarf SEDs extinguished by interstellar dust could recover the \emph{HST/WFC3} measurements to a few millimag. Our proof-of-concept analysis also revealed a previously unknown systematic in the photometric zeropoints reported by the Mikulski Archive for Space Telescopes (MAST). Even with the limited number of objects available during our pilot study, we were able to detect a systematic difference between the response of the UVIS1 and UVIS2 chips of \emph{HST/WFC3}. This difference had been masked by the flat-fielding procedure used in the \texttt{calwf3} pipeline, which produces the processed images available through MAST. This systematic trend has been accounted for in current MAST data products~\citep{Deustua16, Deustua17}, which improved the flat-fielding procedure and corrected for a systematic bias in the previous calibration, which led to a 4\% underestimate of fluxes on UVIS2. 

The exercise revealed the magnitude of the challenge posed by establishing highly accurate spectrophotometric standards: any program aiming to reduce systematic errors arising from photometric calibration must itself be robust against systematic errors arising from photometric calibration. To achieve 1\% level calibration of ground-based surveys, our network of DA white dwarf stars must be calibrated to \emph{sub-percent} accuracy.

This work refines all aspects of the analysis procedure used in \citetalias{Narayan16}. In particular, we develop a new methodology to fit our \emph{HST} photometry and ground-based spectroscopy \emph{simultaneously} to infer the underlying DA white dwarf SED. This work produces a calibrated network of northern and equatorial DA white dwarfs with accurately measured SEDs tied directly to the three CALSPEC primary standards, GD71, GD153, and G191-B2B. This network of faint spectrophotometric standards is suitable for use by wide-field surveys from the ultraviolet to the near infrared. Detailed descriptions of the observations and the procedures to obtain and process them are presented in \citet[][hereafter \citetalias{Calamida18}]{Calamida18}, a companion paper to this work.

\subsection{Structure of This Paper}
\begin{figure}[htpb]
\begin{centering}
\includegraphics[width=0.47\textwidth]{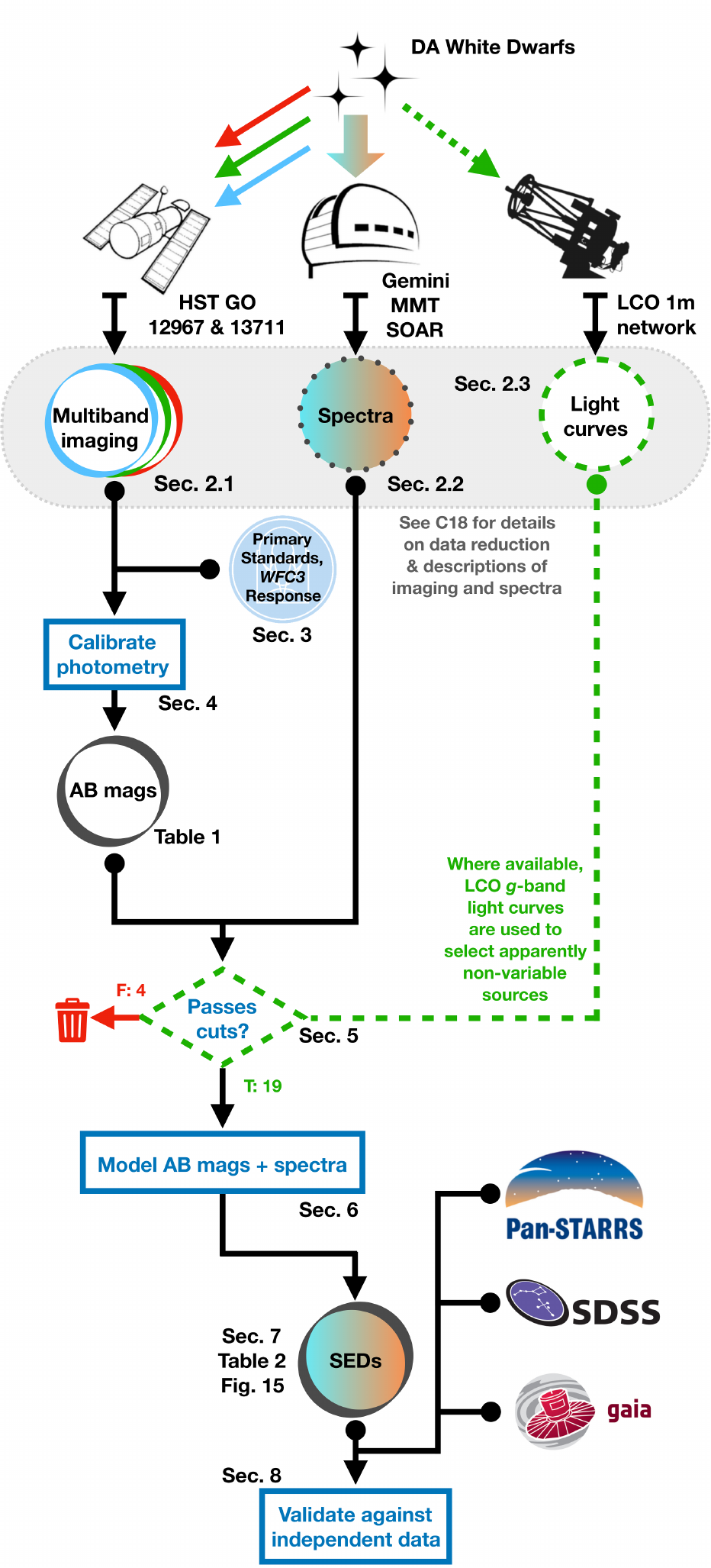}
\caption{A flowchart of the analysis presented in this work. Data products are indicated by circles, while discrete stages of our analysis are indicated with boxes. Arrows that originate in small dots indicate that the preceding data product is an input to the subsequent analysis. The text labels indicate the sections of this work that correspond to the data product or analysis. We include a brief description of the primary data products produced by our observing program (rounded grey rectangle) that are presented in \citetalias{Calamida18}, a companion paper to this work. Derived data products (outputs from blue rectangles) are available from our \href{https://github.com/gnarayan/WDdata}{archive} (see footnote~\ref{footnote:archive}).}\label{fig:schematic}
\end{centering}
\end{figure}

We provide a schematic overview of this work in Fig.~\ref{fig:schematic}. In \S\ref{sec:obs}, we provide a brief overview of our observational program, while in \S\ref{sec:changes} we highlight the improvements made to our procedure from cycle 20 to cycle 22. We use a Bayesian hierarchical model to combine our multi-cycle \emph{HST} observations and to generate a combined photometric catalog that serves as the primary input for our analyses; we describe this model in \S\ref{sec:phot}. In \S\ref{sec:excluded} we impose selection cuts to eliminate objects that do not meet the stringent requirements of our program. We develop a Bayesian model for a joint analyses of the spectroscopic and photometric measurements to infer SEDs for our DA white dwarfs in \S\ref{sec:wdmodel}. The results of our analysis of the data are presented in \S\ref{sec:WDmodelResults}, together with a detailed examination of potential sources of systematic error. We test that the SED models are consistent, both internally and with constraints from external data sets in \S\ref{sec:validation}. Finally, in \S\ref{sec:futurework}, we summarize our analysis and outline our plan to complete this program. Tables of photometric and spectroscopic observations, as well tables of inferred model parameters, photometric residuals, and the final calibrated SEDs are available through our archive\footnote{\url{https://github.com/gnarayan/WDdata}\label{footnote:archive}}.


\section{Design of the Observing Program}\label{sec:obs}

Our observational program consists of three components: i) multi-band \emph{HST} imaging using \emph{WFC3}, ii) ground-based spectroscopy using a variety of low-dispersion long slit spectrographs on large aperture telescopes, and iii) temporal monitoring using the Las Cumbres Observatory (LCO) network of robotic 1~m telescopes. All three components are critical to our program. Here, we provide an overview of the different observations, focusing on \emph{why} each component is necessary for our analysis. A detailed description of \emph{what} observations were obtained and \emph{how} the data is processed is provided in \citetalias{Calamida18}. This work focuses on the analysis of these observations.

\subsection{\emph{HST} Photometry}\label{sec:HSTphot}
The key observational component that distinguishes our program from other efforts to establish spectrophotometric standards, such as \citet[][]{Prieto09}, is our pan-chromatic \emph{HST/WFC3} photometry: we can set the flux scale for each object from the UV to the NIR without systematic effects from atmospheric extinction. Ground-based programs to establish spectrophotometric standards face inherent limitations:

\begin{enumerate}
\item{The transmissivity of the atmosphere in the UV and NIR makes sufficiently high S/N ground-based observations of faint standards prohibitive.}
\item{There are significant grey and chromatic variations (arising from aerosol scattering, Rayleigh scattering, as well as oxygen, ozone and precipitable water vapor absorption) on short angular and temporal scales in the optical.}
\item{Many stars with a long legacy of spectrophotometric measurements that are employed as standards, such as Vega and BD+17\degree4708, are not ideal reference calibrators and have complex, hard-to-model spectral energy distributions, or exhibit measurable variability. Most of the remaining well-measured spectrophotometric standards that could be used as a reference for our new standards also saturate the instruments on modern large aperture facilities --- one of the motivations for establishing our network of faint standards in the first place.}
\end{enumerate}
These effects have limited the absolute flux calibration accuracy of ground-based surveys to the few percent level~\citep{Stubbs06}. The resulting calibration \emph{errors} are one of the leading systematic effects afflicting dark energy studies with SNIa that rely on a comparison of the brightness of distant supernovae with nearby supernovae, often observed by a completely different survey~\citep{Scolnic14}. 

To overcome these limitations, we obtained photometry using \emph{HST}, avoiding the time-variable atmosphere of the Earth. This choice has various ancillary benefits. The \emph{HST/WFC3} throughput was precisely determined pre-flight, and staff at the Space Telescope Science Institute (STScI) have continually monitored the system to tie \emph{WFC3} measurements to other \emph{HST} instruments, particularly the Space Telescope Imaging Spectrograph (\emph{STIS}) and the Advanced Camera for Surveys (\emph{ACS}). 

Our use of \emph{HST} makes the three CALSPEC primary standards, GD71, GD153, and G191-B2B, the natural choice for our network's spectrophotometric reference. These standards have the advantage of being the same class of astrophysical source as our targets. Moreover, all have an extensive set of observations with a wide variety of instruments; \citetalias{Bohlin2014} used these observations to constrain their line-of-sight interstellar extinction to be consistent with $< 2$~mmag in the optical. 

While planning our \emph{HST/WFC3} observations, we avoided sources with a detectable close companion in archival SDSS and Pan-STARRS images and selected objects with low line-of-sight Galactic reddening by limiting the color excess $E(B-V) < 0.2$~mag. We obtained observations in \textit{F275W, F336W, F475W, F625W, F775W,} and \textit{F160W} to establish the flux scale and constrain any interstellar extinction. Given the median exposure times of our observations, stacking both individual exposures taken for cosmic ray rejection yields  5$\sigma$ limiting magnitudes in \textit{\{F275W, F336W, F475W, F625W, F775W, F160W\}} of $\{22.8, 23.1, 24.1, 24.3, 24.6, 25.4\}$ AB mag. These deep observations are sufficient to rule out O--K star companions. The presence of such companions was strongly disfavored from inspection of SDSS and Pan-STARRS images, but our \emph{HST} observations extend further into the UV and the IR, and are deeper than these ground-based surveys. We discuss our selection cuts in detail in \S\ref{sec:excluded}, and eliminate one object, SDSS-J041053 from our sample. This object is now known to be a DA+M:E binary system~\citep{Kleinman13}, and we find that it exhibits significantly different colors from the other targets in our sample. This is consistent with \citet[][]{Eisenstein06}, which demonstrates that the temperature inferred from the spectra of these DA-subdwarf M binary systems have a significantly different correlation with SDSS colors than isolated DA white dwarfs, and they form a different locus in SDSS color-color diagrams. 

Our near infrared (NIR) observations are critical to establishing our DA white dwarfs as useful spectrophotometric standards for major future space-based observatories including the James Webb Space Telescope (\emph{JWST}) and \emph{WFIRST}. The utility of these standards is not limited to space-based observatories. By establishing our network of faint standards directly on the CALSPEC white dwarf scale, we extend \emph{HST} heritage of precision calibration to \emph{all} current and future observatories. 

\subsection{Ground-based Spectroscopy}\label{sec:spectroscopy}
While the \emph{HST} photometry is sufficient to establish our targets as standards in our program passbands, we wish to establish spectrophotometric standards that can be used by any ground and spaced-based facilities. This requires that we construct a SED model for our targets, which in turn requires that we infer any intrinsic and extrinsic properties of standards that correlate with their flux. Fully radiative pure-hydrogen DA white dwarf models have been used to establish SEDs for over three decades \citep[see Sec.~4 of ][ and references therein]{Bohlin14}. Only two parameters are necessary to model their SED: temperature $T_{\text{eff}}$ and surface gravity $\log g$. The second component of our observational program is to obtain a high S/N (> 20) spectrum of our targets using large aperture ground-based facilities to constrain these two parameters.

The depths of the hydrogen Balmer lines of DA white dwarf stars are strongly correlated with $T_\text{eff}$, while the line widths are very sensitive to $\log g$, although all line features depend on both parameters, and the line shapes are impacted by non-ideal effects arising from proton and electron perturbations~\citep{Tremblay09}. The shape of the DA white dwarf continuum is largely determined by $T_\text{eff}$ and interstellar reddening. Accounting for reddening is critical for unbiased inference of the intrinsic SED parameters. Our requirement for faint targets implies that our DA white dwarfs will be significantly more distant than the CALSPEC primary standards, and therefore subject to extinction. Within our Galaxy, the interstellar reddening can be modeled by extinction curves such as those of \citet[][hereafter \citetalias{Fitzpatrick99}]{Fitzpatrick99} and \citet{ODonnell94}. Given these curves, the extinction at any wavelength can be described by two parameters: the extinction in the \textit{V} band $A_{V}$, and the ratio of $A_{V}$ and the color excess $E(B-V)$, denoted by $R_{V}$. 

The intrinsic $T_\text{eff}$ model parameter and extrinsic $A_V$ model parameter are strongly inversely correlated as lowering the temperature or increasing the line of sight extinction both lead to a redder continuum. While it is in principle possible to constrain all these model parameters solely from multi-band photometry, the photometry can only weakly constrain the surface gravity, and the number of independent multiband observations spanning a range in wavelength long enough to break the degeneracy between temperature and reddening makes a purely photometric analysis impractical. Spectroscopy is critical to disentangling these distinct physical processes as they affect the Balmer line shapes differently, and care must be taken in obtaining and reducing these observations.

\begin{figure}[htpb]
\centering
\includegraphics[width=0.452\textwidth]{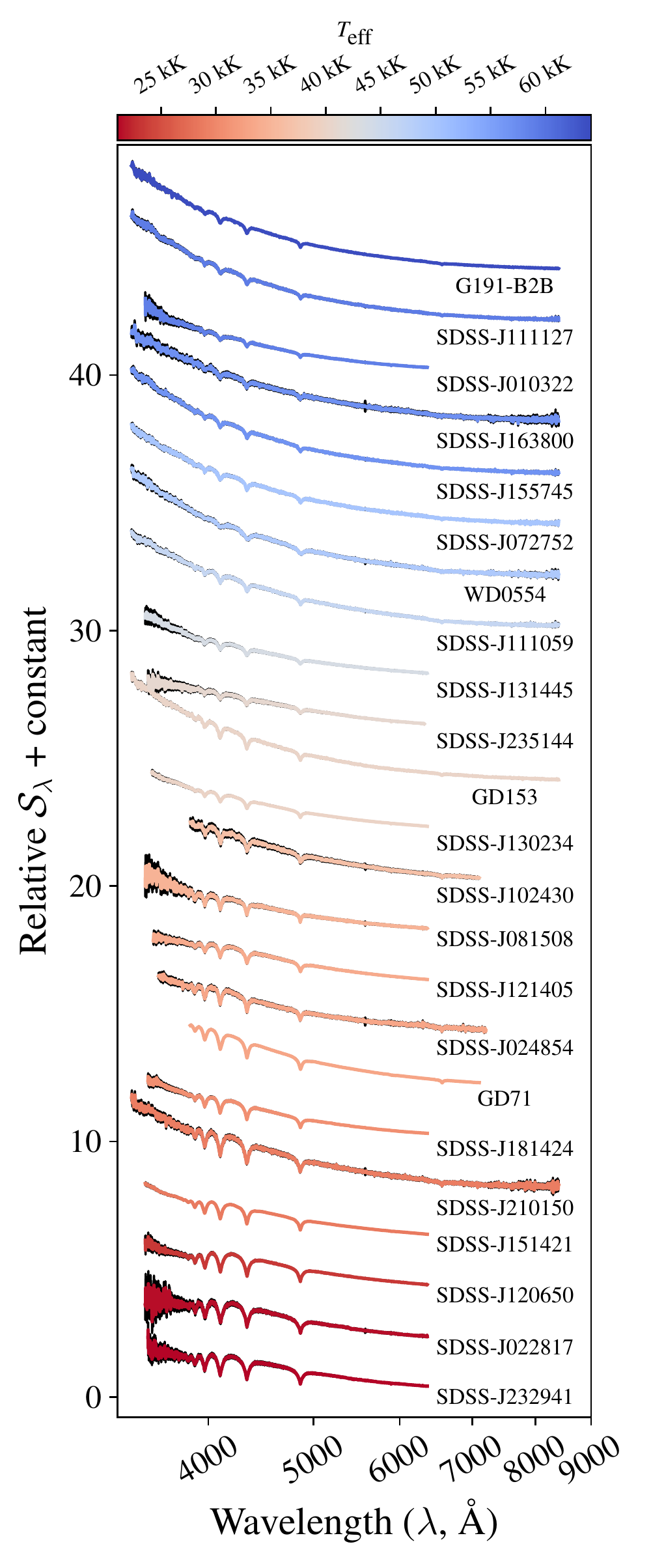}
\caption{Sequence of observed spectra $\mathcal{S}_{\lambda}$ of our DA white dwarfs, ordered by inferred $T_{\text{eff}}$ (indicated by color, with uncertainties on the flux shown in the in black shaded region about each trace). The depths of the hydrogen Balmer lines are sensitive to the temperature of the photosphere, while the line widths are sensitive to $\log g$. The shape of the continuum is sensitive to temperature and interstellar reddening and can exhibit correlated errors with wavelength arising from errors in flux calibration of the spectrum.}\label{fig:spectralseq}
\end{figure}

The wide hydrogen Balmer features and faintness of our targets makes low-dispersion ($R \sim 300-1000$~lp/mm), long slit instruments the optimal solution for our program's spectroscopic requirements. As our flux scale is independently set by the \emph{HST} photometry of the CALSPEC primary standards, we use relatively narrow slits compared to most spectrophotometric efforts in order to reduce the sky background and to ensure that the instrumental resolution is smaller than the Balmer line widths. We employ the optimal extraction algorithm of \citet{Horne86} to extract the 1-D trace from the 2-D spectrum images and recover the flux at both the blue, where the drop in the detector quantum efficiency (QE) dominates the increase in the DA white dwarf flux, and the red where both the QE and white dwarf flux drop precipitously, while the amplitude of fringing increases. The narrow slit width also make us relatively insensitive to small centering errors that would propagate into velocity shifts. We remove cosmic rays, and use telluric corrections determined from observations of two standards with a smooth blue and red continuum respectively to remove the atmospheric absorption lines from ozone and water vapor. We illustrate one spectra $\mathcal{S}$ for each of our targets in Fig.~\ref{fig:spectralseq}. 

Throughout this work, variables representing a spectral flux density are denoted with a subscript $\lambda$ when reported per unit vacuum wavelength, and with a subscript $\nu$ when reported per unit vacuum frequency. The equations in \S\ref{sec:synphot} can be used to transform between the two conventions. The data products provided with this work are reported as $F_\lambda$.

The spectroscopy, without any additional photometric constraints, can be used to determine $T_\text{eff}$ and $\log g$ to a few percent. This inference is sufficient to impose weak selection cuts. We select objects with $T_\text{eff} > 20,000$~K to ensure their atmospheres are fully radiative. Theory suggests that the white dwarf luminosity in the $T_\text{eff} > 20,000$~K regime is too high for circumstellar dust grains to survive~\citep{Koester14}. 

Even though we have only a few distinct epochs of spectroscopy, these observations do place limits on the variability of our sources. Our spectra do not exhibit any absorption bands to indicate the presence of M dwarf or subdwarf companions such as those of \citet[][]{Silvestri06}. At least 10\% of white dwarfs exhibit magnetic fields of $10^{6}$--$10^{9}$~Gs level~\citep{Liebert03}. Our spectroscopic observations allows to rule out targets exhibiting strong magnetic fields, which would be apparent in Zeeman splitting of the Balmer lines, or in the presence of the Minkowski bands --- shallow and broad absorption bands near $3,650$~\AA, $4,135$~\AA, and $4,466$~\AA. White dwarfs with weaker magnetic fields cannot be excluded on the basis of the spectroscopy alone, as the high surface gravity broadens absorption lines, making Zeeman splitting undetectable without high resolution observations.

Our analysis in \S\ref{sec:wdmodel} describes the white dwarf with a NLTE pure-hydrogen model atmosphere, and we look for deviations from this assumption. Our observation methodology and our extraction procedures optimize the S/N of any narrow non-hydrogen features such as the \ion{Ca}{2} doublet and MgN species, occasionally seen in DA white dwarf spectra (see for example \citealt{Farihi12}). While these features have largely been observed in objects cooler than our temperature cutoff, the fraction of hot DA white dwarfs exhibiting such features is only weakly constrained by existing observations. Metal lines have been seen in high-resolution ($R = 2.3 \times 10^5$) UV spectra of G191-B2B~\citep{Bohlin14} but the number fraction of elements relative to hydrogen is $\lesssim 10^{-6}$~\citep[inferred abundances for identified species are given in Table 8 of ][]{G191B2Bmodel}. The presence of these trace metal lines makes a $<2$~mmag difference in \textit{F275W} relative to a pure hydrogen model with the same extinction as the \citet{Bohlin14} model. Metal lines have not been identified in high resolution spectra of either GD71 or GD153.

Following this initial analysis, we combine our spectroscopy and photometry to infer the SED for each DA white dwarf. While we use the \emph{shape} of the continuum for inference, the absolute normalization of the ground-based spectroscopy is typically only accurate to a few percent, and is often tied to spectrophotometric standards that are less precisely established than our target DA white dwarfs. The noise model we employ in \S\ref{sec:specnoisemodel} can account for broad correlated errors in the continuum shape and prevents biases in the inferred intrinsic parameters.

\subsection{Temporal Monitoring}\label{sec:temporal}
The final component of our observing program is designed to detect any photometric variability exhibited by our DA white dwarfs. This includes extrinsic variability (binary companions or debris disks) as well as weak intrinsic variability over short and long time-scales (stochastic outbursts or pulsation). We are obtaining repeated observations of our targets using the Las Cumbres Network of robotic 1~m telescopes. These time-series data allows us to look for variability over timescales of a few hours to several months. We also look for evidence of variability in archival data from other ground-based surveys such as the Pan-STARRS 1 3$\pi$ Survey. This program is still in progress, and not all of DA white dwarfs have sufficient observations ($\gtrsim$35 epochs with approximately logarithmic spacing) to exclude variability comprehensively at present. \citetalias{Calamida18} present results from the our temporal monitoring to date. Based on the current observations, we exclude WD0554 and SDSS-J203722 as variable, as described in \S\ref{sec:excluded}. More detailed results from our temporal monitoring will be presented in future analysis, and we will continue to obtain observations of our targets to rule out variability on timescales longer than a year. 

Despite this component of our observing program not being complete, there are several reasons to expect the fraction of objects in our network that exhibit variability to be small ($< 2$\%). Our \emph{HST} photometry and spectroscopy is sufficient to exclude main sequence stars and subdwarf companions. Our selection criteria also excludes cool DA white dwarfs in the ZZ Ceti instability strip \citep{Gianninas14}, which have convective atmospheres and may exhibit strong variability on timescales as short as a few minutes~\citep{Winget08}. Our spectroscopy excludes strongly magnetic white dwarfs (MWDs), which have large-amplitude photometric variations on timescales of a few hours to a few days (for e.g., see \citealt{Brinkworth13}). 

Studies of nearby white dwarfs can be used to place limits on the expected fraction of objects that exhibit intrinsic variability or have binary companions. \citet[][hereafter \citetalias{Hermes17}]{Hermes17} have carried out a detailed analysis of variability for a sample of 398 high-probability white dwarfs (252 spectroscopically confirmed, 146 photometrically-selected on the basis of their SDSS colors) having \emph{Kepler} \textit{K}$_\text{P} < 19$~mag. The majority of these objects are DA, but the sample also includes helium-dominated DB and DO stars, carbon-dominated DQ stars and continuum-dominated DC stars. Using the \emph{Kepler/K2} 30~min cadence light curves of these sources, \citetalias{Hermes17} identified only 9 that exhibit variability exceeding a peak-to-peak amplitude of 1\% on 1~hr to 10~day timescales, i.e. $>97$\% of their sample of apparently isolated and non-pulsating DA white dwarfs do indeed make good flux standards. The controls adopted for this work likely reduce the percentage of variables even further. Of the 9 sources that \citetalias{Hermes17} found to exhibit variability, four variable sources were photometrically selected, and have colors consistent with $T_{\text{eff}} < $9,000~K, while an additional two spectroscopic targets have $T_{\text{eff}} \sim 100,000$~K and significantly outshine their putative companions. These six objects would have been excluded from our analysis by the selection cuts imposed in \S\ref{sec:spectroscopy}. An additional 15 of the 398 sources exhibit strong magnetic fields and would have been excluded on the basis of their spectroscopy. The \citetalias{Hermes17} analysis suggests that the fraction of objects which exhibit intrinsic variability is likely less than 2\% of the sample. 

\citet[][hereafter \citetalias{Toonen17}]{Toonen17} have modeled the evolution of a nearly complete white dwarf sample within 20~pc with a binary population synthesis approach. Beginning with an initial binary fraction of 50\%, their population synthesis model suggests that the most common outcome for the system is to evolve through Roche lobe overflow, followed by a common envelope phase and a merger, with $\sim65$--80\% of events ending as isolated sources. This conclusion is in good agreement with the observed 78\% fraction of isolated white dwarfs in the local population. The majority of the remaining 20--35\% of white dwarfs are in wide binaries. Even after factoring in the greater distance to our network of DA white dwarfs, which is between 10--50 farther than the local population, these resolved companions would be easily detected in our 0.04\arcsec$/$pix resolution of our UVIS images. We must also consider the possibility of an unresolved companion that contaminates the flux. \citetalias{Toonen17} expect only 0.5--1\% of the local population to consist of unresolved WD-MS companions, which agrees well with the observed fraction of 0.4\% in the local population. Of this small fraction, only late-type companions are of concern, as emission from other stars should be readily evident in our optical spectra and our multi-color photometry.


\section{Improvements Made in the Design of the Observing Program}\label{sec:changes}

The proof-of-concept \citetalias{Narayan16} analysis provided us with several valuable insights into how to reduce various sources of systematic bias that could impact our \emph{HST} photometry and ground-based spectroscopy programs. We used these insights to improve our methodology for our cycle 22 and cycle 25 programs. We provide a brief summary of the changes to our methodology below, focusing on how these improvements affect our analysis in \S\ref{sec:wdmodel}. The changes discussed here have been ordered by our assessment of their impact on the output SEDs that are the result of our forward modeling of the observations. While the differences caused by some of these changes taken individually are small, we emphasize that their effects are \emph{cumulative}.

\subsection{Addition of \textit{F275W} Observations to Constrain Extinction Due to Interstellar Dust}
\begin{figure*}[htpb]
    \centering
    \includegraphics[width=0.96\textwidth]{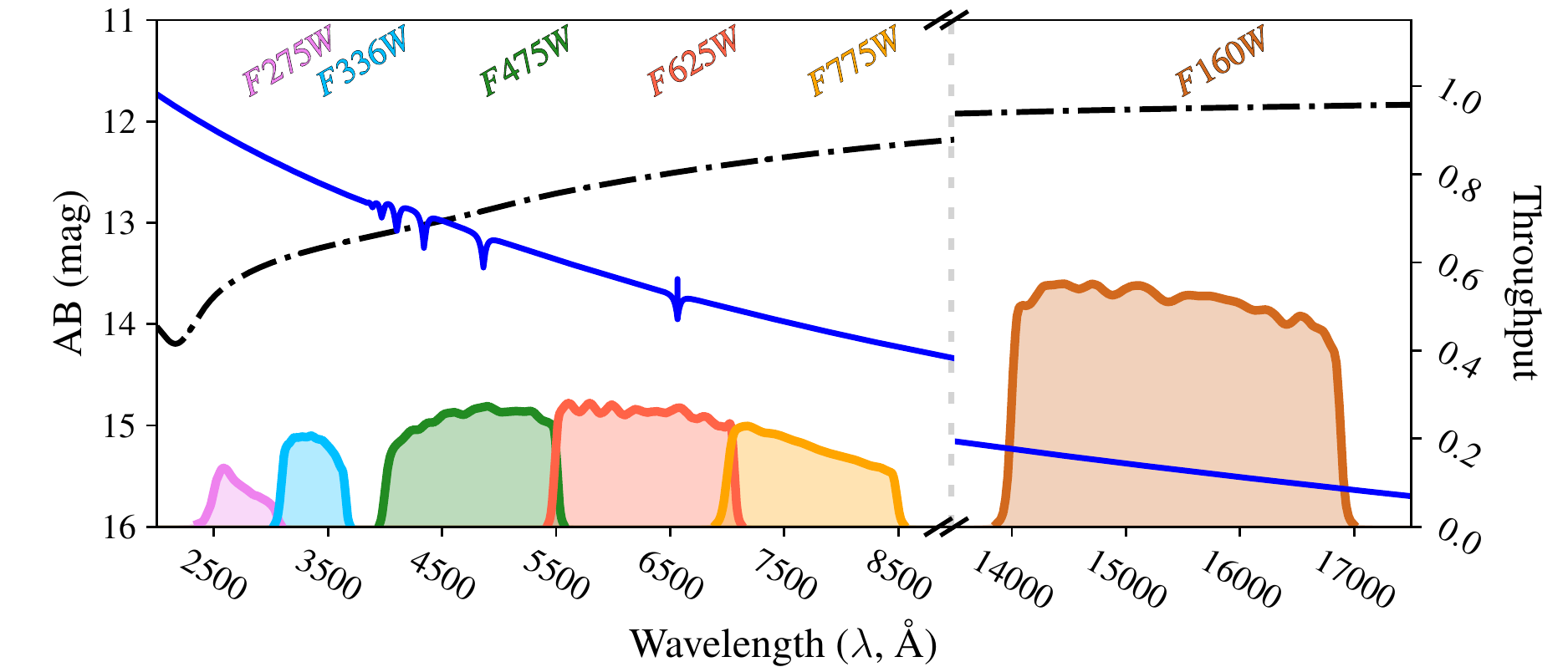}
    \caption{Throughput of the \emph{HST/WFC3} passbands (y-axis on right) used in our program, shown together with the SED of CALSPEC primary standard GD153 in AB mag (blue, y-axis on left). The \citetalias{Fitzpatrick99} transmission (dash-dotted black, y-axis on right) for $E(B-V) = 0.2$ with the canonical Milky Way $R_V = 3.1$ is also shown, as we expect our network of DA white dwarfs to be affected by dust along the line of sight.}
    \label{fig:passbands}
\end{figure*}

In cycle 20, we obtained photometry in \textit{F336W}, \textit{F475W}, \textit{F625W}, \textit{F775W}, and \textit{F160W}. In \citetalias{Narayan16}, we modeled the photometry and spectroscopy independently, using the spectroscopy to constrain $T_\text{eff}$ and $\log g$ and define the unreddened DA white dwarf model atmosphere. We ascribed the difference between the observed photometry and the unreddened synthetic photometry of the model photometry to extinction due to diffuse interstellar dust with $R_V = 3.1$. In cycle 22 we added observations in $F275W$ to improve our ability to determine $A_V$. While $F275W$ does not extend sufficiently far into the UV to cover the 2,100~\AA\ bump in the reddening law and strongly constrain $R_V$, it increases the lever arm in wavelength significantly, reducing the correlation between the inferred surface temperature of the white dwarf and the reddening. The response curves of the passbands\footnote{The specific \texttt{pysynphot} reference files used in this work are listed in \url{https://github.com/gnarayan/WDdata/blob/master/photometry/synphot_obscomponents.txt}, and included with the data. They match the master reference for \emph{HST/WFC3} maintained at \url{http://www.stsci.edu/hst/observatory/crds/SIfileInfo/pysynphottables/current_wfc3_throughput_html}\label{footnote:pbtransmission}} used in our program~\citep{Deustua16b}, the CALSPEC SED of GD153, and the \citetalias{Fitzpatrick99} transmission for $E(B-V) = 0.2$~mag are illustrated in Fig.~\ref{fig:passbands}.

\subsection{Additional Observations to Exclude Cosmic Rays}
Our exposure times were often hundreds of seconds, even in the bluest bands where our faint DA white dwarfs are brightest intrinsically, in order to achieve the requisite signal-to-noise ratio (S/N) for this program. Given these long exposure times and the rate of cosmic-ray events, we split our cycle 20 exposures up into at least two, but typically three repeat exposures per passband to avoid contamination. While fully mitigating cosmic rays requires an even higher number of repeats, we were constrained by the need to avoid overflowing the \emph{WFC3} data buffer before data downlink and to fit a large number of observations for a large number of targets into a modest number of orbits. 

While the low number of repeat exposures proved sufficient for most sources, \citetalias{Narayan16} found that the flux in the bias-corrected, flattened ``\texttt{flt}'' images corrected by the pixel area map (PAM) to be inconsistent in some of the cases with only two repeats. In these cases, it was not possible to determine what the true flux of the source was. In cycle 22, we ensured we obtained at least one additional exposure for any observations in cycle 20 with only two repeats, dithering exposures slightly to aid in cosmic ray and hot pixel rejection, and mitigate the effect of bad pixels. In principle, the process also allows better sampling of the point spread function (PSF), but given the limited number of exposures for image combination, this gain in resolution is not realized.

\subsection{Intra-cycle Monitoring of CALSPEC Primary Standards}
In the \citetalias{Narayan16} analysis we used the \emph{HST/WFC3} zeropoints reported by MAST, but discovered that there were unmodeled systematic differences between the UVIS1 and UVIS2 chips of \emph{WFC3} that we had to correct for. We decided to mitigate this systematic by observing the three CALSPEC primary standards over the duration of cycle 22 in all our program passbands, roughly contemporaneously with observations of our science targets. This allows us to estimate the zeropoint in each passband and tie our faint DA white dwarfs directly to the CALSPEC primary standards. As exposure times of the bright primary standards are short (less than 1 second in some passbands), we follow a recommendation from \citet{sahu2014} and use shutter blade `A' on \emph{WFC3} UVIS channel, rather than the default of alternating between blade `A' and `B' for consecutive exposures. This reduces shutter-induced vibrations and results in a $\sim10\%$ narrower PSF compared to our cycle 20 observations~\citep{Hartig08}.

\subsection{Use of Primary Imaging Data-products to Avoid \texttt{calwf3} Pipeline Systematics}\label{sec:primaryphot}
In \citetalias{Narayan16}, we used the combined ``\texttt{drz}'' images produced by the \texttt{multidrizzle} package, and available through MAST as inputs to the \texttt{SourceExtractor} photometry routine~\citep{sourceextractor}. Measuring the flux solely from the \texttt{drz} images missed instances where the flux measurements from the individual PAM-corrected \texttt{flt} images were in disagreement. Moreover, while the \texttt{multidrizzle} algorithm is designed to conserve flux, the growth curves of some point sources differed significantly from the median growth curve for all point sources on the \texttt{drz} image, but this behavior was not exhibited by the corresponding sources on the individual PAM-corrected \texttt{flt} images. 

For the cycle 22 data presented in \citetalias{Calamida18}, we reprocessed individual exposures through the \texttt{calwf3} pipeline, disabling the default auto-scaling of UVIS2 flux levels to match UVIS1. We applied a correction for charge transfer efficiency (CTE) losses to the \texttt{flt} images, producing ``\texttt{flc}'' images, which we photometered directly, bypassing \texttt{multidrizzle} entirely. Another advantage to this approach is that it avoids combining data taken at very different times during the same cycle directly, allowing us to account for any small changes in the system throughput over the duration of the cycle, by contemporaneously observing the three primary CALSPEC standards, GD71, GD153 and G191-B2B, at the same detector location as our standards. A simple measurement of the relative count rate is necessary to determine apparent magnitudes, thereby avoiding many potential sources of systematic error. 

Using the primary data products does incur a cost --- while the \texttt{drz} images created by \texttt{multidrizzle} undergo cosmic ray and hot pixel rejection, our \texttt{flc} images do not. We attempted to apply the cosmic ray rejection algorithm directly to the \texttt{flc} images but found this process could change the shape of the PSF, particularly in \textit{F625W} and \textit{F775W} where point sources are mistakenly classified as cosmic rays due to the narrow full width at half maximum. As the fraction of measurements of our program stars likely to affected by cosmic rays is small, we elected to not apply any cosmic ray rejection to the images, and instead account for outliers in our photometric catalogs in \S\ref{sec:photmodel}. The \texttt{drz} image combines the multiple dithers to subsample the PSF and improve the resolution across the field-of-view, but as noted previously, this gain is not realized given their limited number. It may be possible to adjust the numerous settings of the \texttt{multidrizzle} algorithm to eliminate instances where magnitudes measured from the \texttt{drz} images and the individual cosmic ray and hot-pixel rejected ``\texttt{crj}'' images disagree. For this work, we prefer to work directly with less processed data products presented in \citetalias{Calamida18} in order to avoid introducing systematic errors in the image processing which cannot easily be tracked back to a single exposure. 

\subsection{Optimization of Imaging to Reduce Detector Systematics and Improve Program Efficiency}\label{sec:imagingoptimazation}
In our cycle 20 observations the primary target was placed at the center of the UVIS1 chip, and we applied a post-flash of 12 electrons to increase the background. Charge-Transfer Efficiency (CTE) trails were visible on the \texttt{drz} images, and \citetalias{Narayan16} accounted for these trails by selecting a large enough aperture for photometry to include the majority of the flux. The target was placed at the center of the IR chip for observations in \textit{F160W}.

For the data obtained during cycle 22, we elected to place our target in the UVIS2 readout corner, on the `C' amplifier, to minimize the total number of pixels over which charge has to be transferred before reaching the readout registers and the A/D converter. While minimizing CTE losses benefits observations in all UVIS passbands, this choice particularly benefits observations in \textit{F275W} as the response of UVIS2 is 30\% higher than UVIS1 for wavelengths bluewards of 3,000~\AA. Furthermore, ghosting on the `C' amplifier is lower than the `A' and `D' amplifiers of the \emph{WFC3} UVIS channel. We maintained the post-flash of 12 electrons, as exposure times for the three primary standard images are only a few seconds, and the resulting sky background is low. It is in this regime that the post-flash is effective at mitigating CTE losses. Moving the target to the readout corner and applying corrections that model the CTE losses completely mitigate CTE-related systematic effects. Left uncorrected, these could cause an underestimate of the flux of our targets. 

Moving the primary target to the readout corner in cycle 22 had the additional advantage of allowing us to select just the center of the UVIS2-C512C-SUB subarray, rather than the UVIS1-FIX array used in cycle 20\footnote{The pixel coordinates and sizes of the \emph{WFC3} apertures are defined in \url{http://www.stsci.edu/hst/observatory/apertures/wfc3.html}.}. This significantly reduced the instrument readout and data transfer overhead at the cost of fewer sources on the frame with which to check relative photometry. We judged this trade off to be acceptable, given the small number of secondary sources on the frame in passbands bluer than $F625W$. Additionally, in cycle 22 we moved the target from the IR-UVIS-FIX aperture to the IR-FIX aperture as the former is affected by a row of bad pixels. While our reduction pipeline accounted for this row in the analysis of the cycle 20 data, we prefer to avoid systematic issues prior to image acquisition, rather than relying on data reduction procedures to mitigate their impact.

\subsection{AB as the Spectrophotometric Reference}\label{sec:vegaisawful} 
In \citetalias{Narayan16}, we reported magnitudes for our DA white dwarf stars relative to Vega. \citet[][hereafter, \citetalias{Bohlin2014}]{Bohlin2014} sets the normalization of the three primary standards SED relative to the Vega flux at 5,556~\AA\ (in air, or $5,557.5$~\AA\ in vacuum), $F^{\text{Vega}}(5,556$~\AA$) = 3.44\times10^{-9}$~ergs$\cdot$cm$^{-2}\cdot$\AA$^{-1}\cdot$s$^{-1}$ measured from \emph{HST/STIS} spectrophotometry. This monochromatic flux is a reconciliation of the \citet{Megessier95} flux in the visible with the \emph{Midcourse Space Experiment} (\emph{MSX}) mid-IR fluxes. It is the single tie-point from the CALSPEC primary standards to Vega, and the only location where consistency between the flux of Vega and the flux of the CALSPEC stars is guaranteed. The statistical uncertainty on this monochromatic measurement, and thus the absolute calibration of the primary standards, is 0.5$\%$. Any errors in this absolute calibration propagate to all wavelengths, and so does not affect the shape of the SEDs

There are, though, considerable \emph{systematic} uncertainties in the \emph{HST/STIS}-based SED model of Vega\footnote{Specifically the \citetalias{Bohlin2014} model for the SED of Vega included in CALSPEC: \href{ftp://ftp.stsci.edu/cdbs/calspec/alpha_lyr_stis_008.fits}{\texttt{alpha\_lyr\_stis\_008.fits}}.}. Some of these uncertainties arise from the flux calibration of the \emph{HST/STIS} spectra, which exhibit considerable saturation. \citetalias{Bohlin2014} estimate an additional 0.2--0.5\% systematic uncertainty with wavelength in the visible. There are likely additional systematic errors in the calibration of Vega in the infrared due to the presence of a dust disk. \citetalias{Bohlin2014} report a $ \sim 1$\% discrepancy in the IR flux of Vega from \emph{IRAC} measurements, despite including a 3-component model for the dust disk (with its own uncertainties) that has not been resolved to date\footnote{The best current estimate of the systematic uncertainty of the CALSPEC flux system is determined from the three DA white dwarf primary standards, and is reported in the covariance matrix included with the CALSPEC data products:~\href{http://www.stsci.edu/ftp/cdbs/calspec/WDcovar_001.fits}{\texttt{WDcovar\_001.fits}}}. Establishing high-precision spectrophotometric standards, only to report their measurements with respect to a spectrophotometric reference that itself suffers from known systematic errors, is contrary to the goals of our work.

Ideally, we would report an \emph{absolute} flux in units such as janskys, however our experiment only establishes SEDs \emph{relative} to the three CALSPEC primary standards --- our work \emph{cannot} improve on the \citetalias{Bohlin2014} 0.5\% statistical uncertainty on the absolute calibration of the primary standards. Moreover, absolute fluxes in janskys are not as widely used as magnitudes for optical measurements, and are not accepted as inputs or reported as outputs by many image processing pipelines~\citep[e.g. \texttt{photpipe},][]{photpipe}, and are therefore inconvenient for many purposes. We do not report Vega-based magnitudes in this work despite their widespread use in the white dwarf literature, and instead report AB magnitudes with zeropoints determined from the three CALSPEC standards. This choice implicitly assumes that the primary standards are on the same AB flux scale~\citep{OkeGunn83}. \citet{Fukugita96} defines the relationship between AB magnitudes and physical fluxes such that a source with constant spectral flux density per unit frequency $F_{\nu} = 3,631$~Jy~ has magnitude 0 in all passbands. The AB source and Vega are not merely interchangeable spectrophotometric references that define different photometric systems. The AB source is purely a construct, whereas Vega is a time-variable source on the sky with a complex SED that suffers from wavelength-dependent systematic errors that are likely irreducible. Fundamentally, the DA white dwarfs are better spectrophotometric references than Vega itself. 

It is not possible to validate our assumption that the CALSPEC flux scale is an absolute scale within the framework developed in \S\ref{sec:wdmodel}. Several ground-based photometric catalogs also assume that the CALSPEC flux scale is an accurate absolute flux scale, and we can only measure a relative offset in each passband. Quantifying any error in the absolute calibration of the CALSPEC SEDs requires independent data, and we discuss a framework for establishing such absolute flux standards in \S\ref{sec:futurework} using a combination of the DA white dwarf atmospheres themselves and laboratory standards. With this caveat, AB magnitudes have the great benefit of being directly related to physical units and remaining compatible with widely-used photometry routines used to determine the zeropoints of optical and infrared images, while avoiding any of the systematic issues that arise with Vega as a spectrophotometric reference.

\subsection{Additional Spectroscopy for Testing the Consistency of our Analysis}\label{sec:multispec}
As with our cycle 20 targets, we obtained spectra for our cycle 22 targets using the Gemini Multi Object Spectrograph~\citep[GMOS,][]{Hook04} on both Gemini-North and Gemini-South telescopes. While we reduced our slit width from 1.5\arcsec\ to 1\arcsec\ for Gemini/GMOS spectra obtained for cycle 22 targets as the excellent seeing on Mauna Kea warranted the narrower slit, we made no other major changes to our GMOS observing program. In addition to the GMOS spectra, we obtained a second spectrum for a subset of our targets using the Blue Channel spectrograph on the MMT in cycle 20. We found these additional spectra to be invaluable, as they provided a cross-check on the inference of the intrinsic parameters of the DA white dwarfs presented in \citetalias{Narayan16}. While the MMT/Blue Channel spectra are typically lower resolution than the Gemini/GMOS spectra (2--3~\AA/pix vs 0.92~\AA/pix) and lower S/N on average as we had less allocated observing time, they cover a longer range in wavelength. This allows us to model the H$_{\alpha}$ feature which is truncated in our Gemini/GMOS spectra. 

The MMT/Blue Channel spectra can be flux calibrated more accurately than the Gemini/GMOS spectra due to a combination of factors. i) On MMT/Blue Channel, the trace is dispersed across a single CCD readout by a single amplifier, whereas the trace is dispersed across 3 CCDs, each readout by 4 amplifiers on Gemini/GMOS, requiring careful gain-matching and ad-hoc flux adjustment corrections. ii) The Gemini/GMOS observations were executed in queue mode, and a standard spectrum was not guaranteed to be observed on the same nights as our targets. With MMT/Blue Channel, we always obtained multiple observations of at least one, and typically 2--3 blue spectrophotometric standards, spaced across the night for calibration of our DA white dwarfs. iii) We typically obtained a quartz flat and HeNeAr lamp spectrum at each pointing with MMT/Blue Channel. Finally, we obtained several spectra of the primary standard GD71 with the Goodman spectrograph on the Southern Astrophysical Research telescope (SOAR) to flux calibrate spectra of our cycle 25 targets. These high S/N spectra are useful as they not only test the internal consistency of the analysis, but also test the consistency of the SED of GD71, inferred independently from each spectrum, with the CALSPEC SED of GD71.

\section{Deriving Cross-cycle Photometric Catalogs}\label{sec:phot}

In this section, we develop a model to derive the apparent AB magnitudes of our program stars from their instrumental measurements. Combining \emph{HST} photometric measurements from any multi-cycle program requires care as the detector sensitivity evolves non-monotonically with time. In \citetalias{Narayan16}, we asserted that as our measurements were from a single cycle 20 program and the evolution of the sensitivity over the cycle was likely to be small, we could rely on the zeropoints measured by the ongoing \emph{HST} Servicing Mission Observatory Verification (SMOV) calibration program and reported on MAST. \citetalias{Narayan16} unearthed systematic differences between the sensitivity of UVIS1 and UVIS2 that were unmodeled at the time; even if the zeropoints were precisely measured, they were not accurate for the UVIS2 chip. These sensitivity differences have now been accounted for by chip-dependent calibrations in the \texttt{calwf3} pipeline and MAST data products~\citep{Deustua16, Deustua17}. 

Nevertheless, the fiducial zeropoints provided by Space Telescope staff are time-averaged and are not suitable for experiments that require high-precision photometric calibration of multi-cycle imaging such as ours. This shortcoming motivated us to add intra-cycle 22 monitoring of the three primary standards to our program, allowing us to tie our targets directly to the CALSPEC system without any intermediate transformations. In order to combine our data from cycle 20 and 22, we must account for the stochastic and systematic effects in \emph{HST/WFC3} while modeling the primary standard observations together with our DA white dwarf observations. We begin by examining these effects from independent archival \emph{HST} datasets, and then develop a model that accounts for these effects in order to derive the calibrated photometry of our DA white dwarfs. 

\subsection{Accounting for \emph{HST/WFC3} Sensitivity Evolution}\label{sec:zp_evolve}
\begin{figure}[htpb]
\centering
  \includegraphics[width=0.47\textwidth]{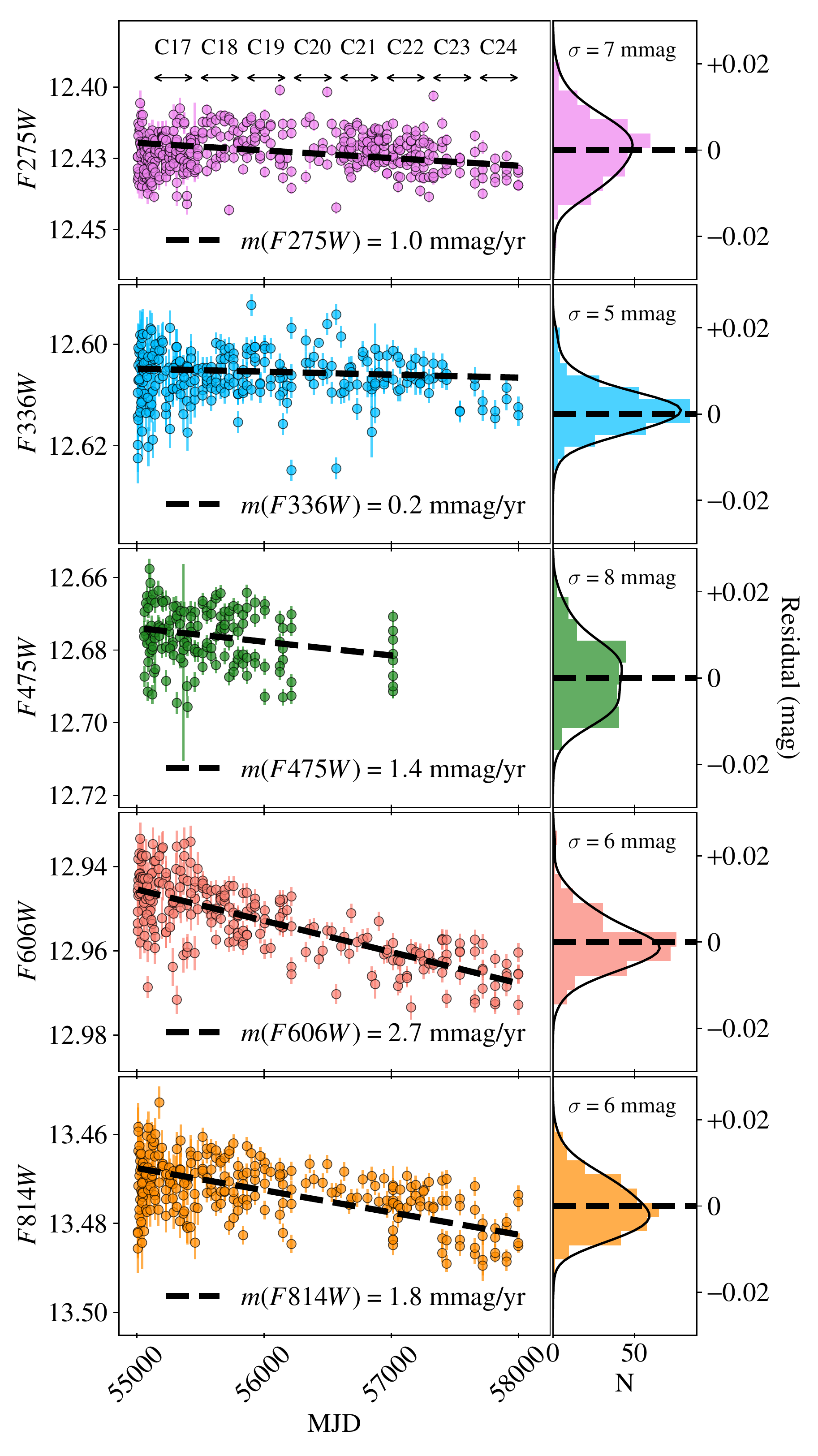}
\caption{Sensitivity evolution of \emph{HST/WFC3} UVIS2 determined by STScI staff from repeated measurements of the standard GRW70, assuming a constant zeropoint. The star has been monitored starting with cycle 17, when \emph{WFC3} was added to the \emph{HST} payload as part of the SMOV program. \emph{HST} cycle numbers corresponding to MJD ranges are indicated in the top panel. The mean trend in each passband can be characterized as a linear decline of typically $< 2$~mmag yr$^{-1}$. The standard deviation of the residuals after removing the linear trend is 5--8~mmag. This dispersion is larger than the photometric uncertainties, and is readily apparent among even closely-spaced observations. We note that the GRW70 photometry is determined from images reduced by the \texttt{calwf3} pipeline, and it is likely that some of the dispersion is a result of the sub-optimal image processing. In addition, there are sources of dispersion intrinsic to \emph{HSTWFC3}, such as spacecraft breathing (optics), charge persistence (detector), instabilities in the amplifiers (electronics), and flat-fielding errors as a function of time (image processing). This figure illustrates the extent and timescale of secular response variations. While we do not use the GRW70 data directly, we account for this dispersion in the model for our observations in \S\ref{sec:photmodel}. We note that the standard error of the mean can be much smaller than the dispersion of the individual measurements. 
}\label{fig:zp_evolve}
\end{figure}

Tracking the change in sensitivity of an instrument over time requires the continuous monitoring of an external reference with the same configuration. GRW+70\degree5824 (hereafter, GRW70) was originally established as a standard by \citet{Oke90} (itself concerned with establishing a faint network of spectrophotometric stars --- 5.5 magnitudes brighter in \textit{V} on the mean than the objects in this work). Space Telescope Staff have been monitoring the white dwarf standard GRW70 in several \emph{HST/WFC3} UVIS passbands~\citep{Shanahan17} since cycle 17. Using archival \texttt{calwf3} processed images, we measured the flux of GRW70 using a 10~pix radius from exposures where the star was located on the same subarray as our targets. The dithers between different observations are small, so we do not expect errors in flat-fielding with position. Fig.~\ref{fig:zp_evolve}, show the measured magnitudes of GRW70 over time, assuming constant sensitivity. 

The sensitivity function inferred from the GRW70 observations in all the passbands is complex, exhibiting a steep increase right after launch, when \emph{WFC3} was added to the \emph{HST} instrument payload, likely due to outgassing. Over time, the sensitivity decreases, but there are sharp changes within each cycle that do not correlate strongly with any physical parameters of the spacecraft. We model the long-term sensitivity decrease with a simple linear trend. The standard deviation of the residuals about the linear trend is 5--8 mmag in the observed passbands, 2--3 times larger than the photometric uncertainties. A single measurement may have a low error, but may still differ significantly from a subsequent measurement as the \emph{repeatability} of the \emph{HST/WFC3} detector is not captured by the photon noise. This excess scatter motivates the introduction of an intrinsic dispersion in \S\ref{sec:photmodel} to explain our \emph{HST/WFC3} observations. 

We model the additional 2--4 mmag of dispersion as intrinsic to \emph{HST/WFC3}. This is reasonable as this additional dispersion is a result of a combination of spacecraft breathing, persistence, instability in the amplifier electronics, and flat-fielding errors as a function of time. We cannot disentangle these effects with presently available data. While the linear decrease in the sensitivity ($< 2$~mmag/yr) and the intrinsic dispersion are small, both effects must be modeled in order to combine our cycle 20 and cycle 22 measurements. Unfortunately, measurements of GRW70 are not available in all of our program's passbands. Given the limitations of this dataset, we do not use the GRW70 observations to set zeropoints when deriving apparent magnitudes from our multi-cycle observations, but we do include an intrinsic dispersion to account for measurement repeatability.

\subsection{Modeling Cycle 20 and 22 Photometry and Tying Measurements to CALSPEC}\label{sec:photmodel}
We construct a hierarchical model to describe our instrumental measurements from cycle 20 and 22 in each passband. The model parameters are i) the zeropoint in each passband, ii) an intrinsic dispersion in each passband to model the scatter in the measurements that is in excess of the estimate from the photon noise, iii) a single offset between the cycle 20 and cycle 22 observations in each passband, iv) the number of degrees of freedom of a Student's $t$-distribution, used to model the small fraction of instrumental measurements that are afflicted by cosmic rays, and are outliers, and v) the desired apparent magnitudes of each source. We discuss each of these parameters below:

\begin{enumerate}
\item Determining the zeropoint in each passband $Z$ requires stars with measured instrumental magnitudes and \emph{known} apparent magnitudes. We determine the zeropoints using our observations of three CALSPEC primary standards, and their synthetic magnitudes derived from the CALSPEC SEDs\footnote{The specific CALSPEC SED models used are\\GD71:~\href{ftp://ftp.stsci.edu/cdbs/calspec/gd71_mod_010.fits}{\texttt{gd71\_mod\_010.fits}},\\GD153:~\href{ftp://ftp.stsci.edu/cdbs/calspec/gd153_mod_010.fits}{\texttt{gd153\_mod\_010.fits}},\\G191-B2B:~\href{ftp://ftp.stsci.edu/cdbs/calspec/g191b2b_mod_010.fits}{\texttt{gd191b2b\_mod\_010.fits}}.\label{footnote:standards}}. As the CALSPEC primary standards are also DA white dwarfs with very similar colors to our program stars, and as they were observed contemporaneously with the same instrumental configuration, we can apply the zeropoint derived from the three primary standards directly to our program stars.

\item As discussed in \S\ref{sec:zp_evolve}, the observed scatter of the multi-cycle GRW70 measurements is underestimated by the photometric errors. This additional scatter likely arises from multiple effects, and it is not possible to disentangle each of them. The residuals between the GRW70 observations and a linear trend to account to zeropoint evolution over time are normally distributed. This motivates the introduction of a single parameter to model the excess dispersion in each passband $\sigma_{\text{int}}$. While this intrinsic dispersion magnitude can be constrained from instrumental measurements of all the stars, we must also account for any systematic difference in the instrumental response between \emph{HST} cycles.

\item In optimizing our observing program between cycle 20 and cycle 22 (see \S\ref{sec:imagingoptimazation}), we changed the on-detector location of our targets from UVIS1-FIX to UVIS2-C512-C-SUB, and from IR-UVIS-FIX to IR-FIX. Consequently, we must account for the difference in zeropoint caused by the change in the system response arising from the change in location, and from any evolution in the overall throughput between cycle 20 and 22. We introduce a single parameter in each passband $\Delta Z_{\text{C20}}$ to model this offset. This offset can only be constrained using stars that were observed in both cycle 20 and 22. In particular, it is not necessary to model the instrumental observations in \textit{F275W}, as all these measurements were obtained in cycle 22. 

\item We measure the brightness of our sources from the individual PAM and CTE-loss corrected \texttt{flc} images, but do not apply any cosmic ray rejection, as we found this affects the shape of the PSF (see \S\ref{sec:primaryphot}). Without cosmic ray rejection, some fraction of the measurements will be impacted by cosmic rays within the aperture, and assuming the measurements are normally distributed would bias the estimate of the zeropoints. Our model must therefore account for non-normally distributed outliers. \citetalias{Calamida18} uses sigma-clipping of each star's measurements and a bi-weight estimator to determine mean instrumental magnitudes for the three primary standards. \citet{Hogg10} cautions against the sigma clipping procedure as it does not optimize an objective function, and the results are dependant on the initial guess. 

We express the probability density of the data given the model parameters as a Student's $t$-distribution, as an alternative to the normal distribution~\citep[see Ch. 17.2 of ][]{gelmanbda04}. In the limit of the number of degrees of freedom $\nu$ growing to infinity, the Student's $t$-distribution is equivalent to the normal distribution, but for smaller values, the $t$-distribution has heavier tails, allowing robust inference in the presence of outliers. As the fraction of outliers is small, and there nearly 200 observations across all stars in each passband, we leave the number of degrees of freedom as a free hyperparameter of the model.

\item Finally, in addition to these four hyperparameters of our hierarchical model in each passband, we introduce a parameter for the apparent magnitude of each of our program DA white dwarf stars. These are the latent variables of the model that are needed to calibrate the SED of each program star in \S\ref{sec:wdmodel}.
\end{enumerate}

The probability density of our observations of the $p$th CALSPEC primary standard given the model parameters is the likelihood function:
\begin{equation}
    P(\bm{\widehat{m}_{p}} |\, m_p,Z,\sigma^2_\text{int}, \nu ) = \prod_{i=1}^{N_{p}} T(\widehat{m}_{p,i} |\, m_p - Z, (\sigma^2_\text{int} + \widehat{\sigma}^2_{p,i})^{-1}, \nu )
\label{eqn:likelihoodprimary}
\end{equation}
where $\widehat{m}_{p,i}$ is the $i$th observed instrumental magnitude with photometric measurement error described by an estimated variance $\widehat{\sigma}^2_{p,i}$ for CALSPEC standard $p$ from the set of $N_p = 3$ primary standards. We discuss the derivation of the synthetic magnitudes in each passband $m_p$ from the SEDs of the CALSPEC primary standards in the following section (\S\ref{sec:synphot}). $T(y|\, \mu, \lambda, \nu)$ generically denotes a Student's $t$-distribution in $y$ centered on $\mu$ with inverse scale parameter $\lambda$ and $\nu$ degrees of freedom. As we began monitoring the primary standards in cycle 22, their observations only set the zeropoint $Z$, and do not constrain the zeropoint offset between cycle 20 and 22 $\Delta Z_\text{C20}$. The synthetic magnitudes of the CALSPEC primary standards are determined for an infinite aperture, and the zeropoints implicitly incorporate the aperture correction from a radius of 7.5~pix in the UVIS bands and a radius of 5~pix in the \textit{F160W}. 

In the limit of $\nu$ growing to infinity, the Student's $t$-distribution reduces to a normal distribution. Conditional on knowing the value of $\sigma_\text{int}$, the zeropoint $Z$ can be estimated from the conventional weighted mean difference between the apparent magnitudes and the mean instrumental magnitudes of the three CALSPEC primary standards:
\begin{equation}
    \hat{Z} = \Big(\sum_{p=1}^3 \sum_{i=1}^{N_p} w_{p,i}\Big)^{-1} \sum_{p=1}^3 \sum_{i=1}^{N_p} w_{p,i} \cdot (m_p - \widehat{m}_{p,i} )
\label{eqn:zeropoint}
\end{equation}
where the weights are the inverse variances: $w_{p,i}^{-1}  = \sigma_\text{int}^2 + \widehat{\sigma}_{p,i}^2$. Within our model, the measurements of the secondary standards inform us of both the values of $\sigma_\text{int}$ and finite $\nu$, which in turn influence our posterior estimate of $Z$, and its uncertainty. Hence, our approach jointly models the primary and secondary standards in a single coherent Bayesian model, with the marginal posterior estimate of $Z$ obtained by computationally marginalizing the full posterior, Eqn.~\ref{eqn:posterior}, over all other parameters using MCMC. This process fundamentally ties the \emph{HST} photometry of our DA white dwarfs to a flux scale defined by the weighted mean of the three CALSPEC primary standards. This pan-chromatic \emph{HST} photometry is used to normalize our inferred SED models in \S\ref{sec:wdmodel}.  

For the secondary standards, we model the probability density of our observations given the parameters with the likelihood function: 
\begin{equation}
\begin{split}
    P(\bm{\widehat{m}_{s}} & |\, m_s, Z, \Delta Z_\text{C20}, \sigma^2_\text{int}, \nu ) = \\
    & \prod_{i=1}^{n^\text{C22}_s} T(\widehat{m}_{s,i} |\, m_s - Z, (\sigma^2_\text{int} + \widehat{\sigma}^2_{s,i})^{-1}, \nu ) \\ 
    \times & \prod_{j=1}^{n^\text{C20}_s} T(\widehat{m}_{s,j} |\, m_s - Z - \Delta Z_\text{C20}, (\sigma^2_\text{int} + \widehat{\sigma}^2_{s,j})^{-1}, \nu ) \\
\end{split}
\label{eqn:likelihoodsecondary}
\end{equation}
where $\widehat{m}_{s,i}$ is the $i$th observed instrumental magnitude with photometric measurement error described by an estimated variance $\widehat{\sigma}^2_{s,i}$ for a DA white dwarf star $s$ from the set of $N_S$ standards $\{s : s = 1,...,N_S\}$ with $n^\text{C20}_s$ observations in cycle 20 and $n^\text{C22}_s$ observations in cycle 22. The latent (true) magnitude of this star $m_s$ is related to the instrumental magnitudes in cycle 22 through the additive zeropoint $Z$, and the cycle 22 zeropoint is related to the cycle 20 zeropoint through an additive offset $\Delta Z_\text{C20}$. The intrinsic dispersion $\sigma_\text{int}$ accounts for the failure of the photometric uncertainties $\widehat{\sigma}_{s,i}$ to capture the full variance of the data $\widehat{m}_{s,i}$. 

\begin{figure}[tpb]
\centering
\includegraphics[width=0.47\textwidth]{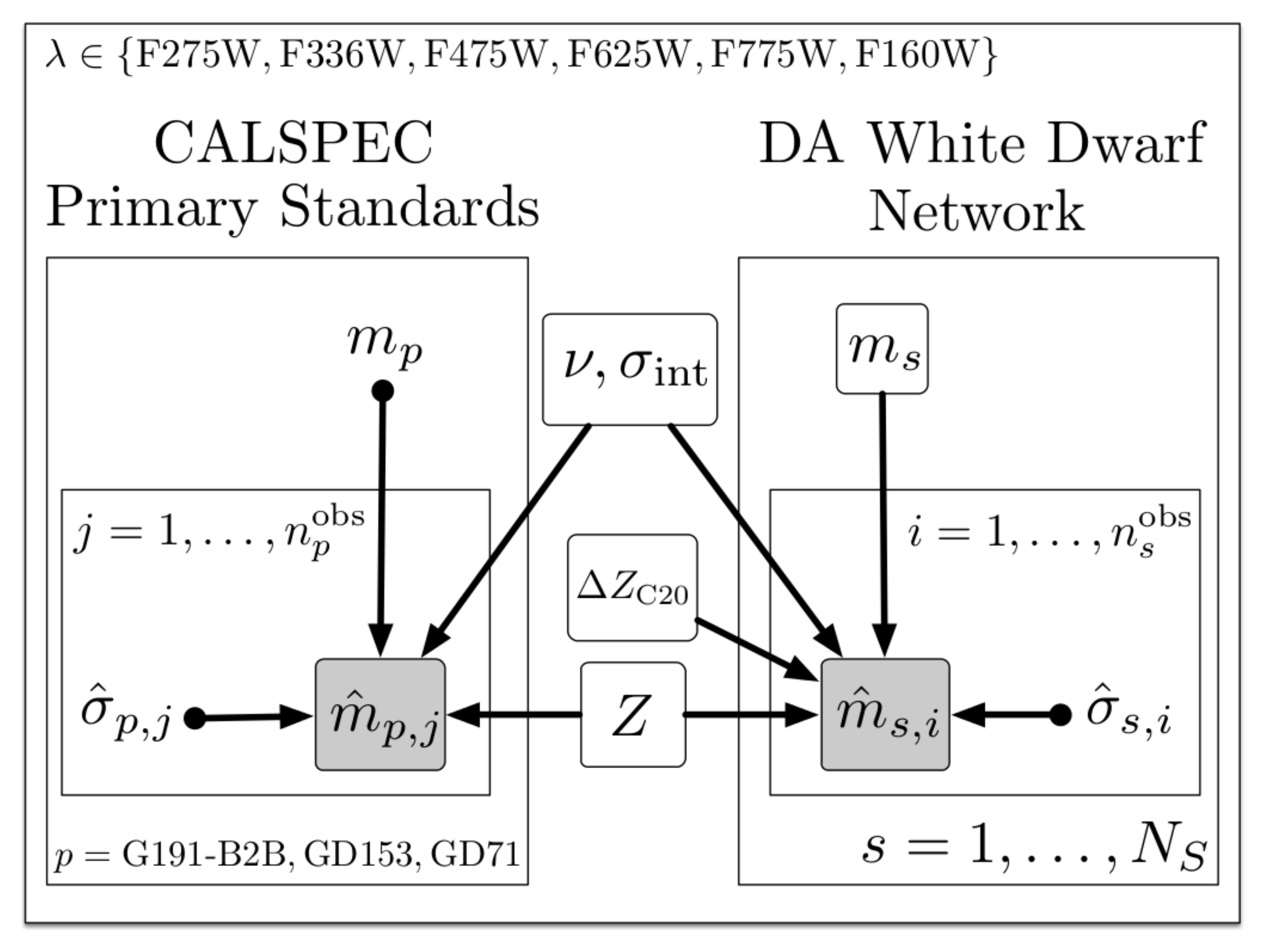}
\caption{Directed acyclic graph depicting the hierarchical model for our instrumental photometry. Clear rounded rectangles denote model parameters, while shaded rounded rectangles denote measurements. Arrows that originate at circles denote inputs that are combined with the measurements. Flat rectangular ``plates'' with subscripts indicate the product of likelihoods over the indicated variables $\{i, j, p, s, \lambda \}$. The instrumental magnitudes $\widehat{m}$ and their uncertainties $\widehat{\sigma}$ for the CALSPEC primary standards $p$ and the DA white dwarfs $s$ are modeled by a Student's $t$-distribution with $\nu$ degrees of freedom about the difference between each star's apparent magnitudes $m$ and the zeropoint $Z$. The synthetic apparent magnitudes of the CALSPEC primary standards $m_p$ set the zeropoint, which is transferred to determine the apparent magnitudes of the DA white dwarf network stars $m_s$. The intrinsic dispersion $\sigma_\text{int}$ accounts for the photometric measurements underestimating the variance of the measurements, while the zeropoint offset $\Delta Z_\text{C20}$ accounts for the difference in the zeropoint between cycle 20 and cycle 22 from changing the detector location and time-evolution of the sensitivity.}\label{fig:nutella}
\end{figure}

We depict the model in Fig.~\ref{fig:nutella} as a directed acyclic graph, a probabilistic graphical representation of our hierarchical Bayesian model. The probabilistic graphical model illustrates how the unknown apparent magnitudes $m_s$ of each of our program DA white dwarf stars (labeled by an index $s$) are related to the measurements of the CALSPEC primary standards (labeled by an index $p$) and the model hyperparameters in each passband: the zeropoint $Z$, the offset between cycle 20 and cycle 22 $\Delta Z_\text{C20}$, the intrinsic dispersion $\sigma_\text{int}$ and the number of degrees of the Student's $t$-distribution used to model photometric outliers.

\subsection{Synthetic Photometry}\label{sec:synphot}
Evaluating the likelihood of the hierarchical model described in the previous section requires the magnitudes of the three CALSPEC primary standards in our program passbands. These are derived using synthetic photometry of the CALSPEC SEDs (see footnote \ref{footnote:standards}) through the model of the transmission for each passband (see footnote \ref{footnote:pbtransmission}). We briefly summarize the synthetic photometry procedure here. The same procedure is used in \S\ref{sec:wdmodel} when determining the SED of each DA white dwarf using the observed spectroscopy and the apparent magnitudes derived in this section.

The synthetic flux of a source with spectral flux density $F^{p}(\lambda)$ through the photon response function of each of the \emph{HST/WFC3} passbands $R(\lambda)$ is defined as:
\begin{equation}
    \langle F^{p}_{\lambda} \rangle = \frac{\int_{0}^{\infty} \lambda \cdot F^{p}(\lambda) \cdot R(\lambda) \cdot d\lambda}{\int_{0}^{\infty} \lambda \cdot R(\lambda) \cdot d\lambda } 
    \label{eqn:synflux}
\end{equation}

The apparent magnitudes of the primary standards $m_p$ are inputs derived from the ratio of the synthetic flux of their CALSPEC SEDs with the synthetic flux of the spectrophotometric reference: 
\begin{equation}
    m_p = -2.5 \log\left( \frac{\int_{0}^{\infty} \lambda \cdot F^{p}(\lambda) \cdot R(\lambda) \cdot d\lambda}{\int_{0}^{\infty} \lambda \cdot F^{\text{AB}}(\lambda) \cdot R(\lambda) \cdot d\lambda } \right)
    \label{eqn:synphot}
\end{equation}
where $F_{\lambda}^{\text{AB}}$ is the spectral flux density of the fiducial AB source expressed in ergs cm$^{-2}$ s$^{-1}$ \AA$^{-1}$. Throughout the text and the data products provided together with this work, synthetic magnitudes are computed using Eqn.~\ref{eqn:synphot}, consistent with our convention of reporting the observed spectral flux density per unit wavelength $F_\lambda$. $F_{\lambda}^{\text{AB}}$ is related to $F_{\nu}^{\text{AB}} = 3,631$~Jy by:
\begin{equation}
  F_{\lambda}^{\text{AB}} = F_{\nu}^{\text{AB}} \cdot \frac{\nu^{2}}{c}
  \label{eqn:jacobian}
\end{equation}
for $\lambda = c/\nu$. 

With Eqn. \ref{eqn:jacobian}, it is also possible to define the synthetic flux following \citet[][]{Koornneef86} as:
\begin{equation}
    \langle F^{p}_{\nu} \rangle = \frac{\int_{0}^{\infty} \nu^{-1} \cdot F^{p}(\nu) \cdot R(\nu) \cdot d\nu}{\int_{0}^{\infty} \nu^{-1} \cdot R(\nu) \cdot d\nu } 
    \label{eqn:synfluxfreq}
\end{equation}
and the synthetic magnitude as 
\begin{equation}
\begin{split}
    m_p & = -2.5 \log\left( \frac{\int_{0}^{\infty} \nu^{-1} \cdot F^{p}(\nu) \cdot R(\nu) \cdot d\nu}{\int_{0}^{\infty} \nu^{-1} \cdot F^{\text{AB}}(\nu) \cdot R(\nu) \cdot d\nu } \right) \\
        & = -2.5 \log\left( \frac{\int_{0}^{\infty} \nu^{-1} \cdot F^{p}(\nu) \cdot R(\nu) \cdot d\nu}{F^{\text{AB}} \cdot \int_{0}^{\infty} \nu^{-1} \cdot R(\nu) \cdot d\nu } \right) \\
        & = -2.5 \log \left( \frac{ \langle F^{p}_{\nu} \rangle }{1~\text{ergs} \cdot \text{cm}^{-2} \cdot \text{s}^{-1} \cdot \text{Hz}^{-1} } \right) -48.60~\text{mag}
\end{split}
\label{eqn:synphotnu}
\end{equation}
where the last simplification to the form provided by \citet{OkeGunn83} is possible as the spectral flux density of the AB standard is constant per unit frequency, i.e. $ F^{\text{AB}}_{\nu} \equiv 3.631 \times 10^{-20}$~ergs$\cdot $cm$^{-2} \cdot $s$^{-1} \cdot$Hz$^{-1}$. If the synthetic flux is expressed in janskys, the appropriate zeropoint is instead $2.5\log \left( \frac{3,631~\text{Jy}}{1~\text{Jy}} \right) = 8.90$~mag.

Eqn. \ref{eqn:jacobian} holds for all monochromatic values of the flux density, and is also valid for the total flux through a passband, i.e., the integral of the flux density in Eqn. \ref{eqn:synphot} when computed at the ``pivot'' wavelength $\lambda_{\mathrm{pivot}}$~\citep{Koornneef86}:
\begin{equation}
  \lambda_{\mathrm{pivot}} = \sqrt{ \frac{\int_{0}^{\infty} \lambda \cdot R(\lambda) \cdot d\lambda}{\int_{0}^{\infty} R(\lambda)/\lambda \cdot d\lambda}}
  \label{eqn:pivot}
\end{equation}
The pivot wavelength is a characteristic of the passband, and can be used to convert our AB-based magnitudes to the synthetic flux derived from a source reported with spectral flux density per unit wavelength $\langle F_{\lambda} \rangle$ without knowledge of the underlying source SED. The pivot wavelength differs from the effective wavelength $\lambda_{\mathrm{eff}}$:
\begin{equation}
  \lambda_{\mathrm{eff}} = \frac{\int_{0}^{\infty} \lambda^{2} \cdot F^{p}(\lambda) \cdot R(\lambda) \cdot d\lambda}{\int_{0}^{\infty} \lambda \cdot F^{p}(\lambda) \cdot R(\lambda) \cdot d\lambda }
  \label{eqn:effwave}
\end{equation}
which is source-dependent and is the mean wavelength of photons detected through a passband. The effective wavelength is necessary when comparing apparent magnitudes of a source observed through a passband directly with the calibrated model SED of the source.

\subsection{The Probability Density of the \emph{HST} Observations}\label{sec:photprob}
The full probability density of the data $\mathcal{D} \equiv \{\bm{\widehat{m}_{p}},\bm{\widehat{m}_{s}} \}$ given the model parameters $\bm{\Phi} \equiv \{ \{\bm{m_s}\}, Z, \Delta Z_\text{C20}, \sigma^2_\text{int}, \nu\}$ and the CALSPEC magnitudes $\{m_p\}$ is the product of the likelihoods for the DA white dwarf network stars (Eqn.~\ref{eqn:likelihoodprimary}) and the primary standards (Eqn.~\ref{eqn:likelihoodsecondary}): 
\begin{equation}
\begin{split}
    P(\mathcal{D} |\,\bm{\Phi},\{m_p\}) = &\prod_{p=1}^{N_p} P(\bm{\widehat{m}_{p}} |\, m_p,Z,\sigma^2_\text{int}, \nu ) \\
    \times & \prod_{s=1}^{N_S} P(\bm{\widehat{m}_{s}} |\, m_s, Z, \Delta Z_\text{C20}, \sigma^2_\text{int}, \nu ) \\
\end{split}
\label{eqn:likelihood}
\end{equation}

\subsection{Priors}
We model the prior as separable functions on each of the model parameters. We follow \citet{gelman2006} and use weakly informative priors on all the parameters of our hierarchical model. We use a uniform prior on the apparent magnitudes $m$ for 8~mag $ \le m \le $25~mag. This weakly-informative prior spans a large enough range to encompass all of the observed magnitudes of the primary and program standards in all of our observed passbands. We use a normal distribution, denoted by $N(\mu, \sigma)$ with standard deviation $\sigma = 1$~mag and mean $\mu$ equal to the average difference of the input apparent magnitude and the measured instrumental magnitudes for the three primary standards $\langle m_p - \widehat{m}_{p,j} \rangle_{p,j}$. This prior is only weakly informative as the true zeropoints have uncertainties on the order of a few millimag, rather than width of $1$~mag we have used in our prior. We use a $N(0, 1~\text{mag})$ distribution as the prior on the zeropoint offset between cycle 20 and 22 $\Delta Z_{C20}$, which is weakly informative as our analysis in \S\ref{sec:zp_evolve} constrains the change to be at most a few millimag with similar uncertainties. We use a half-Cauchy distribution with $\beta = 1$~mag on the intrinsic dispersion $\sigma_\text{int}$, as this quantity must always be positive and is known to be at most a few millimag from our analysis in \S\ref{sec:zp_evolve}. Finally, we employ a half-Cauchy distribution with $\beta = 5$ on the number of degrees of freedom of the Student's $t$-distribution $\nu$, as this parameter must always be positive, and we expect the fraction of outliers to be small. Assuming prior independence of all the parameters, the full joint prior distribution is the product of the marginal priors on each of the parameters:
\begin{equation}
\begin{split}
    m_s \, \sim & \, U(8 \; \text{mag}, 25 \; \text{mag}) \\
    Z \, \sim & \, N(\mu = \langle m_p - \widehat{m}_{p,j} \rangle_{p,j}, \sigma = 1 \; \text{mag}) \\
    \Delta Z_{\text{C20}} \, \sim & \, N(\mu = 0~\text{mag}, \sigma = 1 \; \text{mag}) \\
    \sigma_\text{int} \, \sim & \, HC(x_0 = 0~\text{mag},\, \beta = 1~\text{mag}) \\ 
    \nu \, \sim & \, HC(x_0 = 0,\, \beta=5) \\
    P(\bm{\Phi}) \, = & \, P(Z) \cdot P(\Delta Z_{\text{C20}}) \cdot P(\sigma^2_\text{int}) \cdot P(\nu ) \cdot \prod_{s=1}^{N_S} P(m_s).
\end{split}
\label{eqn:priors}
\end{equation}

As there are alternative parameterizations in the literature, we define the probability density of a half-Cauchy random variable $x\sim HC(x_0, \beta)$ as 
\begin{equation}
       P(x | \, x_0, \beta) = \frac{2}{\pi} \cdot \left[ \beta \cdot \left(1 + \frac{(x-x_0)^2}{\beta^2} \right) \right]^{-1}
\end{equation}
for $x \ge0$, and zero otherwise.

The hierarchical model is conditioned on the apparent magnitudes $m_p$ of the primary standards, which are used to infer the zeropoint $Z$ given the instrumental magnitudes of the primary standards. This zeropoint is then coherently propagated to our network of stars within the hierarchical model, thus incorporating the covariance of the zeropoint and the apparent magnitudes of our DA white dwarfs.

\subsection{Posterior Distribution and Estimation}\label{sec:nutellaposterior}
The full posterior distribution of the model $\bm{\Phi}$ given the data $\mathcal{D}$ is proportional to the product of the likelihood (Eqn.~\ref{eqn:likelihood}) and the prior (Eqn.~\ref{eqn:priors}) 
\begin{equation}
    \begin{split}
         P( \bm{\Phi} & |\, \mathcal{D}, \{m_p\} ) \propto P( \mathcal{D} |\, \bm{\Phi}, \{m_p\} ) \cdot P (\bm{\Phi}) \\ 
        = \; & \prod_{p=1}^3 P(\bm{\widehat{m}_{p}} |\, m_p,Z,\sigma_\text{int}, \nu ) \\
    \times & \Big[ \prod_{s=1}^{N_S} P(\bm{\widehat{m}_{s}} |\, m_s, Z, \Delta Z_\text{C20}, \sigma_\text{int}, \nu ) \cdot P(m_s)\Big] \\ 
    \times & P(Z) \cdot P(\Delta Z_{\text{C20}}) \cdot P(\sigma_\text{int}) \cdot P(\nu ) \\
    \end{split}
    \label{eqn:posterior}
\end{equation}

The model is solved for the $N_S + 4 $ parameters in each of program passbands independently. As we did not obtain \textit{F275W} observations in cycle 20, there is no zeropoint offset defined for this passband. \citetalias{Calamida18} report measurements using three separate photometry packages: \texttt{DAOPHOT}~\citep{daophot}, \texttt{SourceExtractor}, and \texttt{ILAPH}, a custom interactive aperture photometry developed by one of us (AS). \citetalias{Calamida18} find that the \texttt{DAOPHOT} and \texttt{ILAPH} measurements show reasonable agreement, while both differ systematically from the \texttt{SourceExtractor} measurements. They determine this to be a result of how \texttt{SourceExtractor} treats the sky background. \texttt{DAOPHOT} is also known to exhibit systematic differences~\citep{Bajaj17} to careful aperture photometry performed with the \texttt{PhotUtils} Python package~\citep{photutils} or the \texttt{APER} routine available through the IDL Astronomy Library\footnote{\url{https://idlastro.gsfc.nasa.gov/}}. We analyze the \texttt{ILAPH} measurements in this work as these \texttt{IDL} procedures were developed specifically for measuring count rates from our \texttt{flc} images.

\begin{table*}
\scriptsize
\begin{centering}
\begin{tabular}{l|cc|cccccc}
\hline
\hline
Object & R.A. & $\delta$  & \textit{F275W} & \textit{F336W} & \textit{F475W} & \textit{F625W} & \textit{F775W} & \textit{F160W} \\
 & hh:mm:ss &  \degree:\arcmin:\arcsec & \multicolumn{6}{c}{AB mag (mmag)} \\
\hline
G191-B2B & 05:05:30.61 & $+$52:49:51.96 & 10.4904 (1) & 10.8902 (1) & 11.4988 (1) & 12.0307 (1) & 12.4514 (1) & 13.8853 (2) \\
GD153 & 12:57:02.34 & $+$22:01:52.68 & 12.2016 (2) & 12.5679 (1) & 13.0998 (2) & 13.5976 (1) & 14.0017 (1) & 15.4139 (2) \\
GD71 & 05:52:27.61 & $+$15:53:13.75 & 11.9888 (1) & 12.3360 (1) & 12.7988 (1) & 13.2790 (1) & 13.6720 (1) & 15.0676 (2) \\
\hline
SDSS-J010322 & 01:03:22.19 & $-$00:20:47.73 & 18.1952 (4) & 18.5268 (5) & 19.0833 (5) & 19.5686 (5) & 19.9648 (6) & 21.3552 (12) \\
SDSS-J022817 & 02:28:17.17 & $-$08:27:16.41 & 19.5183 (8) & 19.7152 (10) & 19.8151 (7) & 20.1690 (7) & 20.5014 (6) & 21.7371 (17) \\
SDSS-J024854 & 02:48:54.96 & $+$33:45:48.30 & 17.8285 (4) & 18.0400 (6) & 18.3696 (3) & 18.7459 (3) & 19.0773 (2) & 20.3400 (6) \\
SDSS-J041053\tablenotemark{*} & 04:10:53.63 & $+$06:30:27.75 & 18.1162 (9) & 18.4041 (4) & 18.8796 (5) & 19.2537 (3) & 19.3936 (5) & 19.4982 (5) \\
WD0554\tablenotemark{*} & 05:57:01.30 & $-$16:35:12.00 & 16.7760 (5) & 17.1531 (3) & 17.7271 (5) & 18.2205 (3) & 18.6172 (5) & 20.0431 (7) \\
SDSS-J072752 & 07:27:52.76 & $+$32:14:16.10 & 17.1636 (3) & 17.4715 (3) & 17.9933 (3) & 18.4567 (2) & 18.8370 (3) & 20.2166 (7) \\
SDSS-J081508 & 08:15:08.78 & $+$07:31:45.80 & 18.9505 (6) & 19.2635 (8) & 19.7162 (5) & 20.1838 (5) & 20.5794 (6) & 21.9616 (24) \\
SDSS-J102430 & 10:24:30.93 & $-$00:32:07.03 & 18.2606 (18) & 18.5143 (4) & 18.9042 (5) & 19.3174 (4) & 19.6649 (10) & 20.9905 (13) \\
SDSS-J111059 & 11:10:59.43 & $-$17:09:54.10 & 17.0406 (3) & 17.3544 (4) & 17.8668 (3) & 18.3135 (2) & 18.6887 (2) & 20.0566 (5) \\
SDSS-J111127 & 11:11:27.30 & $+$39:56:28.00 & 17.4429 (4) & 17.8298 (6) & 18.4206 (3) & 18.9390 (4) & 19.3441 (3) & 20.7975 (9) \\
SDSS-J120650 & 12:06:50.41 & $+$02:01:42.46 & 18.2397 (4) & 18.4888 (4) & 18.6719 (4) & 19.0601 (3) & 19.4112 (7) & 20.7027 (9) \\
SDSS-J121405 & 12:14:05.11 & $+$45:38:18.50 & 16.9401 (2) & 17.2827 (2) & 17.7606 (2) & 18.2362 (3) & 18.6292 (2) & 20.0378 (4) \\
SDSS-J130234 & 13:02:34.44 & $+$10:12:39.01 & 16.1879 (2) & 16.5216 (2) & 17.0364 (2) & 17.5140 (2) & 17.9037 (2) & 19.3031 (4) \\
SDSS-J131445 & 13:14:45.05 & $-$03:14:15.64 & 18.2577 (4) & 18.5969 (5) & 19.1018 (5) & 19.5668 (5) & 19.9553 (9) & 21.3284 (12) \\
SDSS-J151421 & 15:14:21.27 & $+$00:47:52.79 & 15.1100 (2) & 15.3907 (2) & 15.7090 (2) & 16.1202 (2) & 16.4712 (1) & 17.7870 (4) \\
SDSS-J155745 & 15:57:45.40 & $+$55:46:09.70 & 16.4999 (2) & 16.8766 (2) & 17.4702 (3) & 17.9917 (2) & 18.3880 (2) & 19.8343 (5) \\
SDSS-J163800 & 16:38:00.36 & $+$00:47:17.81 & 18.0158 (8) & 18.3177 (4) & 18.8399 (5) & 19.2808 (3) & 19.6605 (5) & 20.9963 (9) \\
SDSS-J172135\tablenotemark{*} & 17:21:35.98 & $+$29:40:16.00 & 20.3714 (13) & 20.0782 (16) & 19.6559 (5) & 19.6699 (3) & 19.7678 (3) & 20.5520 (21) \\
SDSS-J181424 & 18:14:24.13 & $+$78:54:02.90 & 15.7913 (2) & 16.1213 (2) & 16.5441 (2) & 17.0056 (2) & 17.3926 (1) & 18.7857 (2) \\
SDSS-J203722\tablenotemark{*} & 20:37:22.17 & $-$05:13:03.03 & 18.2568 (7) & 18.5438 (4) & 18.9428 (6) & 19.3504 (12) & 19.6718 (10) & 20.9790 (23) \\
SDSS-J210150 & 21:01:50.66 & $-$05:45:50.97 & 18.0681 (4) & 18.3344 (4) & 18.6560 (3) & 19.0636 (2) & 19.4140 (4) & 20.7396 (8) \\
SDSS-J232941 & 23:29:41.33 & $+$00:11:07.80 & 17.9434 (4) & 18.1090 (4) & 18.1607 (6) & 18.4697 (3) & 18.7753 (7) & 19.9949 (6) \\
SDSS-J235144 & 23:51:44.29 & $+$37:55:42.60 & 17.4494 (4) & 17.6619 (3) & 18.0751 (3) & 18.4595 (3) & 18.7868 (2) & 20.0747 (4) \\
\tableline
Parameter & & &  \textit{F275W} & \textit{F336W} & \textit{F475W} & \textit{F625W} & \textit{F775W} & \textit{F160W} \\
 & & & \multicolumn{6}{c}{AB mag (mmag)} \\
\tableline
Zeropoint $Z$ &  &  & 24.0596 (1) & 24.5899 (1) & 25.5774 (1) & 25.4056 (1) & 24.7189 (1) & 25.8116 (1) \\
Offset $\Delta Z_{\text{C20}}$ & &  & --- (---) & $-$0.0326 (3) & $-$0.0091 (4) & $-$0.0139 (2) & $+$0.0089 (4) & $-$0.0125 (5) \\
Dispersion $\sigma_{\text{int}}$ &  &  & 0.0031 (1) & 0.0017 (1) & 0.0026 (1) & 0.0015 (1) & 0.0006 (1) & 0.0045 (1) \\
$\nu$ (Dimensionless) &  &  & 2.451 (0.435) & 1.626 (0.212) & 2.275 (0.406) & 3.186 (0.714) & 1.380 (0.151) & 2.978 (0.542) \\
\hline
\end{tabular}
\footnotesize{\tablecomments{Coordinates are reported with epoch J2000. Apparent magnitudes measured through each passband are reported in columns with names corresponding to the passband names, followed by the uncertainties parenthetically. All measurements are rounded to a tenth of a millimag. Magnitudes are tabulated on the AB system, and are followed by the parameters of our model of the photometric observations. The zeropoint in each passband $Z$ are determined from the difference between the synthetic magnitudes of the three CALSPEC primary standards and their measured instrumental magnitudes as described in \S\ref{sec:photmodel}. The offsets between cycle 20 and 22 $\Delta Z_{\text{C20}}$ are determined from all stars with observations in both cycles. All observations are used to infer the dispersion $\sigma_{\text{int}}$ and the degrees of freedom $\nu$ of the Student-$t$ distribution, which describe the photometric repeatability and the outliers caused by cosmic rays respectively in each passband. Higher values of $\nu$ indicate increasing Gaussianity. The parameter $\nu$ is dimensionless, and its value and error are reported to three decimal places. }
\tablenotetext{*}{\citetalias{Narayan16} showed that SDSS-J203722 exhibited time-variable emission in the cores of the Balmer lines and excluded this object from their analysis. Additionally, we exclude SDSS-J041053, WD0554, and SDSS-J172135 in this work - see \S\ref{sec:excluded} for details. Their measured apparent magnitudes are listed here for completeness.}}
\caption{Apparent AB Magnitudes and Photometric Uncertainties of the Network of DA White Dwarfs and CALSPEC Primary Standards\label{table:phot}}
\end{centering}
\end{table*}

We use the No U-Turn Sampler (NUTS)~\citep{Homan14} implemented in the \texttt{pymc3} package~\citep{Salvatier2016} to estimate the posterior distribution. We run four independent Markov Chains initialized to different positions for a 20,000 steps, following a burn-in of 2,000 steps which are discarded. We use a suite of diagnostic tests including visually inspecting the chains for convergence, verifying that the chain auto-correlation lengths are small, determining that the Gelman-Rubin statistic~\citep[Eqn. 20 from ][]{gelman1992} is near unity, and comparison of our inferred zeropoints against the fiducial values determined from the SMOV program and reported on MAST. Additionally, we check that the outliers to the Student's $t$-distribution are all brighter than the mean apparent magnitudes of our DA white dwarfs $m_s$, consistent with cosmic rays. Only 2--8 out of $\sim$190 distinct observations in each passband are flagged as outliers. The intrinsic dispersion $\sigma_{\text{int}}$ is between 1--4~mmag, decreasing from \textit{F275W} to \textit{F775W}. The apparent magnitudes of our DA white dwarf stars are listed in Table~\ref{table:phot}.

\section{Selection Cuts}\label{sec:excluded}

Up to this point, the analysis has been independent of any assumptions about the physical nature of the sources being modeled --- we do not need to specify temperature, surface gravity and extinction to determine the mean apparent magnitudes with respect to the three CALSPEC primary standards. However, in order to infer accurate SEDs from our \emph{HST/WFC3} photometry and spectroscopy of our DA white dwarfs, we must specify a model for the observations. This model fundamentally assumes that our objects are isolated, photometrically stable sources that are well described by NLTE pure-hydrogen atmospheres that do not exhibit strong magnetic fields (see \S\ref{sec:temporal}). Not all of our targets satisfy this condition --- these must be excluded from further analysis. 

\citet{Kleinman13} identifies SDSS-J041053 as a DA+M:E binary system. \citet{Eisenstein06}, \citet{Kleinman13}, and several other sources identify SDSS-J172135 as a DA white dwarf with $T_{\text{eff}} \sim 9,450$~K, well below our cutoff of 20,000~K to exclude pulsating white dwarfs, and even more distant from our initial $T_{\text{eff}}$ estimate of 30,000~K. We attempted to determine how SDSS-J172135 was included in our list of cycle 22 targets, despite being more than 10,000~K cooler than our lower limit in $T_{\text{eff}}$. We determined that the object was initially shortlisted as a possible target in 2012, prior to our cycle 20 program. At that time, either the spectrum of a different object was inspected due to a failure in name resolution --- it is unlikely to be a coincidence that SDSS-J172132 is a DA white dwarf with a $T_\text{eff} \sim 30,000$~K listed directly above SDSS-J172135 in the Montreal White Dwarf Database\footnote{\url{http://www.montrealwhitedwarfdatabase.org/}} --- or the temperature was somehow grossly overestimated. Unfortunately, as the object was shortlisted for our cycle 20 program, it was not scrutinized again prior to cycle 22, and we failed to flag it as an unsuitable target. We find that the SED parameters of both SDSS-J041053 and SDSS-J172135 inferred using the methodology in \S\ref{sec:wdmodel} place them near the edge of our DA white dwarf atmosphere grid with significant extinction, and their photometric residuals are two orders of magnitude larger than the rest of our sample. We cannot exclude SDSS-J041053 and SDSS-J172135 on the basis of our existing temporal observations, but given the binarity of the former and the sub-threshold temperature of the latter, we feel excluding them from our network is justified. 

\citetalias{Calamida18} report that WD0554 is photometrically unstable with a 0.2~mag peak-to-peak amplitude and a Welch-Stetson variability index~\citep{welchstet1993} $I$ of 3.98. The inferred $\log g$ parameter of WD0554 is $\sim9$~dex, near the edge of the grid and more than 1~dex higher than the mean of our sample. To verify the high surface gravity inferred from our MMT/Blue Channel spectrum, we obtained a second high S/N and high resolution spectrum with the Inamori Magellan Areal Camera and Spectrograph (IMACS) on Magellan-Baade. Our inferred surface gravity remains consistent irrespective of which spectrum is used in the analysis. This anomalously high surface gravity and photometric variability suggests that WD0554 has a weak magnetic field with unresolved Zeeman splitting leading to line broadening and an overestimate of the surface gravity. Finally, \citetalias{Narayan16} found that SDSS-J203722 exhibited time-variable emission in the cores of the Balmer lines. \citetalias{Calamida18} also report that this object has several time-resolved observations from the LCO network and is photometrically unstable with $I=3.35$ and a standard deviation of observations $\sigma = 0.04$~mag, well above the mean standard deviation of field stars which have $\sigma \sim 0.01$~mag. \textbf{We therefore exclude SDSS-J041053, WD0554, SDSS-J172135 and SDSS-J203722 from additional analysis.}


\section{Forward-modeling the DA White Dwarf Spectroscopy and Photometry}\label{sec:wdmodel}

In order to establish our DA white dwarfs as spectrophotometric standards on an equal footing with the three CALSPEC primary standards, we must determine robust SEDs, spanning the wavelength range from the UV to the NIR (roughly 1,350~\AA\ -- 2.5~$\mu$m), for each of our stars. The most direct approach to inferring the intrinsic and extrinsic parameters and deriving an SED is to forward-model the observations of each DA white dwarf --- the apparent magnitudes presented in Table~\ref{table:phot} and the spectroscopy presented in \citetalias{Calamida18}. In this section, we describe the \citetalias{Narayan16} analysis and address its shortcomings with an improved methodology developed for this work.

\subsection{Comparison to \citetalias{Narayan16} Methodology}
\citetalias{Narayan16} developed a fitter around the \texttt{Tlusty} synthetic DA white dwarf atmosphere grid to infer the intrinsic parameters $T_\text{eff}$ and $\log g$ from spectra. Using a standard methodology first introduced in \citet[][hereafter \citetalias{Bergeron1992}]{Bergeron1992}, a local linear continuum is fit across each of the hydrogen Balmer lines, and the line profile is extracted and normalized to have a constant continuum equal to unity. The same features are extracted from the model atmosphere and used to fit the spectral lines, with the log-likelihood defined as the sum of the average sum of squared differences between data and model for H$_{\beta}$ through H$_{\zeta}$. The best-fit intrinsic parameters are used to construct an unreddened SED model. \citetalias{Narayan16} attributed the difference between the unreddened synthetic magnitudes and the \emph{HST} observations to reddening, which, once inferred, was used to correct the unreddened SED model. The key feature of this methodology is that it divides the inferences of the model parameters into two discrete steps; inference of the intrinsic DA white dwarf parameters, followed by inferences of the extrinsic reddening parameters.

More detailed analysis of our growing dataset revealed two shortcomings of the \citetalias{Narayan16} analysis when applied to lower S/N spectroscopic data from our faint DA white dwarfs:
\begin{enumerate}
\item \textit{Lack of propagation of uncertainty from the local linear continuum fit to the intrinsic parameters}: Despite the heteroskedasticity of measured error in the spectrum flux, the \citetalias{Narayan16} analysis does not incorporate any notion of uncertainty, largely because the spectroscopic reductions were preliminary. While it is possible to incorporate the measurement errors by modifying the log-likelihood to be the sum of the chi-square statistic of each of the Balmer lines, this is inadequate as the flux calibration errors are also correlated with wavelength and are not independent or identically distributed. The \citetalias{Narayan16} analysis attempted to fit a local pseudo-continuum to each of the Balmer lines and determine the intrinsic DA white dwarf parameters purely from the shape of these features. The local continuum between the Balmer lines becomes less-pronounced or is completely absent for H$_\gamma$ and bluer, and this local linear model exhibits increasing bias with decreasing temperature and increasing surface gravity. The parameters of the local linear continuum model are treated as ``nuisance'' variables and seldom reported, and it is not possible to coherently propagate any uncertainties in these parameters into the SED flux. We also find that the method underestimates the uncertainties on the inferred parameters by as much as a factor of 4--10 for the four stars presented in \citetalias{Narayan16}. 

\item \textit{Not simultaneously modeling the intrinsic DA WD parameters and the extrinsic reddening}: Our DA white dwarfs are impacted by reddening, which affects the shape of the entire spectrum, including the wide Balmer lines. The \citetalias{Bergeron1992} method infers the intrinsic white dwarf parameters solely from the Balmer lines. The reddening was determined serially, asserting that the difference between the synthetic photometry of the unreddened SED and the observations was solely due to reddening. This procedure neglected the reddening of the spectra, and therefore may lead to a biased estimate of the intrinsic white dwarf parameters, particularly for our distant DA WDs where we expect the effect of extinction to be larger than for nearby stars. \citetalias{Narayan16} attempted to mitigate this by using an estimate of the reddening derived from SDSS photometry to deredden the spectrum prior to estimating the intrinsic DA white dwarf parameters. The \citetalias{Narayan16} framework is iterative --- an estimate of the reddening is needed to estimate the unreddened SED, which is in turn used to derive the reddening, and the procedure can be repeated until a predefined convergence criterion is satisfied. While \citetalias{Narayan16} used independent data to derive the initial and final reddening estimate, the methodology could not fully account for the covariance between the intrinsic DA white dwarf parameters and extrinsic reddening, as it followed the \citetalias{Bergeron1992} approach of splitting the inference of these two quantities into different steps. 
\end{enumerate}

We developed the \texttt{WDmodel} methodology to address these shortcomings and coherently forward model all of the observations. We detail the various components of the \texttt{WDmodel} and define the likelihood of the observations given the model in the following sections.

\subsection{The DA white dwarf Atmosphere Grid and Intrinsic Parameters}\label{sec:modelgrid}
\begin{figure}[htpb]
    \centering
    \includegraphics[width=0.5\textwidth]{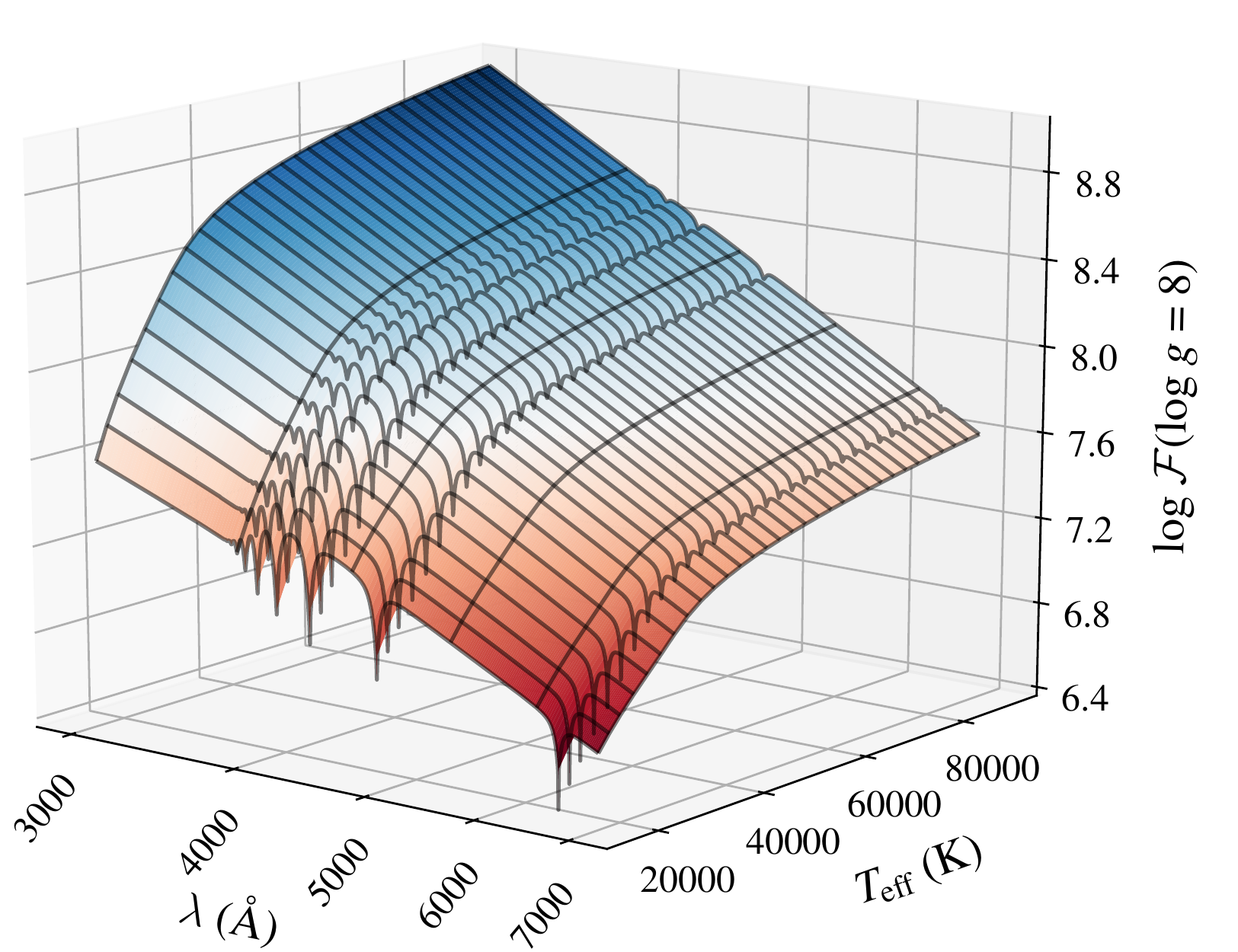}
    \caption{A slice with $\log g = 8$ through the \texttt{Tlusty} v202 DA white dwarf atmosphere model grid used in this work. We interpolate the grid with logarithmic scaling of wavelength and flux $\mathcal{F}$ to reduce errors. The model clearly exhibits the increasingly blue continuum and decreasing equivalent width of the Balmer lines with increase in $T_{\text{eff}}$ exhibited by real observations in Fig.~\ref{fig:spectralseq}.}
    \label{fig:grid}
\end{figure}

We describe the unreddened DA white dwarf SED with two parameters, $T_\text{eff}$ and $\log g$. We use the same \texttt{Tlusty}~\citep{Hubeny95} v202 NLTE model atmosphere grid\footnote{\url{http://nova.astro.umd.edu/Tlusty2002/tlusty-frames-refs.html}} as \citetalias{Narayan16}. The grid has 31 uneven steps in $T_\text{eff}$ from 16,000--90,000~K, with a spacing of 2,000~K from 16,000--20,000~K and 2,500~K from 20,000--90,000~K. The grid has 6 even steps in $\log g$ from 7--9.5 dex, with 0.5 dex spacing. The grid covers a wavelength range of 1,350~\AA\ -- 2.7~\micron, in 1~\AA\ steps from 1,350~\AA\ $\le \lambda \le$3,000~\AA, 0.5~\AA\ steps from 3,000~\AA\ $\le \lambda \le$~7,000~\AA, and 5~\AA\ steps for $\lambda >$~7,000~\AA. A slice through the grid is shown in Fig.~\ref{fig:grid}. The grid uses all available hydrogen line profiles from \citet{Tremblay09}. The treatment of level dissolution and pseudocontinua follow \citet{Hubeny94}.

While there are DA white dwarf stars with intrinsic parameters outside the range covered by our grid, we do not use them as spectrophotometric standards at this time. In particular, we limit our consideration to DA white dwarfs with  $T_\text{eff} > 16,000$~K, as there are still model uncertainties about the treatment of convection and the mixing length prescription for cooler DA WDs~\citep{Bergeron92}. Some of these sources may exhibit variability as described in \S\ref{sec:temporal}, and such sources are not suitable for use as spectrophotometric standards. This selection cut is conservative, but is justified as our primary goal is to establish an all-sky network of standards with a minimal set of theoretical assumptions. We emphasize that our \texttt{WDmodel} code \emph{should not} be used to model DA white dwarfs cooler than 16,000~K, and will report erroneously high values of $A_V$ as the only way to redden the model below the lower limit of the grid is to add extinction. A new model atmosphere grid that extends to lower temperatures could in principle be used together with our \texttt{WDmodel} code to fit cool ($T_\text{eff} < 16,000$~K) DA white dwarfs. When modeling spectroscopy together with photometry to establish spectrophotometric standards, though, we make the fundamental assumption that the white dwarf is not variable and the model is \emph{stationary}. This assumption does not hold unless the spectroscopy and photometry are obtained contemporaneously.

The logarithm of SED model flux is tri-linearly interpolated at any $T_\text{eff}$, $\log g$ and $\log(\lambda)$ within the bounds of the grid, and then exponentiated to determine the SED model $\mathcal{F}_s(T_\text{eff}, \log g, \lambda)$ for a DA white dwarf $s$. \citet{Levenhagen17} have made a higher resolution grid of atmosphere models available, but this new grid exhibits a discontinuity where it transitions from LTE atmospheres for $T_\text{eff} < $~34,000~K to NLTE atmospheres for $T_\text{eff} \ge $~34,000~K, whereas our \texttt{Tlusty} grid is NLTE throughout.

\newpage
\subsection{The Extrinsic Reddening Parameters}
We adopt the \citetalias{Fitzpatrick99} model to describe the wavelength dependence of the extinction due to interstellar dust. \citetalias{Fitzpatrick99} is defined for 1,150~\AA\ $\le \lambda \le$\ 6~\micron, and uses the \citet{Fitzpatrick90} model of interstellar extinction for $\lambda \le~$2,700~\AA, and a spline model above. The extinction at any wavelength is specified by two parameters, $A_V$ and $R_V$. We apply extinction $A_s(A_V, R_V, \lambda)$ to the unreddened SED model $\mathcal{F}_s(T_\text{eff}, \log g, \lambda)$ to determine the reddened SED model $\widetilde{F}_s$ for a DA white dwarf $s$.
\begin{equation}
    \begin{split}
\widetilde{F}_s(T_{\text{eff}}, & \log g, A_V, R_V, \lambda) = \\
& \mathcal{F}_s(T_{\text{eff}}, \log g, \lambda) \cdot 10^{-0.4 \cdot A_s(A_V, R_V, \lambda)} \\
    \end{split}
\end{equation}

The \texttt{WDmodel} code supports any reddening law implemented in the \texttt{extinction}~\citep{extinctionpy} Python module. As in \citetalias{Narayan16}, we do not find any significant difference in our results when using the \citet{ODonnell94} reddening law.

\subsection{The Normalization of Synthetic Photometry}
We compute synthetic magnitudes from the reddened SED $\widetilde{F}_s$ using Eqn.~\ref{eqn:synphot}. We add a single achromatic normalization parameter $\mu$ to the synthetic reddened magnitudes in all passbands to account for the distance and radius of the DA white dwarf, and thereby model the apparent magnitudes inferred from the observations in \S\ref{sec:phot}. Normalizing the flux of the reddened SED $\widetilde{F}_s$ with $\mu$ results in the final calibrated SED $\mathbf{F}$ tied to the CALSPEC system. 

\begin{equation}
    \begin{split}
\mathbf{F}_s(T_{\text{eff}}, & \log g, A_V, R_V, \mu, \lambda) = \\
& \widetilde{F}_s(T_{\text{eff}}, \log g, A_V, R_V, \lambda) \cdot 10^{-0.4 \cdot \mu}\\
    \end{split}
\end{equation}

\begin{figure}[htb]
    \centering
    \includegraphics[width=0.47\textwidth]{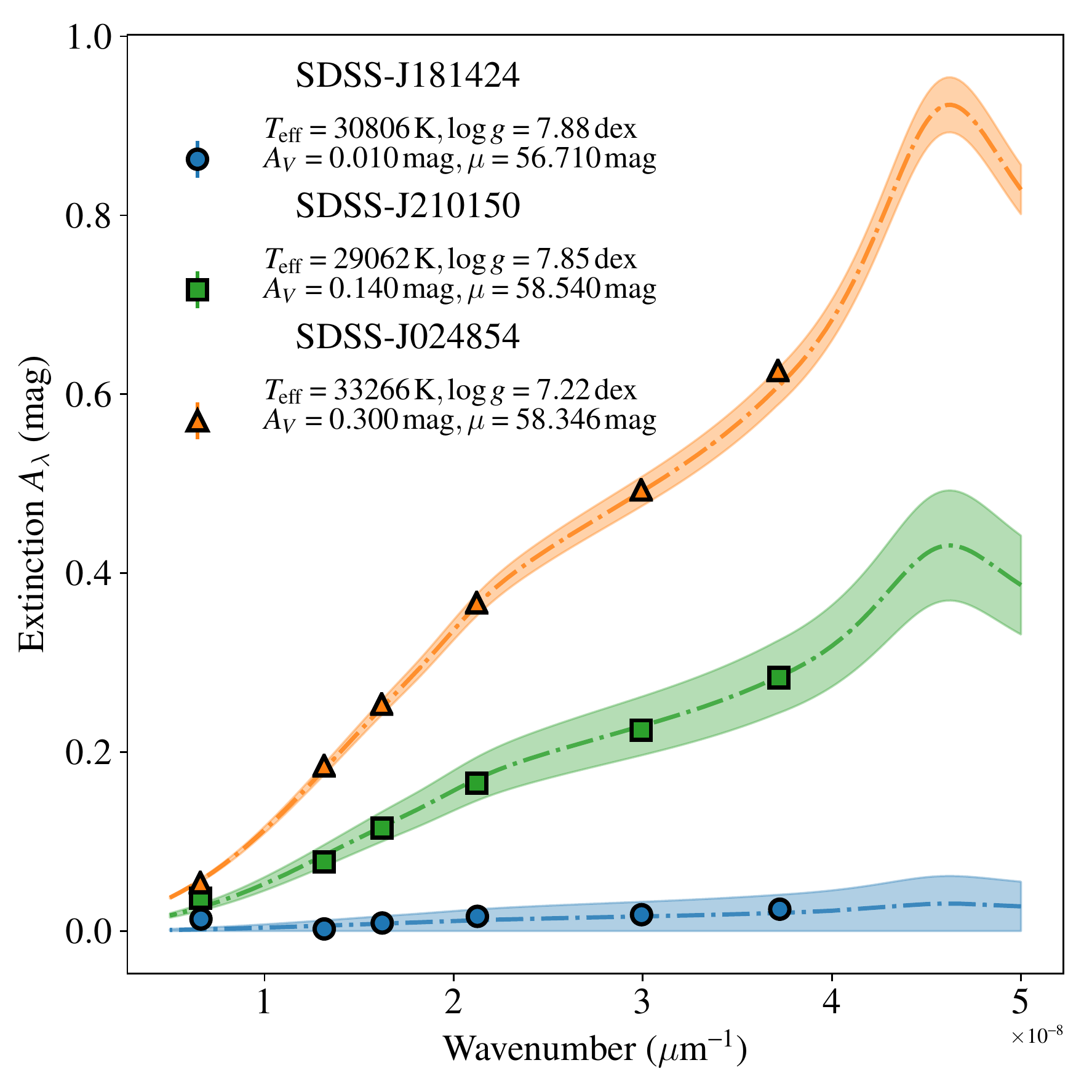}
    \caption{\citetalias{Fitzpatrick99} extinction curves (dashed-dotted lines) for our inferred values of $A_V$ with $R_V=3.1$ for three of our DA white dwarfs. The markers show the difference between the observed \emph{HST/WFC3} photometry and the synthetic magnitudes of the unreddened model in each passband at the effective wavelength of the reddened SED. Model parameters are indicated in the legend. The shaded region encompasses the 1$\sigma$ uncertainty on the observations and the model parameters. Despite the stars having comparable $T_{\text{eff}}$, there is a wide range in the line-of-sight extinction that must be accounted for when modeling the observations of each star.}
    \label{fig:extinction_curve}
\end{figure}

While $\mu$ is not the distance modulus, as the absolute magnitude of each white dwarf is not known without imposing a mass-radius-luminosity relation, it is functionally equivalent. The difference between the \emph{HST/WFC3} photometry and the unreddened synthetic magnitudes for three of our DA white dwarfs are illustrated together with the inferred extinction curves in Fig.~\ref{fig:extinction_curve}.

\subsection{The Probability Density of the \emph{HST} Apparent Magnitudes}
The probability density of the set of photometric data $\{m_s\}$ reported in Table \ref{table:phot} given the model parameters $ \{T_{\text{eff}}, \log g, A_V, R_V, \mu \}$ is the likelihood function:
\begin{equation}
    \begin{split}
        P(\{m_s\} & |\, T_{\text{eff}}, \log g, A_V, R_V, \mu) = \\
        & \prod_{\lambda = 1}^{N_{\text{PB}}}\, N(m_{s,\lambda} |\, M_{s,\lambda}(T_{\text{eff}}, \log g, A_V, R_V) + \mu, \sigma_{s,\lambda}) \\
    \end{split}
    \label{eqn:photlikelihood}
\end{equation}
where $m_{s,\lambda}$ is the apparent magnitude for a DA white dwarf star $s$ with photometric measurement error described by an estimated standard deviation $\sigma_{s\lambda}$, and $M_{s,\lambda}(T_{\text{eff}}, \log g, A_V, R_V)$ is the synthetic magnitude of the reddened SED $\widetilde{F}_s$ through \emph{HST/WFC3} passband $\lambda \in \{$\textit{F275W, F336W, F475W, F625W, F775W, F160W}$\}$. 

\subsection{Accounting for Spectral Resolution and Overall Normalization}
Modeling the observed spectra $\mathcal{S}_s$ of DA white dwarf $s$ is more complex. Minimally, i) a normalization parameter applied to the reddened SED to match the observed flux, and ii) the resolution of the reddened, normalized SED model must be degraded to the resolution of the observed spectrum to account for the seeing at the observatory and the configuration of the spectrograph. We normalize the extinguished model SED for a DA white dwarf $s$ using a single parameter $d_L$:
\begin{equation}
    \begin{split}
F_s(T_{\text{eff}}, & \log g, A_V, R_V, d_L, \lambda) = \\
& \frac{\widetilde{F}_s(T_{\text{eff}}, \log g, A_V, R_V, \lambda)} {4\pi \cdot d^{2}_L } \\
    \end{split}
\end{equation}
where $F_s$ is the extinguished, normalized SED constructed from the reddening SED model $\widetilde{F}_s$. As with $\mu$, this parameter is not the true luminosity distance to each white dwarf, as the absolute flux of each white dwarf is not known without imposing a mass-radius-luminosity relation.

\begin{figure}[htp]
    \centering
    \includegraphics[width=0.48\textwidth]{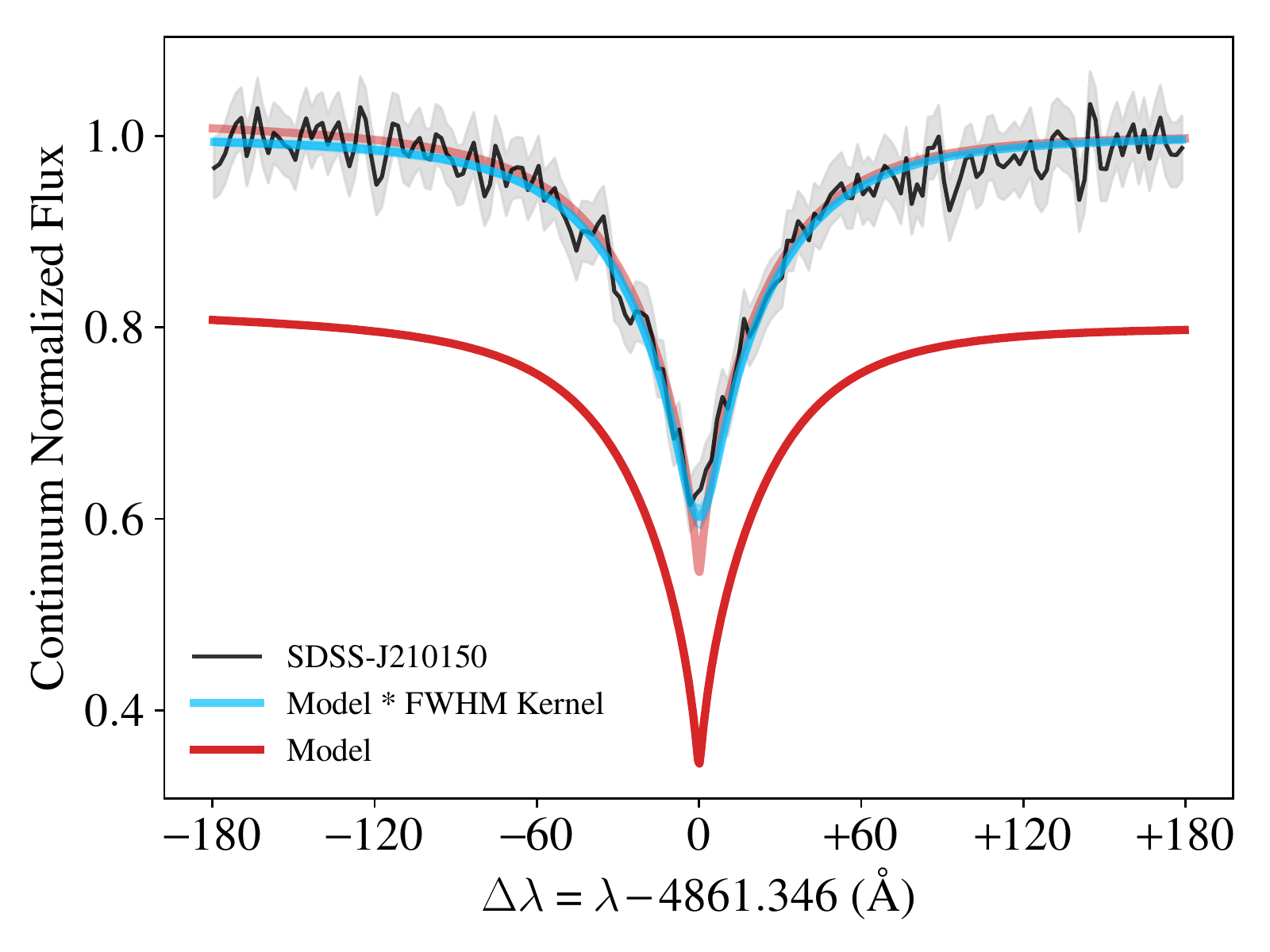}
    \caption{The model $F$ (in red) has a higher spectral resolution than the data $\mathcal{S}$ (in black), and the effect is particularly evident in the cores of the Balmer features (light red, offset to match the data), such as H$_\beta$ for one of our DA white dwarfs. This can lead to biases in the inferred intrinsic DA white dwarf parameters, as $T_{\text{eff}}$ and $\log g$ are sensitive to the shapes of the Balmer lines. We convolve the model with a Gaussian kernel with standard deviation proportional to the $\text{FWHM}$ (in blue) to match the observed spectrum.}
    \label{fig:FWHM}
\end{figure}

We then convolve $F_s$ with a Gaussian kernel to model the observed spectral resolution, as illustrated in Fig.~\ref{fig:FWHM}. The standard deviation $\sigma_{R}$ of the kernel is related to the full width at half maximum $\text{FWHM}$ of the spectrum by: 
\begin{equation}
\sigma_{R} \equiv \frac{\text{FWHM}}{R \cdot \sqrt{8 \cdot \ln(2)}}
\end{equation}
where $R$ is the median resolution of the observed spectrum in \AA\ per spectral unit, while the remaining factor in the denominator is ratio of the $\text{FWHM}$ to the standard deviation of a normal distribution. The convolution kernel is truncated at $\pm 4\sigma_{R}$. The normalized model of the observed spectrum is:
\begin{equation}
\begin{split}
f_s(& T_{\text{eff}}, \log g, A_V, R_V, d_L, \text{FWHM}, \lambda) = \\
 & F_s(T_{\text{eff}}, \log g, A_V, R_V, d_L, \lambda) \ast N(1, \sigma_R(\text{FWHM})) \\
 \end{split}
\end{equation}

\subsection{Accounting for Correlated Errors in the Flux Calibration of the Spectrum}\label{sec:specnoisemodel}
If the \emph{shape} of spectra were free of error, these two parameters would suffice, however, as we note in \S\ref{sec:spectroscopy}, flux-calibration of ground-based spectroscopy is fraught with potential sources of systematic error. The SEDs of the standard stars that are used to flux calibrate the spectra of our DA white dwarfs are often less well determined than the DA WDs themselves. In addition, the flux corrections applied to the blue ($\le$~4,200~\AA) and red ends ($\ge$~7,000~\AA) of the spectra are large, as the throughput drops sharply. The difficulty of determining an unbiased flux correction in the red is compounded by the intrinsic faintness of the DA white dwarfs at these wavelengths, and the fringing in the detector. The flux correction is typically modeled with low-order splines or piece-wise polynomials. These functions are multiplied with the instrumental flux of the spectrum to produce the calibrated flux. Errors in the flux correction can exhibit ringing arising from polynomials and splines overfitting the data. The flux correction is typically only accurate to a few percent, and any error in the calibration procedure results in a non-monotonic error in the shape of the spectrum. 

The residual vector describes the correlated error between the flux of observed spectrum $\mathcal{S}_s$ and the flux of the model spectrum $f_s$:
\begin{equation}
    r(\lambda_i) = \mathcal{S}_s(\lambda_i) - f_s(T_{\text{eff}}, \log g, A_V, R_V, d_L, \text{FWHM}, \lambda_i)
\end{equation}
We model this correlated error with a Gaussian process~\citep[see][and references therein for a detailed background.]{RW2006}. We construct the covariance kernel of the Gaussian process as the sum of a Mat\'ern $3/2$ kernel to describe the correlated error, and a white noise kernel to describe the dispersion in the observed flux of the spectrum that is underestimated by the reported uncertainty on the flux. This kernel can be expressed as:
\begin{equation}
\begin{split}
& k(f_{\sigma}, \tau, f_{\omega}, \Delta \lambda_{ij}) = \\
 & (f_{\sigma} \cdot \mean{\sigma}_{\mathcal{S}})^2 \cdot \left[1 + \frac{\sqrt{3} \cdot \Delta \lambda_{ij}}{\tau} \right] \cdot \exp{\left( - \frac{\sqrt{3} \cdot \Delta \lambda}{\tau} \right)} + (f_{\omega} \cdot \mean{\sigma}_{\mathcal{S}})^2 \cdot \delta_{ij} \\
\end{split}
\end{equation}
where $\Delta \lambda_{ij} \equiv \left| \lambda_i - \lambda_j \right|$ is the absolute value of the difference between any two wavelengths $\lambda_i$ and $\lambda_j$, $\tau$ is the characteristic length-scale of correlations in the spectrum, $f_{\sigma}$ and $f_{\omega}$ are the scale-free amplitudes of the correlated error and white noise components respectively, and $\delta_{ij}$ is the Kronecker delta. The scale-free amplitudes are scaled by the median of the reported uncertainty in the spectroscopic flux $\mean{\sigma}_{\mathcal{S}}$ to model an observed spectrum of a DA white dwarf $\mathcal{S}_s$. 

\begin{figure*}[hptb]
    \centering
    \includegraphics[width=\textwidth]{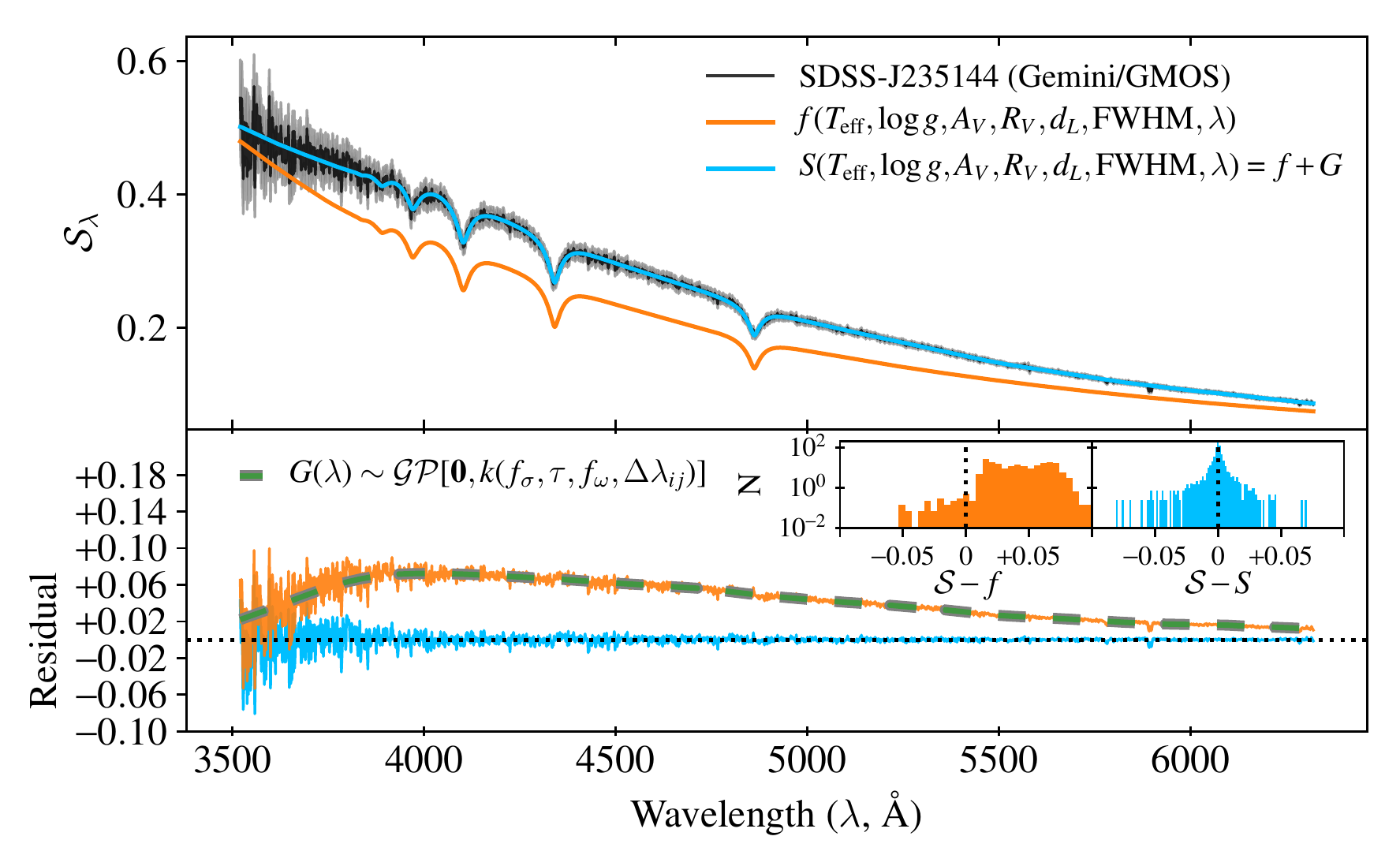}
    \caption{Observed spectrum $\mathcal{S}$ (black, with reported errors in grey) of SDSS-J235144 compared to the normalized model spectrum $f(T_{\text{eff}}, \log g, A_V, R_V, d_L, \text{FWHM})$ (orange). The difference between the model and data is the result of a pronounced miscalibration in the flux from 3,500--6,500~\AA. The inferred normalized model $f$ does not account for this flux calibration error. We parametrize the flux calibration error with a Gaussian process kernel composed of the sum of a Mat\'ern $3/2$ kernel to account for the correlation with wavelength, and a white noise kernel to account for the underestimated error. The corrected spectrum $S$ (blue) agrees well with the observed spectrum $\mathcal{S}$. Without the Gaussian process correction $G$ (the posterior mean is indicated in dashed green in the lower panel), the residuals to the model (indicated with color corresponding to the model) show significant bias from 0 (indicated by a dotted black line). This bias is readily apparent in the inset log-scaled histograms of the residuals, and would lead to a bias in the inferred model parameters if left uncorrected.}
    \label{fig:gpspec}
\end{figure*}

The Mat\'ern $3/2$ kernel is the simplest singly differentiable functional in the Mat\'ern covariance family. These functionals can be represented as the product of a polynomial and exponential for all half-integer values of the kernel order. In the limit of order of the kernel tending to infinity, the functional reduces to a squared exponential kernel. As the squared exponential is infinitely differentiable, the Gaussian process becomes sensitive to structure on any length scale, including sharp noise spikes. The low order Mat\'ern $3/2$ kernel can describe the large scale errors in the flux calibration of the spectrum and incurs the lowest computational cost, while remaining insensitive to sharp features that arise because of the finite S/N of our observations.  The model error is a realization of a zero mean Gaussian process $\mathcal{GP}$ with the kernel:
\begin{equation}
    G(\lambda) \sim \mathcal{GP}[\mathbf{0},k(f_{\sigma}, \tau, f_{\omega}, \Delta \lambda_{ij})] 
\end{equation}

For a spectrum of length $N_{\lambda}$, the full $N_{\lambda} \times N_{\lambda}$ covariance matrix $\mathbf{C}$ incorporates both the reported errors in the observed flux of the spectrum $\sigma_{\mathcal{S}}$, and the model error parametrized by the Gaussian process. The covariance matrix can be expressed as:
\begin{equation}
    C_{ij} = \sigma^{i}_{\mathcal{S}} \cdot \delta_{ij} + k(f_{\sigma}, \tau, f_{\omega}, \Delta \lambda_{ij})
\end{equation}
We model the corrected spectrum $S$ as the sum of the normalized model spectrum $f_s$ and the spectral flux calibration error, represented by a realization of a GP with the above kernel:
\begin{equation}
    \begin{split}
        S_s( & T_{\text{eff}}, \log g, A_V, R_V, d_L, \text{FWHM},\lambda ) = \\
         & f_s( T_{\text{eff}}, \log g, A_V, R_V, d_L, \text{FWHM}, \lambda) + G(\lambda) \\
    \end{split}
\end{equation}
We infer the parameters of the latent GP jointly with the other model parameters. We marginalize over the latent GP to propagate this uncertainty into the parameter inference. We illustrate the effectiveness of the Gaussian process at modeling the correlated error in the flux calibration in Fig.~\ref{fig:gpspec}. We find that the Gaussian process corrects for the errors in the flux calibration for DA white dwarf spectra observed with a variety of different instruments including MMT/Blue Channel, Gemini-N/GMOS, SOAR/Goodman. Unlike other ad hoc methods that we attempted to use including smoothing splines and various families of polynomials, the Gaussian process does not require fine tuning, and can incorporate the reported uncertainties on the flux of the spectrum.

\citet{Czekala15} also use a Gaussian process to model the correlated offsets between observed and model spectra. Their \texttt{Starfish}\footnote{\url{http://iancze.github.io/Starfish/}} package is designed to work with high-resolution Echelle spectra that are not flux calibrated, and the Gaussian process is used to capture the additional variance in observed spectral features that is not captured by stellar atmosphere template libraries. While we cannot employ \texttt{Starfish} package directly for this work, its approach to inferring a stellar atmosphere to model an observed spectrum is conceptually similar to ours. 

\subsection{The Probability Density of the Ground-based Spectrum}\label{sec:speclikelihood}
The natural logarithm of the probability density of the ground-based spectrum $\mathcal{S}_{s}$ for a DA white dwarf $s$ given the model parameters $\{ T_{\text{eff}}, \log g, A_V, R_V, d_L, \text{FWHM}, f_{\sigma}, \tau, f_{\omega} \}$ is the log-likelihood function:
\begin{equation}
    \begin{split}
       \ln P & ( \mathcal{S}_s |\, T_{\text{eff}}, \log g, A_V, R_V, d_L, \text{FWHM}, f_{\sigma}, \tau, f_{\omega} ) = \\
          & - \frac{1}{2} \left[ \mathbf{r}^{\text{T}} \cdot \mathbf{C}^{-1} \cdot \mathbf{r} \, + \, \ln \mathopen| 2\pi \mathbf{C} \mathclose| \right]\\ 
    \end{split}
    \label{eqn:speclikelihood}
\end{equation}
where $\mathbf{r}$ is the residual vector.

\subsection{The Full Probability Density of Observations}
\begin{figure}[htb]
\includegraphics[width=0.47\textwidth]{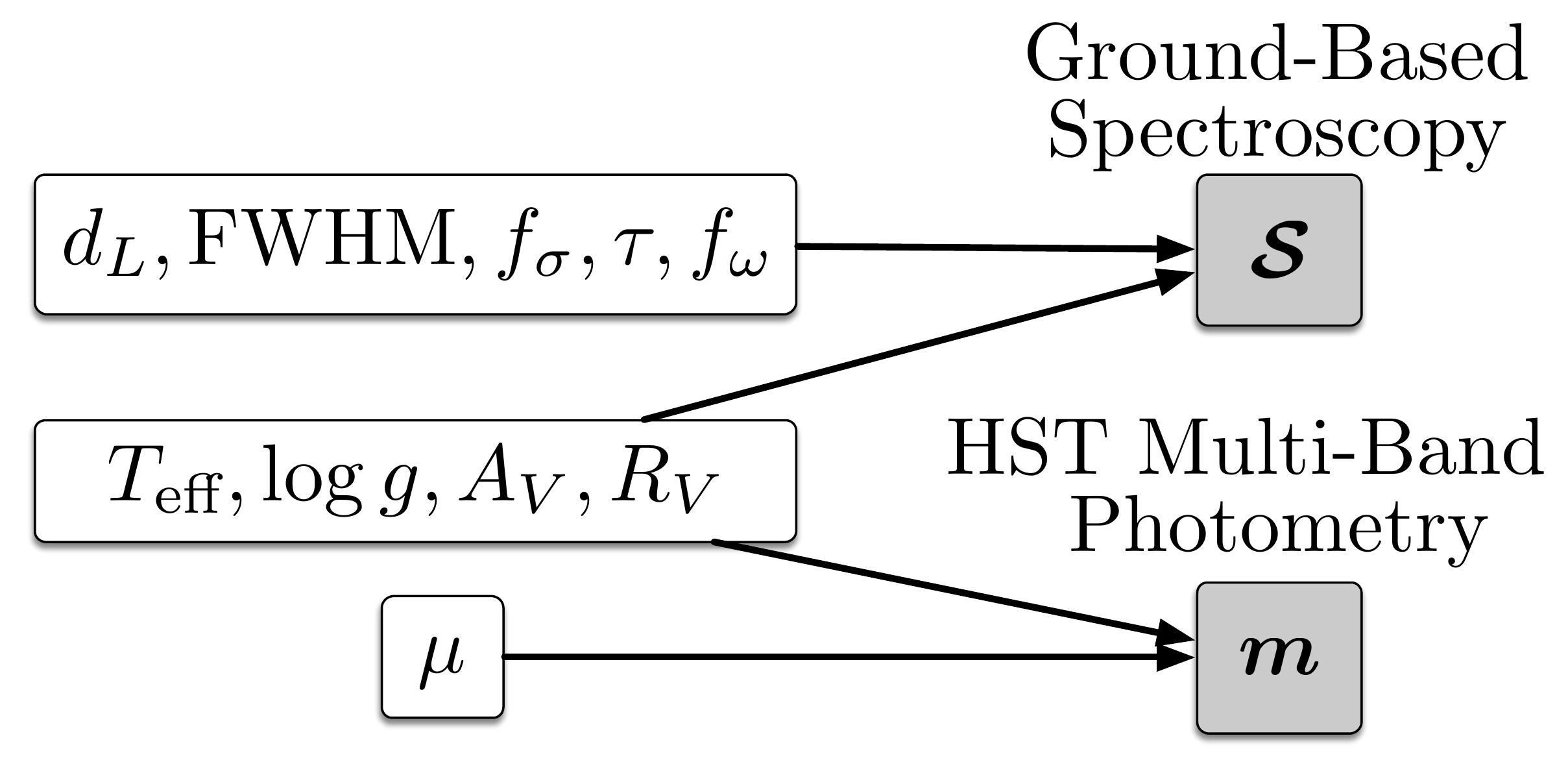}
\caption{Directed acyclic graph depicting the model for our observations of each DA white dwarf $s$ --- the \emph{HST} photometry $\{m_s\}$ tied to the three CALSPEC standards (see \S\ref{sec:HSTphot}), and the ground-based spectroscopy $\mathcal{S}$ presented in \citetalias{Calamida18}. Clear rounded rectangles denote model parameters, while shaded rounded rectangles denote measurements. The latent (true) properties of the DA white dwarf are described by four parameters: two intrinsic parameters that describe the stellar atmosphere, the effective temperature $T_{\text{eff}}$ and the surface gravity $\log g$, and two extrinsic parameters that describe the line-of-sight Galactic extinction, $A_V$ and $R_V$. The synthetic model photometry is matched to the observed \emph{HST/WFC3} photometry using a single normalization constant $\mu$. The observations are modeled as normally distributed about the model magnitudes with variance described by the photometric uncertainty. These five parameters define the latent SED of the DA white dwarf. The spectrum also constrains the intrinsic and extrinsic parameters, but the flux normalization of the spectroscopy is not tied to the normalization of the \emph{HST} photometry. The normalization of the observed flux of the spectrum is parametrized by $d_L$ while the instrumental resolution of the observed spectrum is parametrized by $\text{FWHM}$. Errors in the flux calibration of the observed spectrum is modeled by the posterior mean of a Gaussian process with kernel parametrized by $f_{\sigma}$ and $\tau$ to describe correlated errors with wavelength and $f_{\omega}$ to describe white noise.}
\label{fig:WDmodel}
\end{figure}

The full probability density of the observations of a DA white dwarf $s$, $\mathcal{D}_s \equiv \{ \{m_s\}, \mathcal{S}_s \}$ given the model parameters $\bm{\Phi} \equiv \{ T_{\text{eff}}, \log g, A_V, R_V, d_L, \mu, \text{FWHM}, f_{\sigma}, \tau, f_{\omega} \}$ is the product of the likelihoods for the \emph{HST} photometry (Eqn.~\ref{eqn:photlikelihood}) and the ground-based spectroscopy (Eqn.~\ref{eqn:speclikelihood}):
\begin{equation}
    \begin{split}
        P(& \mathcal{D}_s |\, \bm{\Phi}) = \\
        & P(\{m_s\} |\, T_{\text{eff}}, \log g, A_V, R_V, \mu) \\ 
        \times \; & P ( \mathcal{S}_s |\, T_{\text{eff}}, \log g, A_V, R_V, d_L, \text{FWHM}, f_{\sigma}, \tau, f_{\omega} ) 
    \end{split}
    \label{eqn:wdmodellikelihood}
\end{equation}
This product of the likelihoods ensures the simultaneous and coherent modeling of the \emph{HST} photometry and ground-based spectroscopy, and is a key improvement over the model used in \citetalias{Narayan16}. The directed acyclic graph depicting the model is shown in Fig.~\ref{fig:WDmodel}.

\subsection{Priors}\label{sec:wdmodelprior}
We model the prior as separable functions on each of the model parameters. We follow \citet{gelman2006} and use weakly informative priors on all the parameters of our model. Initial guesses for the intrinsic atmosphere parameters and the spectrum normalization parameter $\{T_{\text{eff}}, \log g, d_L \}$ are determined from a least-squares minimization of the model to the observed spectrum $S$ including only the reported uncertainties in the flux of the observed spectrum along the diagonal of the covariance matrix $C$. For this initial fit, $A_V$ is fixed to 0 and $R_V$ is fixed to 3.1. The FWHM is fixed to an initial input value. The parameter estimates from this initial fit are not accurate enough to model the full dataset, but are sufficient to set weakly informative priors on these parameters. We use normal distributions with variance much larger than the expected error in the parameters to define the marginal priors on $\{T_{\text{eff}}, \log g, d_L \}$ as:
\begin{equation}
\begin{split}
    & P(T_{\text{eff}}) \propto N(T_\text{eff} |\, T^0_{\text{eff}}, \sigma = 10,000 \,\text{K}) \times I_{\{16,000 \,\text{K} \le T_{\text{eff}} \le 90,000 \,\text{K}\}}(T_{\text{eff}})\\
    & P(\log g) \propto N(\log g | \,\log g^0, \sigma = 1 \, \text{dex}) \times I_{\{7 \le \log g \le 9.5\}}(\log g) \\
    & P(d_L) \propto N(d_L | \, d^0_L, \sigma = 1,000) \times I_{\{0 < d_L < 10^7\}}(d_L)
\end{split}
\end{equation}
where the indicator function $I_{A}(x)$ returns $1$ when $x \in A$ and $0$ otherwise, limiting the intrinsic DA white dwarf parameters to the extents of the \texttt{Tlusty} model grid (see \S\ref{sec:modelgrid}), and restricting $d_L$ to be strictly greater than 0.

We set the initial guess for the normalization of the \emph{HST/WFC3} photometry $\mu$ to be the mean difference between the observed photometry $\{m_s\}$ and the synthetic photometry of the initial fit SED, determined using Eqn.~\ref{eqn:synphot}:
\begin{equation}
    \mu^0 = \frac{1}{N_{\text{PB}}} \cdot \sum_{\lambda} m_{s,\lambda} - M_{s, \lambda}(T^0_\text{eff}, \log g^0, A^0_V = 0, R^0_V = 3.1) 
\end{equation}
The marginal prior on $\mu$ is:
\begin{equation}
    \mu \sim N(\mu^0, \sigma = 10 \; \text{mag})
\end{equation}

The marginal priors on the extrinsic reddening parameters $A_V$ and $R_V$ are informed by our knowledge of the Galaxy. We infer these parameters over ranges that are much wider than can be reasonably expected for our low-extinction DA white dwarfs. We use the ``glos'' distribution, originally introduced in \citet{wv07} and derived from studies of the line-of-sight extinction to extragalactic sources, as the marginal prior on $A_V$. This marginal prior is expressed as the sum of a decaying exponential and a Gaussian distribution for $A_V \ge 0$:
\begin{equation}
\begin{split}
    P(A_V) &\propto \Bigg[ \frac{1~\text{mag}}{\alpha} \cdot \exp \left( \frac{-A_V}{\alpha} \right) + \frac{1~\text{mag}}{\sqrt{2 \pi \sigma^2} } \cdot \exp \left( \frac{-A^2_V}{2\sigma^2} \right)\Bigg]  \\ &\times I_{\{0 \le A_V \le 2~\text{mag}\}}(A_V) \\
\end{split}
\end{equation}
where $\alpha = 0.4$~mag and $\sigma = 0.1$~mag. We can expect the marginal posterior distribution on $A_V$ to be much narrower than the ``glosz'' distribution, as our sources are within the Galaxy, and the extinction is tightly constrained by our multiband \emph{HST} photometry. Similarly, we define the marginal prior on $R_V$ as:
\begin{equation}
P(R_V) \propto N(R_V |\, R_V = 3.1, \sigma = 0.18) \times I_{\{1.7 \le R_V \le 5.1\}}(R_V)
\end{equation}
where $R_V = 3.1$ is the canonical value of extinction ratio in our Galaxy for the \citetalias{Fitzpatrick99} reddening model, and the standard deviation of the Gaussian is based on the results of \citet{Schlafly16}.

We define the marginal prior on $\text{FWHM}$ as a normal distribution centered on the initial guess of the parameter $\text{FWHM}^0$ supplied with the observed spectrum $\mathcal{S}$. We use 8~\AA\ as the standard deviation of the marginal prior: 
\begin{equation}
\begin{split}
P(\text{FWHM} ) &\propto N(\text{FWHM}|\, \text{FWHM}^0, \sigma = 8~\text{\AA}) \\
&\times I_{\{0 < \text{FWHM} \le 25~\text{\AA}\}}(\text{FWHM})
\end{split}
\end{equation}
 
The amplitudes of the component kernels of the Gaussian process are the product of the scale-free amplitudes, $f_{\sigma}$ and $f_{\omega}$, and the median reported uncertainty on the flux of the observed spectrum $\mean{\sigma}_{\mathcal{S}}$. We these parameters describe the noise, we define them to be positive and expect them to be at most $\mathcal{O}(1)$. We therefore define the marginal priors on the scale-free amplitudes as half-Cauchy distributions:
\begin{equation}
    \begin{split}
        f_{\sigma} & \sim HC(0, \beta=3) \\
        f_{\omega} & \sim HC(0, \beta=3) \\ 
    \end{split}
\end{equation}
Finally, the marginal prior on the scale of the correlations $\tau$ is defined as:
\begin{equation}
    \tau \sim U(500\, \text{\AA},\, 5,000\, \text{\AA})
\end{equation}
appropriate for the reduced spectra presented in \citetalias{Calamida18}, where the splines used for the flux calibration have knots that are more widely spaced than twice the typical $\sim 200$~\AA\ width of the H$_\beta$ Balmer feature.

The full prior on the model parameters $\bm{\Phi}$ is the product of the marginal prior distributions of each of the model parameters:
\begin{equation}
\begin{split}
P(\bm{\Phi}) \, & = \, P(T_\text{eff}) \cdot P(\log g) \\ 
& \times P(A_V) \cdot P(R_V) \\
& \times P(\mu) \\
& \times P(d_L) \cdot P(\text{FWHM}) \\ 
& \times P(f_{\sigma}) \cdot P(\tau) \cdot P(f_{\omega}) \\
\end{split}
\label{eqn:wdmodelprior}
\end{equation}

\subsection{Posterior Distribution and Estimation}
\begin{figure*}[htb]
    \centering
    \includegraphics[width=\textwidth]{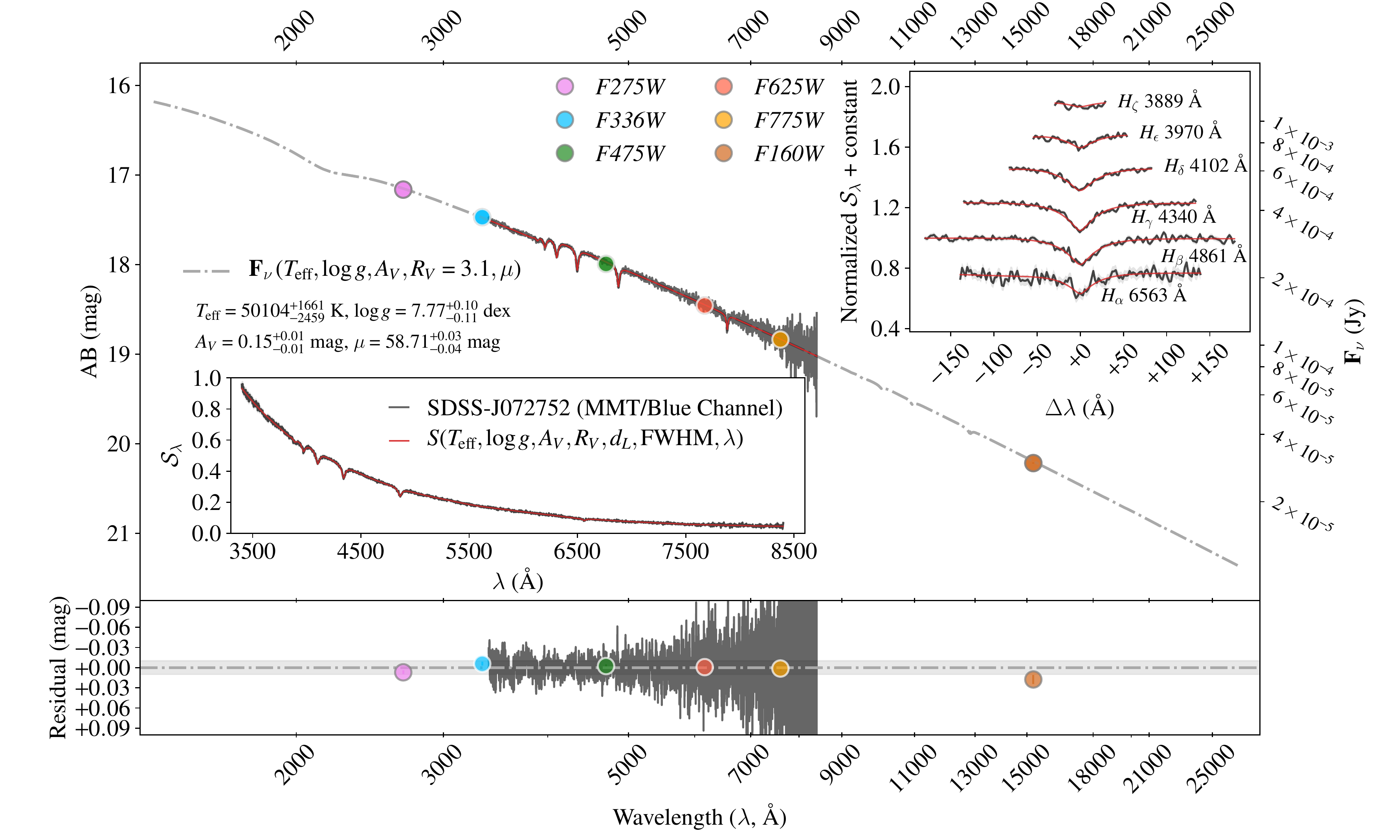}
    \caption{Example of an inferred DA white dwarf SED $\mathbf{F}_{\nu}$ (Top). We model the observed spectrum (Lower Left Inset) and our \emph{HST/WFC3} apparent magnitudes together to infer the parameters of the DA white dwarf SED $\{ T_{\text{eff}}, \log g, A_V, R_V = 3.1, \mu \}$. The observed spectrum $\mathcal{S}_{\lambda}$ (black) and the model of the inferred model spectrum (red) are in excellent agreement. We normalize the continuum of the spectrum to unity and highlight the region around the individual Balmer lines (Upper Right Inset). While both spectrum and model spectrum are plotted, the two are effectively indistinguishable. The photometric normalization parameter $\mu$ is inferred solely from the \emph{HST/WFC3} photometry (circle markers, plotted at the effective wavelength). The residuals (Bottom) in each of our passband are typically less than 5~mmag for the UVIS channel, and $\sim0.01$~mag for the IR channel. We tie the observed spectrum to the CALSPEC flux scale using the outputs of the \texttt{WDmodel} code. The residuals between the calibrated spectrum and the model are shown in black and are consistent with heteroskedastic white noise.}\label{fig:WDmodelFit}
\end{figure*}

The full posterior distribution of the model $\bm{\Phi}$ given the observations of a DA white dwarf star $s$, $\mathcal{D}_s$ is the product of the likelihood (Eqn.~\ref{eqn:wdmodellikelihood}) and the prior (Eqn.~\ref{eqn:wdmodelprior}).
\begin{equation}
 P(\bm{\Phi} |\, \mathcal{D}_s ) \propto P(\mathcal{D}_s |\, \bm{\Phi}) \cdot P(\bm{\Phi})
\end{equation}

We use the apparent magnitudes of our DA white dwarfs tied to the CALSPEC system determined in \S\ref{sec:HSTphot} and the spectroscopy presented in \citetalias{Calamida18} without any additional preprocessing. As described in \S\ref{sec:wdmodelprior}, we perform an fit of the spectrum to obtain initial guesses for the model parameters $\{ T_\text{eff},\log g, d_L, \mu \}$. Initial guesses and prior distributions for the remaining parameters are either inputs implicitly provided with the observation $\mathcal{D}$, or can be set without reference to the observations. 

We use the Parallel Tempering ensemble Markov Chain Monte Carlo (PTMCMC) algorithm implemented in the \texttt{emcee} python package~\citep{dfm2013}. We run 4 chains at different temperatures, each with 100 walkers. Each walker is initialized to different positions, and we save only every 10\textsuperscript{th} position of each walker as a step to construct the Markov chain to ensure the samples are not correlated. Following an initial burn-in of 500 steps, which are discarded, we save a chain with a thinned length of 5,000 steps. We use a suite of diagnostic tests for convergence, including visually inspecting the mixing of the chains, inspecting the marginalized 2-D joint posterior distributions of the parameters for any artifacts, and verify that the Gelman-Rubin statistic is near unity. Figure~\ref{fig:WDmodelFit} presents an example of a DA white dwarf SED inferred with the \texttt{WDmodel} code, combining all the various elements described in \S\ref{sec:wdmodel}.

\begin{figure}[h!tpb]
    \centering
    \includegraphics[width=0.47\textwidth]{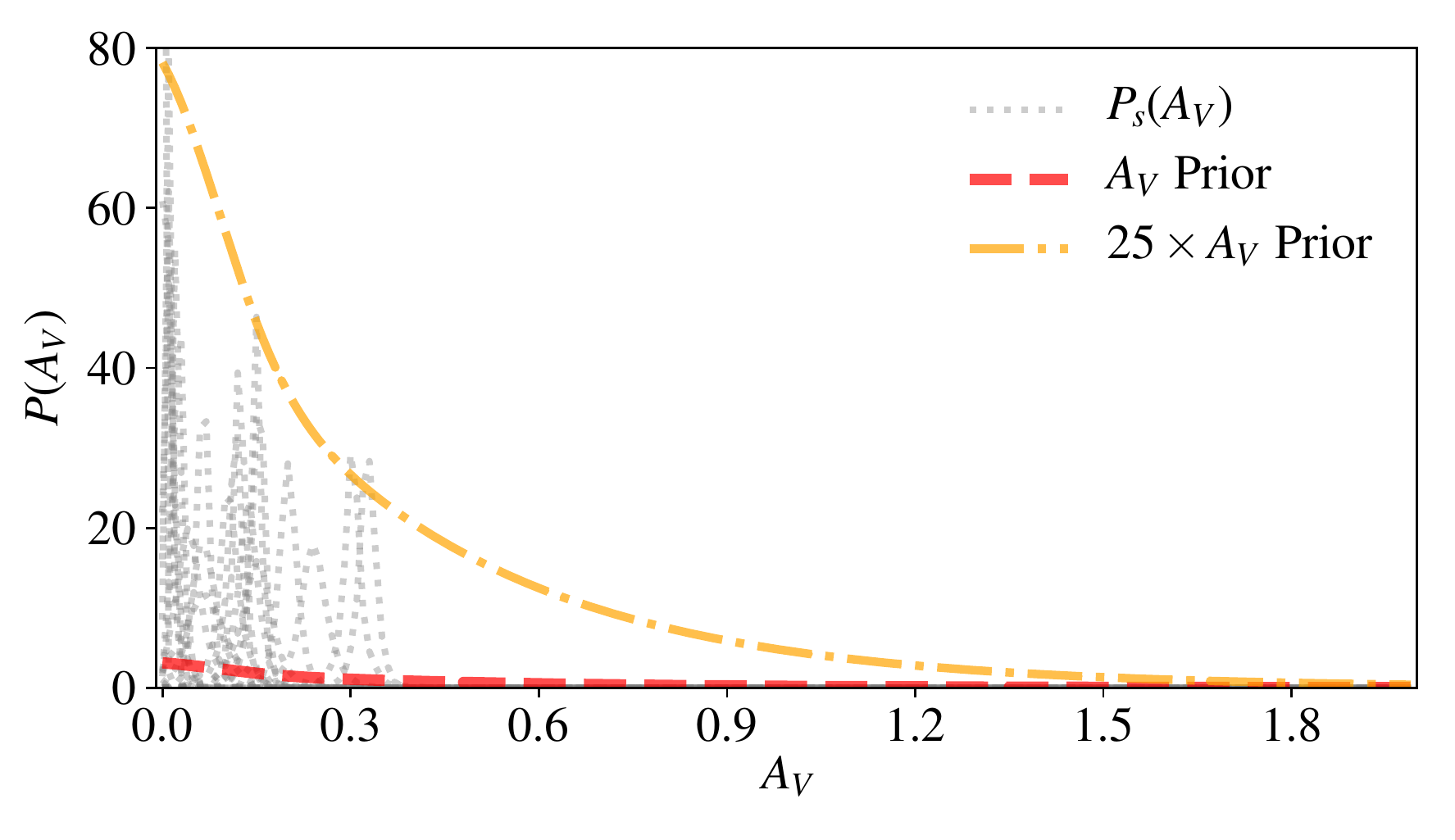}
    \caption{Distributions of inferred $A_V$ for each object (grey, dotted lines) are much narrower than the ``glosz'' prior on $A_V$ (red, dashed line) determined by \citet{wv07}. The shape of the individual marginal posterior distributions is much narrower than the prior (multiplied by a factor of 25 in orange for comparison), and we are justified in treating the ``glosz'' distribution as a weakly informative prior.}
    \label{fig:av_distrib}
\end{figure}

\begin{figure}[h!tpb]
    \centering
    \includegraphics[width=0.47\textwidth]{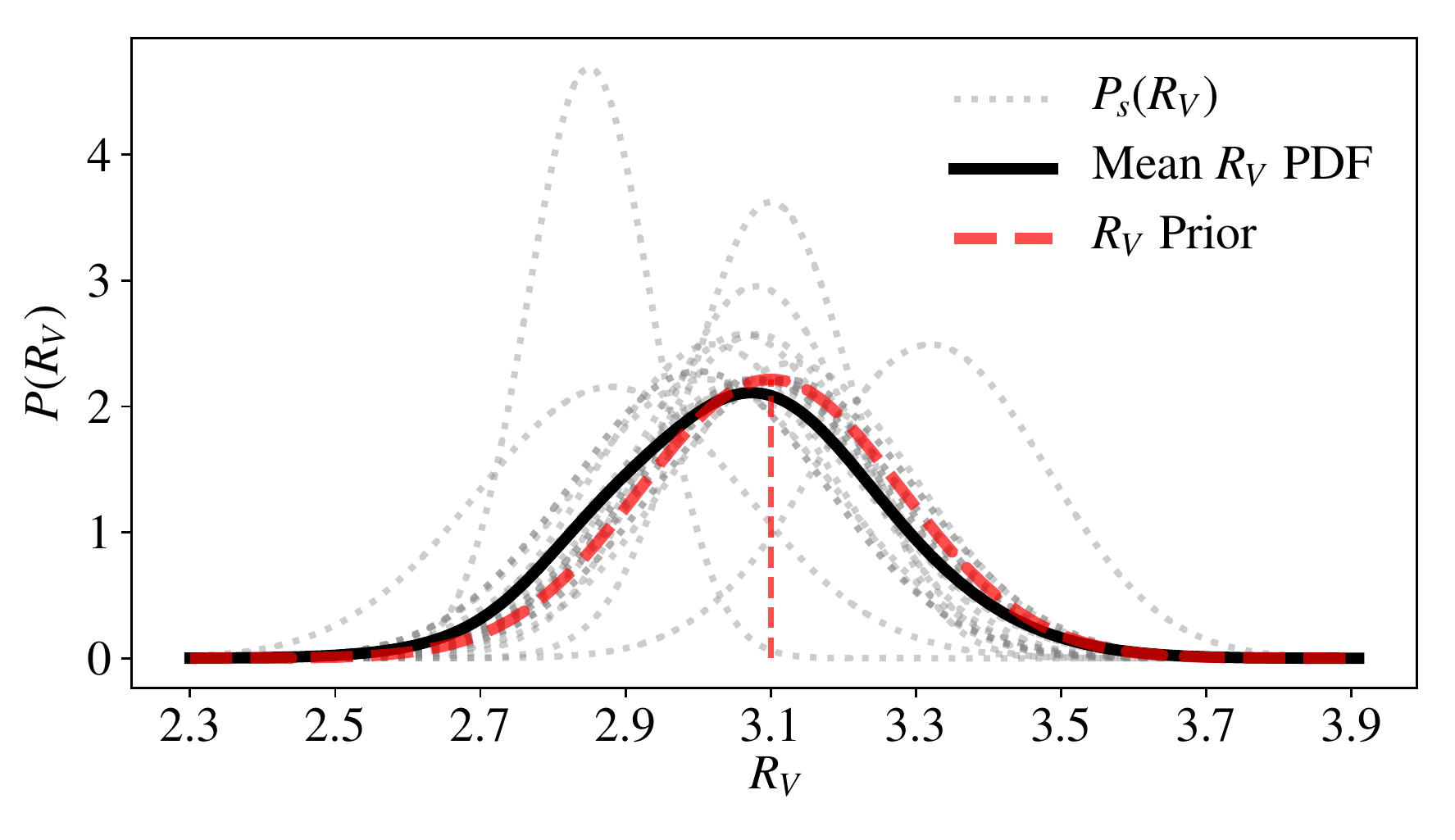}
    \caption{Distributions of inferred $R_V$ for each object (grey, dotted lines) and the mean distribution of the entire sample (black line) is consistent with the prior on $R_V$ centered at 3.1, with a width of 0.18 (red, dashed line) determined by \citet{Schlafly16}. This behavior arises as our DA white dwarfs were chosen to be in low-extinction environments, consistent with diffuse interstellar dust. We have fixed $R_V = 3.1$ for the analysis in this work.}
    \label{fig:rv_distrib}
\end{figure}

While our prior is chosen to be weakly informative, the marginal priors on the extinction parameters $A_V$ and $R_V$ are physically motivated and therefore merit comparison with the inferred marginal posterior distributions. This posterior predictive test checks if our inferred parameters are strongly affected by the choice of prior. In Fig.~\ref{fig:av_distrib}, we show the inferred marginal posterior distribution of $A_V$ for each star compared with the prior. It is evident that the posterior distributions of each star are much narrower than the prior distribution, and our treatment of the ``glosz'' prior as weakly informative is justified. Our initial fits left $R_V$ as a free parameter, but the inferred posterior distribution in $R_V$ is not significantly different from the prior. We illustrate this in Fig.~\ref{fig:rv_distrib} which compares the recovered mean $R_V$ marginal posterior distribution with the prior. This behavior is not surprising as our DA white dwarfs were selected to be in low line-of-sight extinction regions of our Galaxy, and extinction due to diffuse interstellar dust is well described by a canonical $R_V = 3.1$ \citetalias{Fitzpatrick99} model. We have therefore elected to fix $R_V$ to 3.1 for the results presented in this work. This value is appropriate for diffuse interstellar dust outside the plane of our Galaxy, and there is little justification for allowing $R_V$ to vary, given that the \citetalias{Fitzpatrick99} determination is more precise than the value we can determine from our 6-band photometry of each star.

\begin{figure*}[h!tpb]
    \centering
    \includegraphics[width=0.82\textwidth]{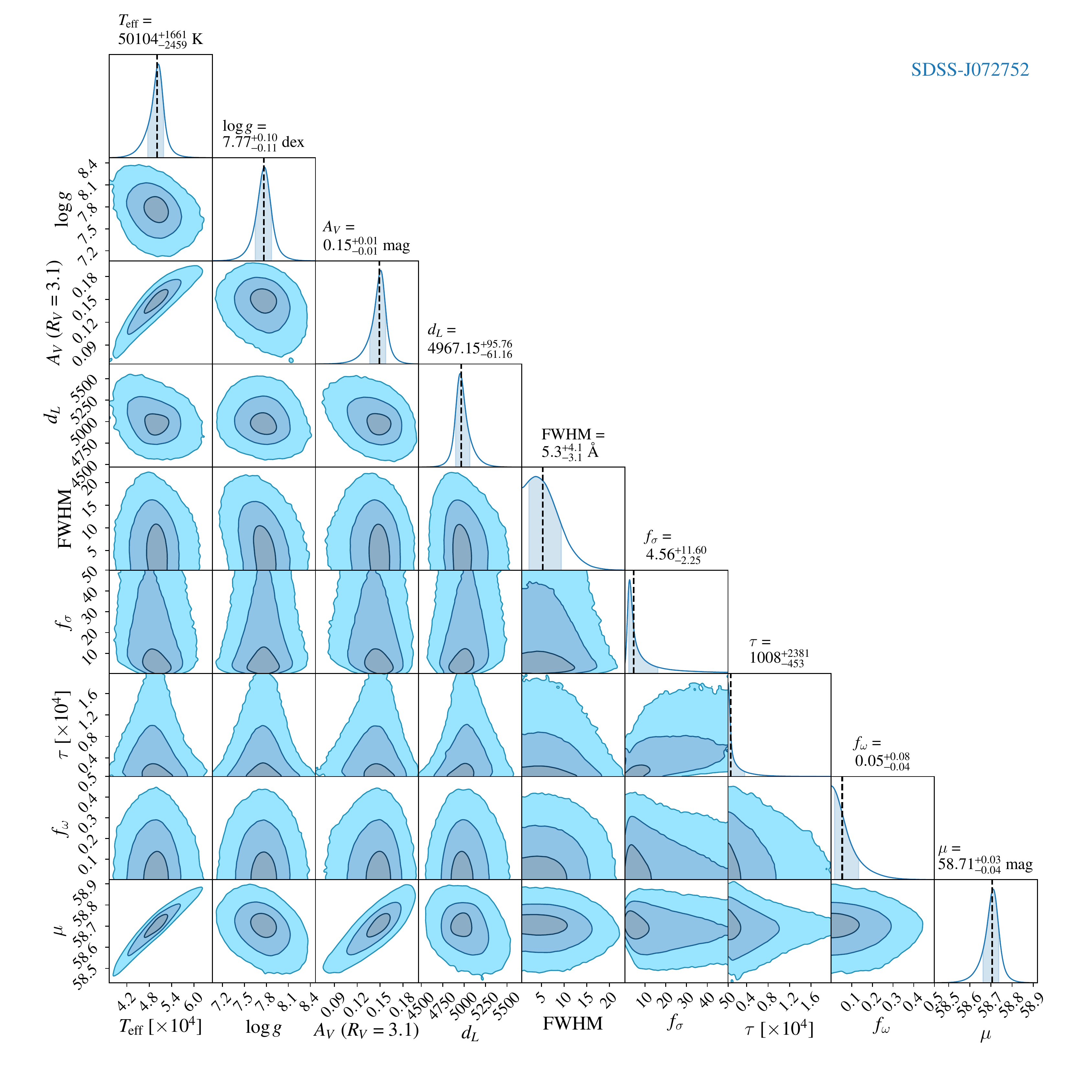}
    \caption{Corner plot for the fit illustrated in Fig.~\ref{fig:WDmodelFit} showing the 2-D joint posterior distributions and 1-D marginalized posterior distributions for each of the model parameters. The 1, 2 and 3$\sigma$ contours are shown with progressively lighter shading. The 50\textsuperscript{th} percentile of the distribution of each parameter is reported above each 1-D marginalization and indicated with a dashed vertical black line.}\label{fig:WDmodelCorner}
\end{figure*}

The 1-D marginal posterior do not capture the correlations between the parameters. Figure~\ref{fig:WDmodelCorner} illustrates the 2-D joint posterior distributions and 1-D marginalized distributions corresponding to the inferred SED in Fig.~\ref{fig:WDmodelFit}. These joint distributions illustrate the strong correlation between $T_{\text{eff}}$, $A_V$ and $\mu$ --- a model can be made hotter (brighter) and still match our \emph{HST/WFC3} observations if the line-of-sight extinction or the distance to the source is increased. These correlations are weakened by the ground-based spectroscopy as the temperature cannot be changed without limit as the shape of the Balmer lines and the continuum slope would no longer agree with the observed spectrum.

The correlations between the intrinsic DA white dwarf parameters and the extrinsic parameters were not captured in the analysis presented in \citetalias{Narayan16}. As a result of accounting for these correlations, the errors reported on the parameters of our model are larger than in the \citetalias{Narayan16} analysis. Our new \texttt{WDmodel} methodology provides a more principled accounting of the uncertainty associated with the measurements of each object and does not exhibit bias when testing when tested with synthetic spectra. We propagate the uncertainty on the parameters $\{ T_{\text{eff}}, \log g, A_V, \mu \}$ to the inferred SED $\mathbf{F}$. The SED is independent of the parameters $\{d_L, \text{FWHM}, f_{\sigma}, \tau, f_{\omega} \}$ as these are only used to model the observed spectrum. We have made our \texttt{WDmodel} analysis package\footnote{\url{http://github.com/gnarayan/WDmodel/}} and detailed documentation\footnote{\url{http://wdmodel.rtfd.io}} public. \texttt{WDmodel} has been successfully used by members of DES to model independently obtained spectra of DA white dwarfs (D. Tucker, D. Guellidge, private communication).

\newpage
\section{Results and Systematics}\label{sec:WDmodelResults}

\begin{figure*}[htpb]
   \centering
    \includegraphics[width=0.99\textwidth]{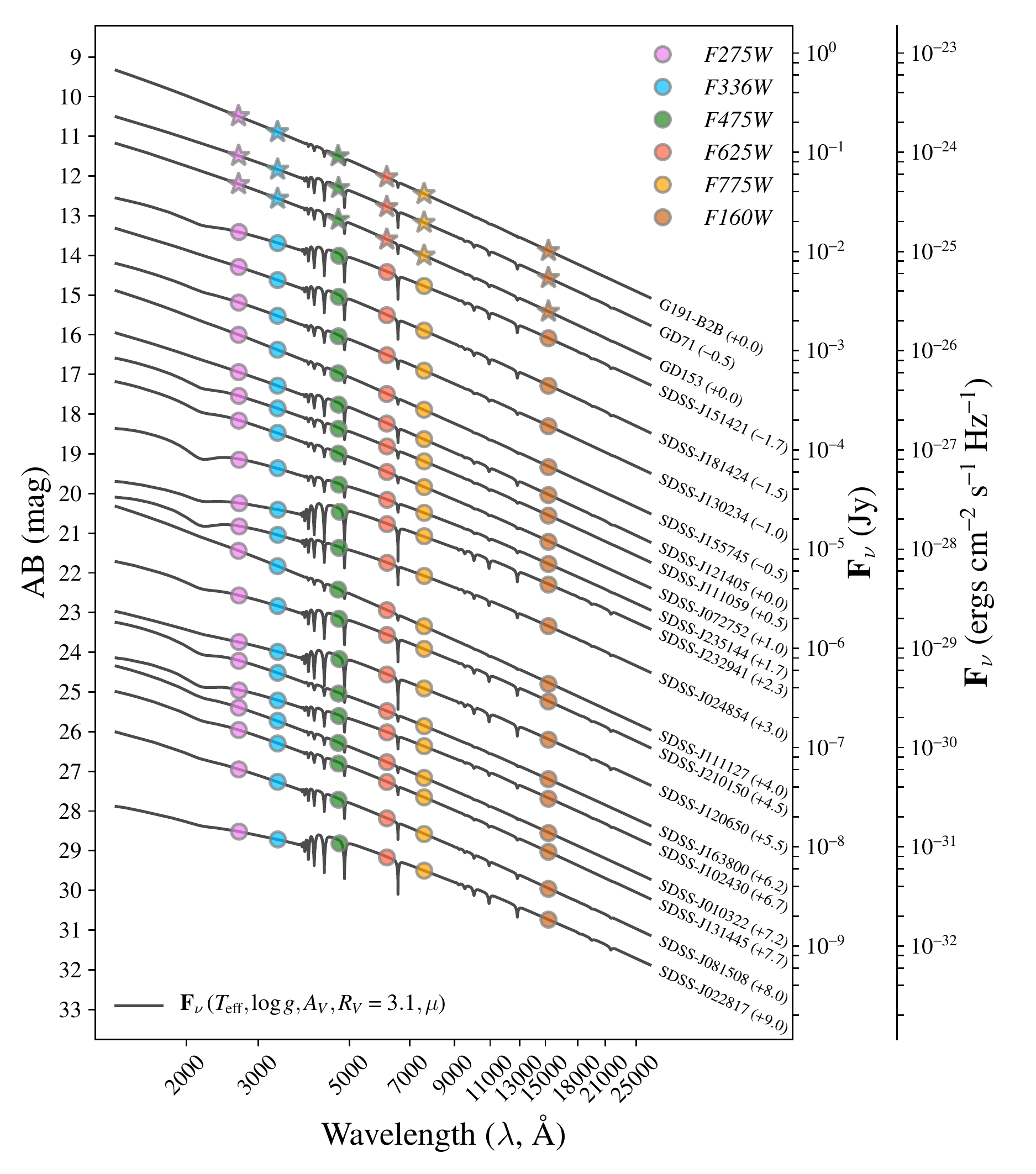}
    \caption{Sequence of calibrated SEDs of our DA white dwarfs shown in AB magnitudes. The SEDs are spaced vertically for clarity by a magnitude offset indicated parenthetically in the label with the object name. The axes on the right specify the SED per unit frequency $\mathbf{F}_{\nu}$, in units of janskys and ergs cm$^{-2}$ s$^{-1}$ Hz$^{-1}$. The observed \emph{HST/WFC3} photometry presented in Table \ref{table:phot} are shown with colored markers (stars for the three CALSPEC primary standards, circles for our program stars) on each SED at the effective wavelength of each source through each passband. The uncertainties on the photometry and the inferred SEDs are smaller than the markers and the lines respectively. The synthetic model magnitudes and residuals between the observations and the synthetic model magnitudes are presented in Table~\ref{table:resid} and Fig.~\ref{fig:WDmodelPhotResid}.}
    \label{fig:WDmodelSED}
\end{figure*}

We list the model parameters $\{ T_{\text{eff}}, \log g, A_V, \mu \}$ describing the SED for each DA white dwarf in our sample in Table \ref{table:WDmodelFit}. The inferred SEDs are tied to the photometric system defined by the three CALSPEC primary standards, and are shown in Fig.~\ref{fig:WDmodelSED}. We also present our inferred SEDs of the three CALSPEC standards. We compare our inferred SEDs for the three primary standards to their original CALSPEC SEDs to quantify the systematic differences between our photometric system and the CALSPEC photometric system in \S\ref{sec:calspeccomp}. 

The model parameters $\{ d_L, \text{FWHM}, f_{\sigma}, \tau, f_{\omega} \}$ are specific to each spectrum and are not directly comparable across objects, and have no effect on the inferred SED $\mathbf{F}$. A machine-readable table including these parameters for our entire sample of DA white dwarfs, together with plots and posterior samples drawn from the Markov chain for each spectrum, is available through our archive (see footnote~\ref{footnote:archive}). We have verified that neither the photometric residuals in any passband, nor the spectroscopic residuals correlate with any of the inferred model parameters, indicating that the posterior distribution is well-sampled and there are no unmodeled correlations. The residuals between the spectra and the model are consistent with white noise. 

\begin{table}[htpb]
\scriptsize
\begin{centering}
\begin{tabular}{l|cccc}
\hline
\hline
Object & $T_{\text{eff}}$ & $ \log g $ & $A_V~(R_V = 3.1) $ & $\mu$ \\
 & K & dex & mag & mag   \\
\hline
G191-B2B \rule{0ex}{1.5ex} & $64161^{+1126}_{-776}$ & $7.57^{+0.08}_{-0.09}$ & $0.00^{+0.00}_{-0.00}$ & $52.64^{+0.02}_{-0.01}$ \\
GD153 & $40087^{+827}_{-497}$ & $7.72^{+0.07}_{-0.07}$ & $0.01^{+0.01}_{-0.00}$ & $53.73^{+0.02}_{-0.01}$ \\
GD71 \rule[-1.5ex]{0ex}{0ex} & $33012^{+417}_{-241}$ & $7.82^{+0.04}_{-0.05}$ & $0.01^{+0.01}_{-0.00}$ & $53.12^{+0.02}_{-0.01}$ \\
\hline
SDSS-J010322  \rule{0ex}{1.5ex} & $59061^{+3389}_{-4122}$ & $7.47^{+0.14}_{-0.15}$ & $0.12^{+0.01}_{-0.02}$ & $60.01^{+0.05}_{-0.06}$ \\
SDSS-J022817 & $21391^{+722}_{-608}$ & $7.94^{+0.08}_{-0.10}$ & $0.07^{+0.04}_{-0.03}$ & $59.14^{+0.03}_{-0.03}$ \\
SDSS-J024854 & $33266^{+873}_{-797}$ & $7.22^{+0.13}_{-0.12}$ & $0.30^{+0.01}_{-0.01}$ & $58.35^{+0.04}_{-0.03}$ \\
SDSS-J072752 & $50104^{+1661}_{-2459}$ & $7.77^{+0.10}_{-0.11}$ & $0.15^{+0.01}_{-0.01}$ & $58.71^{+0.03}_{-0.04}$ \\
SDSS-J081508 & $34735^{+1709}_{-1109}$ & $7.20^{+0.11}_{-0.10}$ & $0.07^{+0.03}_{-0.02}$ & $60.06^{+0.06}_{-0.04}$ \\
SDSS-J102430 & $36691^{+1666}_{-1483}$ & $7.54^{+0.28}_{-0.22}$ & $0.24^{+0.02}_{-0.02}$ & $59.13^{+0.05}_{-0.05}$ \\
SDSS-J111059 & $46298^{+1839}_{-2339}$ & $7.85^{+0.12}_{-0.12}$ & $0.15^{+0.01}_{-0.01}$ & $58.48^{+0.03}_{-0.05}$ \\
SDSS-J111127 & $59422^{+2105}_{-2257}$ & $7.76^{+0.12}_{-0.13}$ & $0.03^{+0.01}_{-0.01}$ & $59.45^{+0.03}_{-0.03}$ \\
SDSS-J120650 & $23434^{+456}_{-408}$ & $7.94^{+0.05}_{-0.05}$ & $0.04^{+0.02}_{-0.02}$ & $58.21^{+0.02}_{-0.02}$ \\
SDSS-J121405 & $33750^{+707}_{-465}$ & $7.96^{+0.08}_{-0.10}$ & $0.01^{+0.01}_{-0.01}$ & $58.11^{+0.03}_{-0.02}$ \\
SDSS-J130234 & $40028^{+1377}_{-1238}$ & $7.94^{+0.08}_{-0.08}$ & $0.06^{+0.01}_{-0.01}$ & $57.59^{+0.03}_{-0.03}$ \\
SDSS-J131445 & $43670^{+2812}_{-2035}$ & $7.65^{+0.18}_{-0.14}$ & $0.11^{+0.02}_{-0.02}$ & $59.72^{+0.05}_{-0.04}$ \\
SDSS-J151421 & $28768^{+297}_{-300}$ & $7.89^{+0.04}_{-0.04}$ & $0.12^{+0.01}_{-0.01}$ & $55.58^{+0.01}_{-0.01}$ \\
SDSS-J155745 & $56760^{+1787}_{-2042}$ & $7.69^{+0.11}_{-0.12}$ & $0.02^{+0.01}_{-0.01}$ & $58.46^{+0.02}_{-0.03}$ \\
SDSS-J163800 & $57181^{+4238}_{-3931}$ & $7.63^{+0.25}_{-0.24}$ & $0.20^{+0.01}_{-0.02}$ & $59.62^{+0.06}_{-0.06}$ \\
SDSS-J181424 & $30806^{+325}_{-258}$ & $7.88^{+0.05}_{-0.05}$ & $0.01^{+0.01}_{-0.01}$ & $56.71^{+0.02}_{-0.01}$ \\
SDSS-J210150 & $29062^{+516}_{-536}$ & $7.85^{+0.09}_{-0.08}$ & $0.14^{+0.02}_{-0.02}$ & $58.54^{+0.02}_{-0.02}$ \\
SDSS-J232941 & $21044^{+445}_{-500}$ & $7.96^{+0.07}_{-0.07}$ & $0.15^{+0.03}_{-0.03}$ & $57.35^{+0.02}_{-0.02}$ \\
SDSS-J235144 \rule[-1.5ex]{0ex}{0ex} & $41058^{+1993}_{-1752}$ & $7.99^{+0.15}_{-0.17}$ & $0.33^{+0.01}_{-0.01}$ & $58.35^{+0.04}_{-0.04}$ \\
\hline
\end{tabular}
\footnotesize{\tablecomments{Parameters are reported as the median of the marginal posterior distributions. The $\pm$34\% percentile interval about the median is reported as superscript and subscript respectively. For objects with multiple spectra, we provide the parameters of the result with the highest log-likelihood.}}
\caption{Parameters of the DA White Dwarf SEDs inferred from spectroscopy and \emph{HST} photometry with the \texttt{WDmodel} code\label{table:WDmodelFit}}
\end{centering}
\end{table}

\begin{figure*}[hptb]
   \centering
    \includegraphics[width=\textwidth]{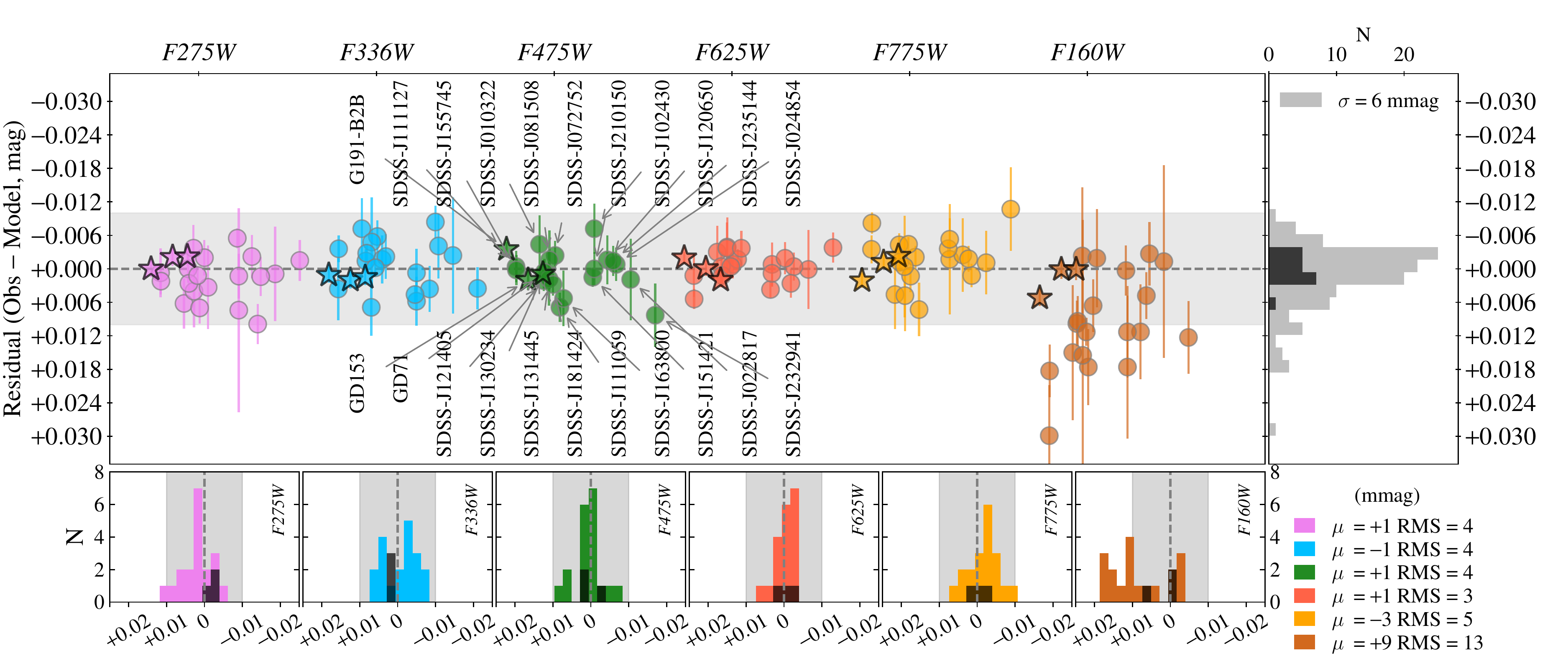}
    \caption{Left, Upper Row: Photometric Residuals (in the sense of observed magnitude $-$ model magnitude) for our network of DA white dwarf stars (circles) and CALSPEC primary standards (stars, not included in the computation of any summary statistics) in each passband. Objects are labelled in the \textit{F475W} passband, and their relative horizontal position is set by an amount proportional to their $g-r$ color, and is the same across all passbands. The light grey region indicates a standard deviation of 0.1~mag. Right, Upper Row: The distribution of residuals of our network DA white dwarfs across all passbands has a standard deviation of 6~mmag. Excluding \textit{F160W} reduces this standard deviation to $\sim3$~mmag. Lower Row: Histogram of residuals in each passband with the mean residual and RMS indicated in the legend. As above, the light grey region indicates a standard deviation of 0.1~mag. The three primary standards are indicated in black in all the histograms. The mean residual in each band $\mu$ is less than 5~mmag in all bands except \textit{F160W}. The RMS and standard deviation of the DA white dwarfs is much larger in \textit{F160W} than in any of the optical passbands.}
    \label{fig:WDmodelPhotResid}
\end{figure*}

The photometric residuals for each DA white dwarf in each passband are presented in Table~\ref{table:resid} and are shown in Fig.~\ref{fig:WDmodelPhotResid}. The residuals of the three CALSPEC standards are also shown but not used in the computation of summary statistics. Our analysis extends the set of DA white dwarfs with SED models that predict observed fluxes to better than 1\% from the 3 CALSPEC standards to 22 objects. This analysis also extends the dynamic range spanned by the stars from $\mathbf{\sim1.5}$~mag to $\mathbf{\sim7.5}$~mag. The standard deviation of the residuals of our DA white dwarfs across all passbands is 6~mmag. 

\begin{table*}[h!tpb]
\scriptsize
\begin{centering}
\begin{tabular}{l|cccccccccccc}
\hline
\hline
Object & \textit{mF275W} & \textit{rF275W} & \textit{mF336W} & \textit{rF336W} & \textit{mF475W} & \textit{rF475W} & \textit{mF625W} & \textit{rF625W} & \textit{mF775W} & \textit{rF775W} & \textit{mF160W} & \textit{rF160W} \\
 & \multicolumn{12}{c}{(AB mag)} \\
\hline
G191-B2B & 10.4903 & \phantom{+}0.0000 & 10.8891 & +0.0011 & 11.5023 & $-$0.0035 & 12.0326 & $-$0.0020 & 12.4493 & +0.0021 & 13.8800 & +0.0052 \\
GD153 & 12.2037 & $-$0.0021 & 12.5659 & +0.0020 & 13.0978 & +0.0020 & 13.5977 & $-$0.0001 & 14.0029 & $-$0.0012 & 15.4138 & +0.0001 \\
GD71 & 11.9910 & $-$0.0022 & 12.3345 & +0.0015 & 12.7979 & +0.0009 & 13.2770 & +0.0020 & 13.6743 & $-$0.0023 & 15.0675 & +0.0001 \\
\tableline
SDSS-J010322 & 18.1890 & +0.0062 & 18.5340 & $-$0.0072 & 19.0877 & $-$0.0044 & 19.5716 & $-$0.0030 & 19.9602 & +0.0046 & 21.3401 & +0.0150 \\
SDSS-J022817 & 19.5174 & +0.0009 & 19.7176 & $-$0.0024 & 19.8132 & +0.0019 & 20.1689 & +0.0001 & 20.5025 & $-$0.0011 & 21.7383 & $-$0.0013 \\
SDSS-J024854 & 17.8272 & +0.0014 & 18.0441 & $-$0.0041 & 18.3704 & $-$0.0008 & 18.7464 & $-$0.0004 & 19.0760 & +0.0012 & 20.3427 & $-$0.0027 \\
SDSS-J072752 & 17.1566 & +0.0070 & 17.4772 & $-$0.0058 & 17.9956 & $-$0.0024 & 18.4575 & $-$0.0008 & 18.8357 & +0.0013 & 20.1990 & +0.0176 \\
SDSS-J081508 & 18.9465 & +0.0040 & 19.2684 & $-$0.0048 & 19.7178 & $-$0.0016 & 20.1877 & $-$0.0039 & 20.5746 & +0.0048 & 21.9462 & +0.0155 \\
SDSS-J102430 & 18.2532 & +0.0074 & 18.5085 & +0.0058 & 18.9113 & $-$0.0072 & 19.3182 & $-$0.0008 & 19.6667 & $-$0.0017 & 20.9730 & +0.0176 \\
SDSS-J111059 & 17.0427 & $-$0.0020 & 17.3561 & $-$0.0017 & 17.8600 & +0.0069 & 18.3152 & $-$0.0017 & 18.6908 & $-$0.0021 & 20.0499 & +0.0066 \\
SDSS-J111127 & 17.4422 & +0.0007 & 17.8262 & +0.0036 & 18.4209 & $-$0.0004 & 18.9378 & +0.0012 & 19.3476 & $-$0.0035 & 20.7676 & +0.0299 \\
SDSS-J120650 & 18.2418 & $-$0.0022 & 18.4852 & +0.0036 & 18.6735 & $-$0.0016 & 19.0620 & $-$0.0019 & 19.4136 & $-$0.0025 & 20.6914 & +0.0113 \\
SDSS-J121405 & 16.9400 & \phantom{+}0.0000 & 17.2841 & $-$0.0014 & 17.7585 & +0.0021 & 18.2366 & $-$0.0004 & 18.6335 & $-$0.0043 & 20.0281 & +0.0098 \\
SDSS-J130234 & 16.1853 & +0.0026 & 16.5244 & $-$0.0028 & 17.0358 & +0.0006 & 17.5128 & +0.0012 & 17.9048 & $-$0.0012 & 19.2938 & +0.0094 \\
SDSS-J131445 & 18.2614 & $-$0.0037 & 18.5900 & +0.0069 & 19.1000 & +0.0018 & 19.5705 & $-$0.0037 & 19.9557 & $-$0.0004 & 21.3307 & $-$0.0023 \\
SDSS-J151421 & 15.1155 & $-$0.0055 & 15.3860 & +0.0046 & 15.7075 & +0.0015 & 16.1165 & +0.0037 & 16.4748 & $-$0.0036 & 17.7867 & +0.0003 \\
SDSS-J155745 & 16.4977 & +0.0022 & 16.8803 & $-$0.0036 & 17.4700 & +0.0002 & 17.9864 & +0.0054 & 18.3962 & $-$0.0082 & 19.8159 & +0.0183 \\
SDSS-J163800 & 18.0125 & +0.0033 & 18.3199 & $-$0.0022 & 18.8346 & +0.0053 & 19.2846 & $-$0.0037 & 19.6531 & +0.0073 & 20.9981 & $-$0.0019 \\
SDSS-J181424 & 15.7902 & +0.0012 & 16.1215 & $-$0.0002 & 16.5413 & +0.0028 & 17.0061 & $-$0.0004 & 17.3971 & $-$0.0044 & 18.7743 & +0.0113 \\
SDSS-J210150 & 18.0669 & +0.0013 & 18.3337 & +0.0007 & 18.6560 & \phantom{+}0.0000 & 19.0629 & +0.0008 & 19.4193 & $-$0.0053 & 20.7283 & +0.0113 \\
SDSS-J232941 & 17.9449 & $-$0.0015 & 18.1055 & +0.0035 & 18.1524 & +0.0083 & 18.4735 & $-$0.0038 & 18.7859 & $-$0.0107 & 19.9826 & +0.0123 \\
SDSS-J235144 & 17.4394 & +0.0099 & 17.6703 & $-$0.0084 & 18.0766 & $-$0.0014 & 18.4569 & +0.0026 & 18.7887 & $-$0.0018 & 20.0699 & +0.0048 \\
\tableline
\end{tabular}
\footnotesize{\tablecomments{Model magnitudes on the AB system computed from synthetic photometry of our inferred SEDs. Model magnitudes are reported in columns with names corresponding to the passband names prefixed with a `m', while the difference between the observed magnitudes reported in Table~\ref{table:phot} and model magnitudes are reported in columns prefixed with an `r'. All quantities are rounded to a tenth of a millimag.}}
\caption{Model AB magnitudes and residuals for the network of DA White Dwarfs and CALSPEC Primary Standards\label{table:resid}}
\end{centering}
\end{table*}

\subsection{Evaluating Sources of Systematic Errors}
We must evaluate any correlated bias in the inference, leading to all the SED models differing systematically from the truth. While the agreement between model and data is better than 1\%, it is critical to control these systematic effects as errors in the spectrophotometric standards can propagate across wide-field surveys. It is likely that some of these systematic errors already affect our inference. A conspicuous feature of the photometric residuals in Fig.~\ref{fig:WDmodelPhotResid} is that the model and data disagree at the few mmag level in \textit{F160W} in the sense that the model overestimates the flux in the NIR. This disagreement is particularly evident in comparison to the \emph{WFC3/UVIS} passbands. The mean residual of our DA white dwarf in each band $\mu$ is less than 4~mmag in all bands except \textit{F160W}. Excluding \textit{F160W} reduces the standard deviation of the residuals to 4~mmag. The mean residual in each passband is consistent with 0 except in \textit{F160W} where there is a 9~mmag bias in the sense of the observed magnitudes being too faint relative to the model prediction. In order to mitigate these systematic errors, we must identify their sources and estimate their impact. There are several elements of our analysis that are common to all objects (see Fig.~\ref{fig:schematic}), and therefore potential sources of systematic error in our SEDs:
\begin{enumerate}
    \item The DA white dwarf atmosphere model grid
    \item The shape of the reddening law of wavelength
    \item The shape of the passband response functions
    \item The overall flux-normalization or zeropoints
    \item The linearity of the \emph{HST/WFC3} detectors
\end{enumerate}
We evaluate the effect of each on the SED models in the following sections, and summarize their impact in \S\ref{sec:syssummary}.

\subsection{Errors in the DA White Dwarf Model Grid}
To estimate any error in the SEDs caused by an error in the DA white dwarf model atmosphere grid, we considered a different atmosphere grid provided by one of us (I. Hubeny). This new grid extends to 30~\micron\ and is based on \texttt{Tlusty} v205~\citep{Hubeny17}. The new grid incorporates Bracket and Pfund series line profiles, and improves the smoothing of the Lyman and Balmer pseudocontinua. Numerically, the NLTE models were constructed using 30,000 internal frequency points to model the discretized mean intensity of radiation, while the original grid used in this work (see \S\ref{sec:modelgrid}) used 5,000. The emergent spectra are constructed with resolution $R = 5000$. Finally, the spacing in $\log g$ was reduced to 0.25~dex from the 0.5~dex spacing used in this work. We found no difference between the new grid and the grid we use in this work. Consequently, the intrinsic parameters and inferred SEDs are in very close agreement.

We also compared our \texttt{Tlusty} model grid vs models generated from the T{\"u}bingen NLTE Model Atmosphere Package~\citep[\texttt{TMAP},][]{Rauch16} grid at the same model parameters $\{ T_{\text{eff}}, \log g \}$. The mean difference between the model atmospheres at the same model parameters is irrelevant as this would be absorbed into the overall flux normalization parameter $\mu$ in \S\ref{sec:wdmodel}. We found the residual differences about the mean offset are 1--2~mmag, significantly smaller than the discrepancy in \textit{F160W} we are seeking to explain. We note that comparison of the two grids at the same values of the intrinsic parameters is a bound on the worst case error, and \citetalias{Narayan16} demonstrated that the best-matching model for any \texttt{TMAP} atmosphere from the \texttt{Tlusty} grid has a slightly different $\{ T_{\text{eff}}, \log g \}$. It is possible that there is a common-mode error in the shape of the continuum of all the DA white dwarf model grids. However, given that the existing residuals between our SED models and data are consistent with white noise in the \emph{WFC3/UVIS} passbands, any such error must be $<1$~mmag. 

All of our DA white dwarf stars have very similar colors, and any error in the grid would likely cause a nearly constant offset in the residuals. In particular, the atmosphere of DA white dwarfs is largely dominated by the Rayleigh-Jeans tail at NIR wavelengths, and is nearly flat and featureless. We do not see any constant offsets in our photometric residuals. It would be a complex proposition to modify the grid in such a way as to cause the residuals to be consistent with zero for some objects, but have significant residuals for others. Nevertheless, we can also rule out errors in the model atmospheres that are prevalent in some regions of parameter space but not others, as the coefficient of correlation between residuals in \textit{F160W} and either of the DA white dwarf intrinsic parameters, $T_{\text{eff}}$ and $\log g$, is consistent with $0$. Errors in the DA white dwarf model grid cannot explain the discrepancy in \textit{F160W}, and the mean of the residuals in the \emph{WFC3/UVIS} passbands are $\sim 1$~mmag. Any systematic error in the DA white dwarf model grid must be below this level. 

\subsection{Errors in the Reddening Law}
As noted previously, we found no significant difference when using the \citet{ODonnell94} reddening instead of \citetalias{Fitzpatrick99}. Additionally, we considered a custom reddening law defined for $R_V = 3.1$ constructed by J. Holberg from \emph{Gaia} DR2 observations, and included with the \texttt{WDmodel} code. This too did not cause a significant difference in the photometric residuals in any passband. Allowing $R_V$ to vary from 3.1 causes the weighted mean of the residuals in \textit{F160W} to decrease from 9~mmag to 7~mmag --- an insignificant change when accounting for the addition of the extra free parameter. Additionally, the difference between $R_V = 3.1$ and the true value of $R_V$ for each object cannot be correlated as our program stars are spread across the sky. Therefore any error in the SEDs induced by fixing $R_V$ to 3.1 is only likely to cause dispersion, rather than a systematic bias in the SEDs. 

It is unlikely that any change to the reddening law can resolve this discrepancy in \textit{F160W} as dust causes less extinction at these wavelengths relative to the optical $(R_H = 0.464)$. Resolving a mean residual difference in \textit{F160W} of 9~mmag with a change in reddening requires $R_H \sim 1.5$ \emph{without changing the coefficient of the reddening law in any other band}. Such a change would be completely unphysical as there is no reason to expect the dust along the line of sight to DA white dwarfs to extinguish the flux more strongly than along other lines of sight in the Galaxy. Any error in the reddening applied to the model is extremely unlikely to be the source of the bias. 

\subsection{Errors in the Passband Model}
We model the passband response using the \texttt{pysynphot} files provided by STScI (see footnote~\ref{footnote:pbtransmission}). The response of the filters were determined pre-flight. While there are periodic adjustments to the overall normalization of the components, these are achromatic. The GRW70 data presented in \S\ref{sec:zp_evolve} have slightly different slopes in each passband. This indicates that the \emph{shape} of the response is evolving with time. This error is likely to be irreducible. When designing our cycle 22 observing program, we determined that it was not possible to have both the \emph{WFC3} grism and passbands in the optical path at the same time, which would allow an in-flight determination of the throughput. The \emph{WFC3} flight spares have not been subjected to the same conditions as the filters on \emph{HST}, and using their transmission as a proxy is likely to introduce new systematic errors. 

Recognizing that a systematic difference between the passband model and the true passband response constitutes a systematic floor for our experiment, we sought to mitigate it in designing our observing program. The photometric zeropoints inferred in \S\ref{sec:phot} are largely determined by the difference between the synthetic and instrumental photometry of the three CALSPEC primary standards. Any error in the passband model would lead to an error in the synthetic magnitudes, and therefore the zeropoints. These zeropoints are applied to the measured instrumental photometry of our program stars, leading to an error in apparent magnitudes. If the SEDs of our program standards differed significantly from those of the CALSPEC primary standards there would be a systematic trend in the residuals with color. As the primary standards are also DA white dwarfs, and have very similar colors to our program stars, we reduce the impact of any error in the specification of the passband response. 

Moreover, the GRW70 data constrains the maximum change in the response to be at most 1--2~mmag per year. Much of this change is dominated by the decrease in sensitivity, rather than the change in passband shape. An error in the passband shape likely to have a $<1$~mmag impact on our SEDs. We emphasize that mitigating this error is not the same as measuring the passband response accurately, and in particular for sources with significantly different SEDs than the primary standards, we can expect a trend in the residuals as a function of color. An observational campaign to measure all the CALSPEC standards with all the \emph{HST/WFC3} passbands could, in principle, be used to determine a correction to each response. Given the constraints of the GRW70 data, such a campaign would have a limited impact on the overall accuracy of our experiment.

A systematic error in the passband model cannot explain the discrepancy in \textit{F160W}. Any adjustment to the shape of the \textit{F160W} passband produces an almost common-mode bias affecting all objects equally as the spectral flux density of DA white dwarf stars is nearly flat per unit wavelength across this passband. This is markedly different than the disagreement between the observations and model in our analysis, where some objects have residuals consistent with 0, while others show significant offsets. There is no change that can be applied to reduce the residual for the largest outliers that does not increase the residual by an almost identical amount for objects where the model agrees with the observations. We find no significant trend in the residual with MJD or with cycle number that might indicate an evolution in the passband response, and rule out passband shape adjustments as an explanation for the discrepancy in \textit{F160W}.

\subsection{Errors in the Overall Flux Normalization}\label{sec:calspeccomp}
The residuals presented in Table~\ref{table:resid} and Fig.~\ref{fig:WDmodelPhotResid} quantify the level of agreement between our measured \emph{HST/WFC3} photometry and our inferred SEDs. However, any error in the passband zeropoints derived in \S\ref{sec:phot} would propagate to all of the apparent magnitudes, and therefore all of the inferred SEDs. Such a systematic error in the passband zeropoints can be considered a difference between the photometric system defined by our inferred SEDs for the DA white dwarf stars and the photometric system defined by the CALSPEC SEDs of the three primary standards. 

\begin{table*}[htbp]
\scriptsize
\begin{centering}
\begin{tabular}{c|c|lllllll}
\hline
\hline
Object & Source & \textit{F275W} & \textit{F336W} & \textit{F475W} & \textit{F625W} & \textit{F775W} & \textit{F160W} & Mean \\
& &  \multicolumn{7}{c}{(AB mag)} \\
\hline
 & Observed & 10.4904 (0.0014) & 10.8902 (0.0014) & 11.4988 (0.0013) & 12.0307 (0.0011) & 12.4513 (0.0013) & 13.8853 (0.0015) &  \\
G191-B2B & CALSPEC & 10.4915 & 10.8917 & 11.4995 & 12.0304 & 12.4491 & 13.8851 &  \\
 & Residual & $-$0.0011 & $-$0.0014 & $-$0.0007 & +0.0003 & +0.0021 & +0.0001 & \phantom{$-$}0.0000 (0.0005) \\
 \hline
 & Observed & 12.2015 (0.0016) & 12.5678 (0.0015) & 13.0998 (0.0018) & 13.5976 (0.0014) & 14.0017 (0.0014) & 15.4139 (0.0017) &  \\
GD153 & CALSPEC & 12.2001 & 12.5662 & 13.0979 & 13.5982 & 14.0040 & 15.4141 &  \\
 & Residual & +0.0015 & +0.0016 & +0.0018 & $-$0.0006 & $-$0.0023 & $-$0.0002 & +0.0001 (0.0006) \\
 \hline
  & Observed & 11.9888 (0.0015) & 12.3360 (0.0014) & 12.7987 (0.0015) & 13.2789 (0.0013) & 13.6720 (0.0012) & 15.0676 (0.0019) &  \\
GD71 & CALSPEC & 11.9811 & 12.3271 & 12.7941 & 13.2749 & 13.6720 & 15.0605 &  \\
 & Residual & +0.0076 & +0.0089 & +0.0046 & +0.0040 & $-$0.0001 & +0.0071 & +0.0049 (0.0006) \\
 \hline
Zeropoint & Mean & +0.0026 (0.0009) & +0.0029 (0.0008) & +0.0016 (0.0009) & +0.0012 (0.0007) & \phantom{$-$}0.0000 (0.0007) & +0.0019 (0.0010) &  \\
\hline
\end{tabular}
\footnotesize{\tablecomments{Apparent magnitudes of the three CALSPEC primary standards on the AB system measured through each passband are reported with Source labelled Observed. These are identical to the values reported in Table~\ref{table:phot}, and are repeated here for convenience. Synthetic magnitudes derived the CALSPEC SEDs are reported with Source labelled CALSPEC to distinguish them from the values reported in Table~\ref{table:resid}. The residuals in the sense of Observed $-$ CALSPEC are reported with Source labelled Residual. The weighted mean of the residuals per standard are reported in the column labelled Mean. The weighted mean in each passband across all standards is reported with Source labelled Zeropoint. Uncertainties on measured or derived quantities are reported parenthetically. All quantities are rounded to a tenth of a millimag.}}
\caption{Comparison of observed \emph{HST/WFC3} magnitudes and synthetic magnitudes derived from the CALSPEC SEDs of the primary standards\label{table:calspeccomp}}
\end{centering}
\end{table*}

Such a difference could be induced because the measurement chain employed by the CALSPEC team differs from that employed in \citetalias{Calamida18} and this work. While we measure the primary standards using the same \emph{HST/WFC3} instrumental configuration as our program stars, the original \citetalias{Bohlin2014} CALSPEC SEDs are determined from \emph{HST/STIS} spectroscopy and \emph{HST/ACS} photometry. As noted previously, our zeropoints are effectively determined by the difference between the CALSPEC synthetic magnitudes and our measured \emph{HST/WFC3} instrumental photometry. Any systematic error in the flux ratios between the CALSPEC SEDs of the three primary standards would propagate into an error in our zeropoints. 

Had we obtained our observations with the same instrumentation as the primary standards, a purely differential measurement would have sufficed to calibrate measured count rates with respect to the primary standards. We would not have needed to use the CALSPEC SED models, and could have determined fluxes relative to the primary standard observations. With such a measurement chain, a single overall achromatic zeropoint suffices to set the flux scale. Such an approach avoids any systematic errors introduced by tying to the CALSPEC SED models. Unfortunately, despite the conceptual attractiveness of a differential measurement, using the same measurement chain for the DA white dwarfs and the CALSPEC primary standards was never a practical option. The WFC channel of \emph{HST/ACS} has low QE below 4,000~\AA\ and above 1~\micron\ and observations at UV and NIR wavelengths are critical to determining the line-of-sight extinction, while exposure times with \emph{HST/STIS} are prohibitive for our faint standards. 

\begin{figure}[h!tpb]
    \centering
    \includegraphics[width=0.47\textwidth]{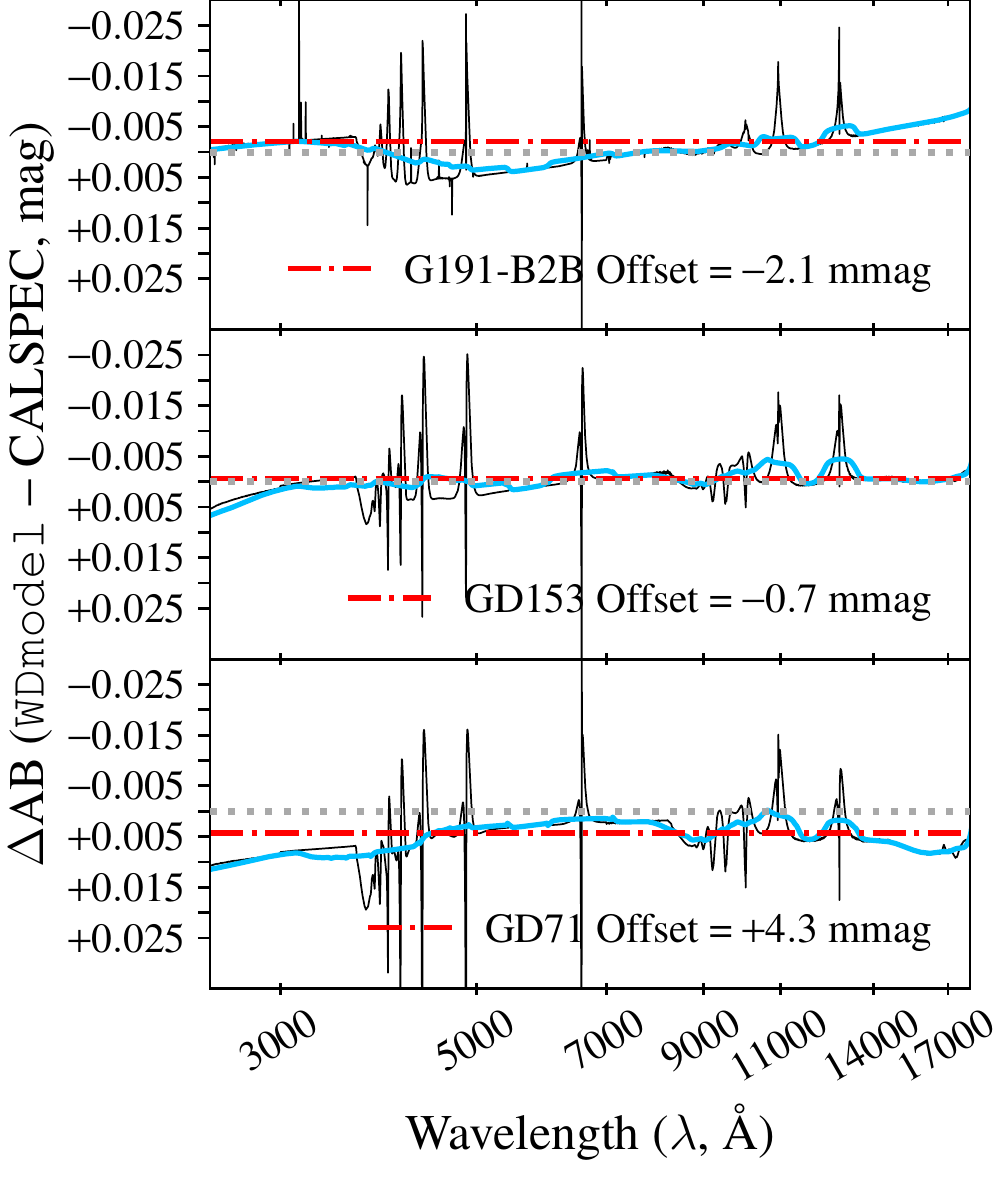}
    \caption{Differences between the CALSPEC SEDs and our inferred SEDs for the three primary standards in AB magnitudes (black). As the CALSPEC SEDs and our SEDs are computed from different DA white dwarf atmosphere grids, there are detailed differences in the line shapes. In blue, we illustrate the differences smoothed with a order 1 Savitzky-Golay filter having a bandwidth of 1,100~\AA\ $\sim$ the width of an optical passband. The median difference (shown in red) between the two SEDs for each star is dominated by the difference in overall normalization arising from the systematic differences in photometry listed in Table~\ref{table:calspeccomp}. The differences in shape of the SED about the median amount to 1--2~mmag.}
    \label{fig:calspeccomp}
\end{figure}

We quantify the systematic difference to the CALSPEC photometric system from our inferred SEDs for the three primary standards in Fig.~\ref{fig:calspeccomp}, as well as from our measured \emph{HST/WFC3} photometry in Table~\ref{table:calspeccomp} --- the first approach incorporates all the systematics from an end-to-end analysis, while the second is independent of the differences between the model atmosphere grids. The residuals between our observed \emph{HST/WFC3} observations and the synthetic CALSPEC magnitudes in Table~\ref{table:calspeccomp} are completely negligible for GD153 and GD191-B2B, but have a mean of $\sim 5$~mmag for GD71. This indicates that there is some tension between the flux ratios of the primary standards \emph{measured} from the \emph{HST/WFC3} images and the flux ratios \emph{defined} by their CALSPEC SEDs. 

With our existing data, we cannot distinguish if the underlying source of this tension arises from the measurement (i.e. a bias in our \emph{HST/WFC3} data), or the model (i.e. an error in the CALSPEC SEDs). As described in \S\ref{sec:imagingoptimazation}, we elected to readout the C512C subarray of \emph{WFC3/UVIS} in cycle 22, when we began monitoring the primary standards. The small images are unsuitable for artificial star injection tests, which would allow us to determine of there is a weak bias in the recovered photometry. Given the repeatability of \emph{HST/WFC3} described in \S\ref{sec:zp_evolve} and illustrated in Fig.~\ref{fig:zp_evolve} is 5--8~mmag, and that observations in all passbands were obtained in the same orbit, it is possible that the non-zero offset of GD71 is simply the result of a correlated statistical fluctuation. At the same time, we do not have observations of the remaining 90 CALSPEC standards with \emph{HST/WFC3} to determine if the CALSPEC GD71 model itself is inconsistent with the observations. The non-zero residuals for GD71 drive the weighted mean difference in the zeropoint between our photometric system and the CALSPEC photometric system, but these biases are small (1--2~mmag). 

The difference between our apparent magnitudes and the CALSPEC synthetic magnitudes is the dominant contribution to the difference between our inferred SEDs and the original CALSPEC SED models. The shape differences about the median offset are dominated by differences in the line profiles. These differences arise from the different Stark broadening prescriptions used by our \texttt{Tlusty} grid and the \texttt{TMAP} models employed by CALSPEC, but have negligible impact on broadband photometry. G191-B2B exhibits the largest difference in shape correlated with wavelength. The CALSPEC SED of G191-B2B has $A_V = 0.0016$~mag whereas we infer an extinction of less than 1~mmag from our \emph{HST/WFC3} photometry and MMT/Blue Channel spectroscopy of this object. This difference in reddening dominates the overall shape of the residual. The CALSPEC G191-B2B SED includes metal lines~\citep{G191B2Bmodel}, while our SED uses a pure hydrogen atmosphere. The difference is evident in the residual between the two SEDs in the UV. However, as the median difference between our inferred SEDs is dominated by the difference in the observed and synthetic \emph{HST/WFC3} photometry, any systematic must be dominated by the difference in the measurement chains, rather than the model grids or the zeropoints.

Conservatively, we adopt 4~mmag, corresponding to the median error on GD71, as our estimate of the systematic error in the overall flux normalization of the SED. Observers using our network of stars to compute synthetic magnitudes derived from our inferred SEDs, and determine zeropoints with these synthetic magnitudes can expect to be on the CALSPEC photometric system to within this amount. This systematic cannot explain the discrepancy in \textit{F160W}, as a zeropoint error would affect all of the apparent magnitudes in a given passband by the same amount. 

We note that the CALSPEC system itself may have non-zero AB offsets. There is no way to quantify the offset between CALSPEC and the true AB system without using external catalogs, most of which are ground-based and less accurately calibrated, and often also tied to CALSPEC themselves. In \S\ref{sec:futurework}, we discuss a more complex framework to put all the CALSPEC standards and the DA white dwarfs presented in this work on a single photometric system. Such an analysis could incorporate data from laboratory or satellite-borne flux standards to set the absolute AB zeropoints. 

\subsection{Identifying the Source of the Residual Bias in \textit{F160W}}
None of the previously considered sources of systematic error can explain the disagreement between model and observations in \textit{F160W} seen in Fig.~\ref{fig:WDmodelPhotResid}. To identify the underlying source of the systematic error, we began looking for correlations between the residuals and other quantities derived in our analysis. The residuals between the observations and model are correlated with the apparent magnitude in \textit{F160W}. Only one component of our measurement chain can be sensitive to the brightness of the source --- the \emph{HST/WFC3} detector itself. 

We do not see this correlation between residuals and apparent magnitude in any of the \emph{WFC3/UVIS} passbands. We find that the residuals remain biased if we use the apparent magnitudes determined from instrumental photometry measured with \texttt{SourceExtractor} or \texttt{DAOPHOT}. The \textit{F160W} exposures are obtained with the MULTIACCUM mode of \emph{WFC3} where the signal is sampled multiple times during an exposure. This is used for both cosmic ray removal as well as reducing the effective read noise. We found that the measurements we obtained by examining the individual reads was the same as determined from ramp-fitting. We manually determined the photometry for a subset of the objects with the largest residuals and found our measurements were in agreement with the \texttt{ILAPH} measurements. 

To exclude any effect arising from an error in the CALPSEC SED of the primary standards, we recomputed zeropoints across all passbands, excluding one of G191-B2B, GD71 and GD153, and performed the entire analysis with the three resulting sets of apparent magnitudes. The residual bias in \textit{F160W} persisted across all three sets of results. We note that the \textit{F160W} residuals of 2 of the three primary standards (GD-71 and GD-153) are completely consistent with 0, while the third is only a 2$\sigma$ outlier. This suggests that the origin of the bias is not in the photometry of our program stars, and not the CALSPEC standards. To verify this, we adjusted the settings employed by \texttt{ILAPH} to produce eight different sets of instrumental photometry. The default \texttt{ILAPH} configuration uses a 5~pix aperture and an annulus from 14--21~pix in \textit{F160W}. We varied the aperture size from 3--9 pix in steps of 2 pix, and used two different annuli, the first from 23--46 pix and the second from 46--62 pix. If the underlying source of the residual bias originates with the instrumental photometry, this test would exhibit a change in the strength of the bias with increasing aperture and annulus size. We found the bias in \textit{F160W} increased (i.e. the apparent magnitudes become fainter and more discrepant with model magnitudes) with increase in both aperture and annulus size above 5~pix, while the residuals in the UVIS passbands remained consistent with 0. The dispersion in \textit{F160W} increases significantly for all aperture sizes at the 3 pix aperture as the IR channel of \emph{WFC3} has a 0.13\arcsec$/$pix scale, and small aperture photometry is extremely susceptible to centroiding errors. These tests suggest that the discrepancy between model and observations in \textit{F160W} are due to some unmodeled systematic effect with our \emph{WFC3/IR} data.

\subsection{Count Rate Non-Linearity}\label{sec:crnl}
The correlation of the residuals with apparent \textit{F160W} magnitude and the sensitivity of the size of the residuals to the number of pixels included in the aperture strongly suggest that the bias is a count-rate non-linearity (CRNL) effect. The \texttt{calwf3} pipeline corrects for non-linearity with the total instrumental counts, and we adjust our exposure times to ensure that our measurements have similar S/N, but no correction is applied based on the count-rate of the source. Previous HgCdTe detectors on \emph{HST} have suffered from a count-rate dependent non-linearity, and the effect has been well characterized for \emph{HST/NICMOS}~\citep{nicmos1, nicmos2}. The effect in \emph{HST/NICMOS} amounts to $-0.1$~mag/dex at \textit{F110W} but is strongly chromatic, decreasing to $-0.03$~mag/dex at \textit{F160W}, where the negative sign indicates that the observations are fainter than what would be measured in the absence of non-linearity. The analysis in \citet{nicmos3} corrects for the non-linearity using a power law in the count-rate, which translates to a linear trend in magnitudes. 

We quantify the CRNL for \emph{WFC3} IR channel by repeating the analysis as described in \S\ref{sec:wdmodel} excluding the \textit{F160W} observations. The resulting SED parameters are consistent with the values inferred from the full data set, but as \textit{F160W} was not included in this second analysis, the synthetic \textit{F160W} model magnitudes and observations are independent. We fit a linear relation between the observed and synthetic magnitudes, accounting for the errors in both quantities, and allowing a dispersion to account for the imperfect fit. While the strength of the CRNL effect and the errors increase with magnitude unlike a flat dispersion at all magnitudes, we cannot justify a more complex noise model with the limited number of observations available at present. The three CALSPEC standards are used to define the zeropoint in \textit{F160W} and the CRNL effect can only be measured with respect to these stars. We therefore fix the intercept in our analysis to the mean \textit{F160W}$ = 14.7889$~mag of the three primary standards

\begin{figure}[htpb]
    \centering
    \includegraphics[width=0.47\textwidth]{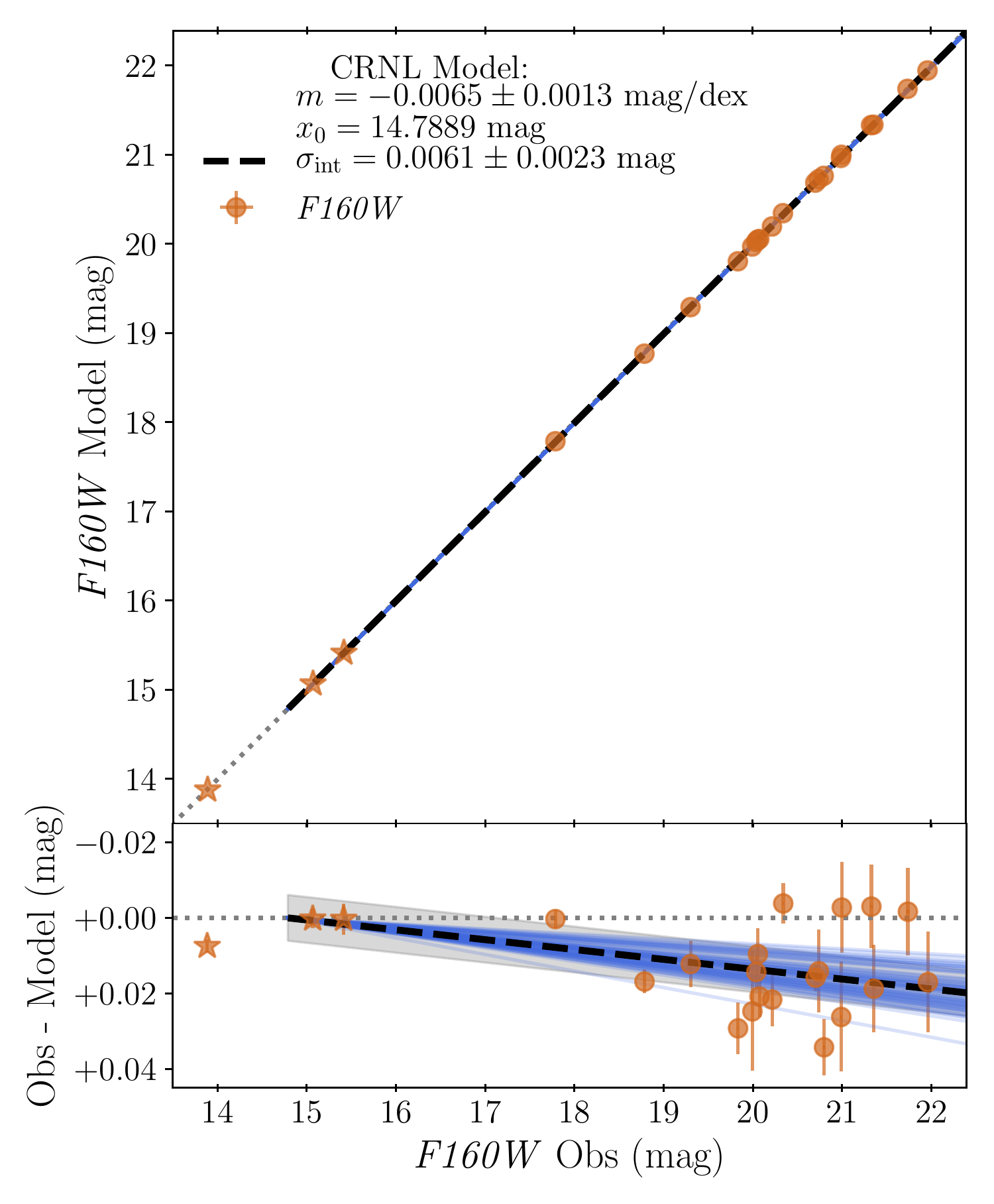}
    \caption{We quantify the CRNL in \textit{F160W} by fitting the \emph{HST/WFC3} UVIS passbands and the spectrum of each object, but excluding \textit{F160W}. We then compute the synthetic \textit{F160W} magnitudes from the inferred SED. The model magnitudes are thus completely independent of the observed \textit{F160W} magnitudes. We model the CRNL as a linear relationship between the observed and synthetic magnitudes (top panel, dashed black line), accounting for the observed and synthetic errors. The CRNL cannot be determined for the three primary standards as these are used to set the zeropoints, and the intercept of the line is fixed to their mean \textit{F160W} magnitude. The deviation from the 1:1 relationship (dotted grey line) is clearly visible in the residuals (lower panel). The range of the dispersion about the mean parameters of the linear fit is indicated by the grey shaded region. Random draws from our MCMC are shown in light blue, to illustrate the range of inferred slope.}
    \label{fig:F160Wcrnl}
\end{figure}

The results of our analysis of the CRNL are shown in Fig.~\ref{fig:F160Wcrnl}. We find that the slope of the CRNL is $-0.0065 \pm 0.0013$~mag$/$dex, leading to observed magnitudes that are fainter than would be predicted from the SED model. The lamp of the \emph{WFC3} IR channel cannot be operated during observations, preventing a direct measurement where the count-rate of sources is enhanced artificially. Previous studies have reported CRNL measurements derived from a comparison of count rates between overlapping passbands imaged with the \emph{WFC3} UVIS and IR channels~\citep{RiessCRNL1}, and from artificially boosting the count rate of sources with Earth limb shine~\citep{RiessCRNL2}. Our CRNL measurement is consistent with these previous limits, both of which find that the effect is $< 0.01$~mag$/$dex. It is also consistent with measurements on the \emph{WFC3} IR channel flight spares carried out at the Goddard Detector Laboratory presented in \citet{RiessCRNL3}. The agreement between these various independent studies leads us to conclude that the residual bias exhibited in \textit{F160W} is the result of count rate non-linearity in the \emph{WFC3} IR channel. 

The complex analysis procedure to transform from our multi-cycle observations to apparent magnitudes in \S\ref{sec:phot} makes it intractable to determine the CRNL directly from the instrumental counts of our images. Our measurement is consistent with previous work, but is also more precise than those determinations, so we have elected not to make a correction based on those independent estimates. While the strong chromatic trend of the CRNL with \emph{HST/NICMOS} leads us to expect a variation with passband, our DA white dwarf program only included observations in \textit{F160W}. Finally, there is no way to correct the CRNL effect for our data using a measurement from our data. While we expect to be able to account for the CRNL in our analysis of the combined observations from cycle 20, 22 and 25, we are forced to incorporate the bias into our error budget for this work. 

The error in our SEDs of the DA white dwarf stars induced by the CRNL is smaller than the magnitude of the effect in \textit{F160W}, as the output SEDs are constrained by \emph{all} of the observed photometry and the spectroscopy, and should be robust against a systematic bias in a single passband. Conversely, even though the CRNL effect is only present in \textit{F160W}, it will have some effect at all wavelengths as all the data are modeled coherently. We can evaluate the bias in the SEDs caused by the CRNL in \textit{F160W} by repeating the analysis in \S\ref{sec:wdmodel} with \textit{F160W} excluded entire;y. Compared to the results shown in Fig.~\ref{fig:WDmodelPhotResid}, excluding \textit{F160W} entirely from the analysis reduces the mean residual in $\{F275W, F336W, F475W, F625W, F775W\}$ to $\{0, 0, 0 +1, -1\}$~mmag. This is a 1--2~mmag change for each of the \emph{WFC3/UVIS} passbands. The RMS in \textit{F625W} and \textit{F775W} is reduced to 2~mmag and 3~mmag respectively, while the standard deviation of the residuals across all passbands is reduced from 6~mmag to 4~mmag. These indicate that the systematic bias caused by our present inability to remove the CRNL is at the 2~mmag level. Because of the CRNL, synthetic \emph{WFC3/IR} magnitudes derived from our SEDs are likely better predictors of the true flux of our DA white dwarfs than observations with the instrument. 

\subsection{The Systematic Error Budget}\label{sec:syssummary}

The best estimate of the effect on our results from random errors is 6~mmag RMS in any one passband, for any one of our stars.  The true effect from random errors is likely closer to 4~mmag, as the largest residuals are in the \textit{F160W}, which are systematically biased because of count rate non-linearity in the \emph{WFC3/IR} channel. In addition to random effects, we have considered five potential systematic effects in \S\ref{sec:WDmodelResults}. We summarize our estimate of their effects in Table~\ref{table:WDmodelSys}. We can divide the potential sources of systematic effects we have considered into two categories:
\begin{enumerate}
    \item Biases that arise because of misspecification of the model. 
    \item Biases that arise because of miscalibration of the data. 
\end{enumerate}
Our analysis is robust against the first of these. We elected to establish DA white dwarfs as standards because the physics that describes their atmospheres is well understood. We chose our objects to be in low-extinction environments, and obtained multiband photometry to tightly constrain the reddening. Consequently, any error in our reddening model has negligible impact, and as our standards are spread across the sky, the errors are extremely unlikely to be correlated. While we cannot rule out an error in the response of the \emph{HST/WFC3}, we elected to use reference standards with almost the same colors as our program stars as this minimizes the effect of our imperfect knowledge of the true system throughput. 
\begin{table}[htpb]
\scriptsize
\begin{centering}
\begin{tabular}{l|c}
\hline
\hline
Underlying Source of Bias & Systematic Effect on SED \\
 & (mmag) \\
\hline
\multicolumn{2}{c}{Effects Caused by Model Misspecification} \\
\hline
Model Atmosphere Grid & $<$1\\
Reddening Model\tablenotemark{*}& $<$1 \\
Passband Model & $<$1 \\
\hline
\multicolumn{2}{c}{Effects Caused by Bias in Observations} \\
\hline
CALSPEC Flux Scale & 4 \\
Count Rate Non-Linearity & 2 \\
\hline
\end{tabular}
\footnotesize{\tablenotetext{*}{An error in the reddening model will cause dispersion rather than bias unless the error is correlated for all our objects. This is extremely unlikely for our all-sky network.}}
\caption{Sources of Systematic Bias and Estimated Effect on the \texttt{WDmodel} SEDs\label{table:WDmodelSys}}
\end{centering}
\end{table}

We are more sensitive to systematic effects arising with inputs over which we have no control. Any error in the flux scale defined by the three CALSPEC primary standards will propagate to all our SEDs. While the synthetic colors of the SED are set by the model and will remain accurate, the overall fluxes can be systematically off by up to 4~mmag, though this is the most conservative estimate of the error possible, and the true error is likely smaller. Additionally, our program has provided the most precise measurement of the count rate non-linearity in the \emph{WFC3/IR} detector. While this non-linearity affects the our measurements in \textit{F160W} only, it has the potential to impact our model at all wavelengths, as this is the most red passband in our program. Despite this, we found that excluding \textit{F160W} only causes achromatic 1--2~mmag shifts in the SEDs. Our model is robust against a bias in a single passband precisely because we coherently forward model all the observations, and their combined statistical weight prevents the biased \textit{F160W} measurements from torquing the SED significantly.

We note that there is an additional up to 0.5\% error arising from how well the CALSPEC flux scale is tied to the true AB flux scale defined by \citet{OkeGunn83}. This error is achromatic and affects both the CALSPEC primary standards and our DA white dwarf SEDs by the same amount in the same direction, and we therefore do not include it in our systematic error budget. Reducing this error further requires a different experiment and analysis from that described in \citetalias{Calamida18} and this work. We consider such an experiment in \S\ref{sec:futurework}.

\section{Verification and Validation}\label{sec:validation}

In this section, we describe various tests of the model described in \S\ref{sec:wdmodel} for internal consistency (verification) as well as against external ``truth'' (validation). 

\subsection{Testing Model Parameters For Objects with Multiple Spectroscopic Observations}
Fifteen of the objects in our sample have more than one spectroscopic observation. 14 of these are objects with a spectrum observed in queue mode with Gemini/GMOS. Despite the high S/N of the spectra, these observations may suffer from subtle systematic biases which are not ideal for the analysis in this work as they were executed without the rotator set at the parallactic angle or without a flux standard observed contemporaneously. Moreover, the GMOS spectra are dispersed across three CCDs and stitched together into a single trace. We found sharp discontinuities in the continuum of the reduced spectra as the gain values were incorrectly set in the pipeline. We re-reduced the spectra for these objects, but the correction for these discontinuities is \emph{ad hoc}. 

While the Gemini/GMOS observations are likely sufficient for most purposes, we elected to obtain at least one MMT/Blue Channel spectrum for 14 objects out of an abundance of caution. As described in \S\ref{sec:multispec}, the MMT/Blue Channel spectra generally have lower S/N and lower resolution than the Gemini spectra but cover a larger range in wavelength, including the H$_\alpha$ feature, and while they yield slightly weaker constraints on the surface gravity $\log g$, they suffer from fewer systematics. Finally, we observed the primary standard GD71 multiple times with SOAR across a range of airmass. These 15 objects with at least two spectra each are a valuable test of the consistency of inference with the model described in \S\ref{sec:wdmodel}. 

\begin{figure*}[htpb]
    \centering
    \includegraphics[width=\textwidth]{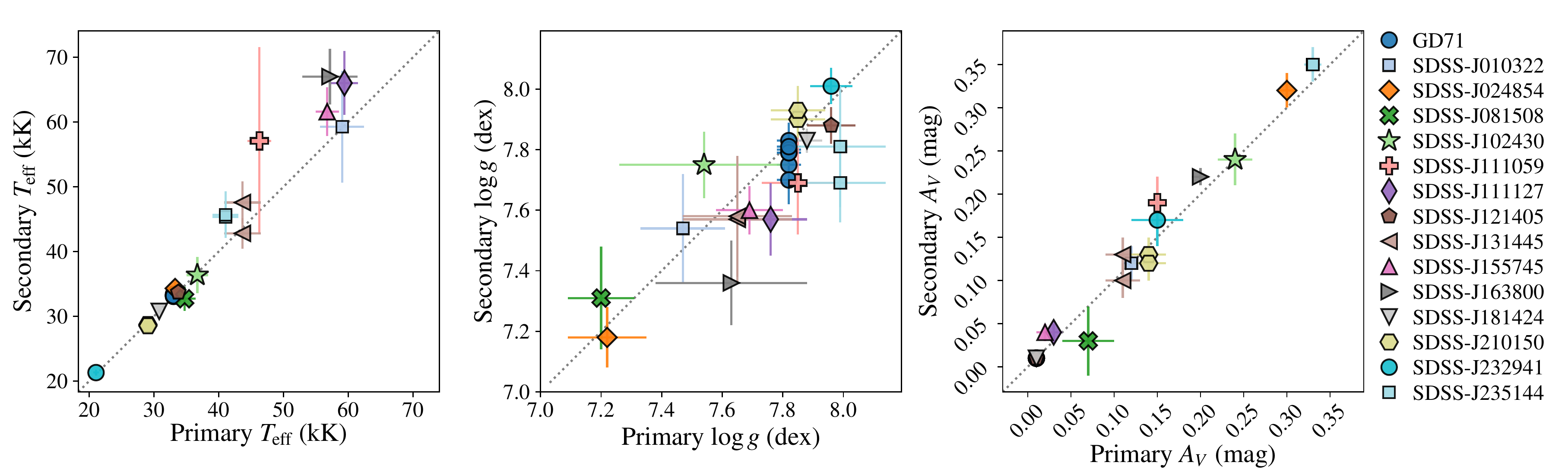}
    \caption{The inferred values of $T_{\text{eff}}$ (left), $\log g$ (center) and $A_V$ (right) for objects (indicated by the marker listed in the legend) with multiple spectra are consistent with each other at the 1.5$\sigma$ level (the grey dotted line indicates a 1:1 relationship). The primary values along the abscissa of both axes are listed in Table~\ref{table:WDmodelFit} and are inferred from our \emph{HST/WFC3} photometry and typically the MMT/Blue Channel spectroscopy. These values have the highest log-likelihood of the photometry, and typically also have the lowest error on the inferred $T_{\text{eff}}$ . The secondary ordinate values are inferred from the same \emph{HST} photometry, together with another spectrum of the same object, typically from Gemini/GMOS. GD71 has multiple spectra from SOAR. We do not compare the overall normalization parameter $\mu$ as it is determined solely by the \emph{HST} photometry.}
    \label{fig:wdmodel_multispec}
\end{figure*}

As described in \S\ref{sec:spectroscopy}, the intrinsic DA white dwarf parameters $T_{\text{eff}}$ and $\log g$ are strongly constrained by the spectroscopic observations. $A_V$ is also constrained by the spectra. Of the two remaining parameters that determine the SED $\mathbf{F}$, $R_V$ is fixed in this work, and $\mu$ is set by our \emph{HST/WFC3} photometry. We use the objects with more than one spectrum to verify that the inferred SED parameters are consistent with the parameters listed in Table~\ref{table:WDmodelFit} determined from the spectra with the highest log-likelihood of the observed photometry. The results of this comparison are shown in Fig.~\ref{fig:wdmodel_multispec}. Despite our spectra being obtained with different instruments, telescopes, sites, conditions and at different epochs by different observers, the inferred SED parameters are entirely consistent with each other. 

\subsection{Correlations between Inferred Extinction and Sodium Absorption Lines in White Dwarf Spectra}
The white dwarf spectra used to determine $T_{\rm eff}$ and log~$g$ provide additional information that can validate our model fitting process. Specifically, the spectra may show evidence for interstellar extinction through the presence of \ion{Na}{1}~D lines. The equivalent width of the sodium absorption features can provide a reasonably accurate, albeit imprecise, estimate for the extinction along the line of sight~\citep[e.g., ][]{poznanski12}. Given the coarse resolution of our spectra, though, we can expect only a rough correlation between equivalent width and extinction.

For the spectra from the MMT, the signal-to-noise ratio (S/N) at the location of the sodium lines is relatively poor as a result of shorter exposure times. The spectra obtained with Gemini have considerably higher S/N at the sodium lines, partly as a result of larger aperture and more exposure time, but also because the GMOS spectrograph is typically more sensitive at these wavelengths than Blue Channel spectrograph. For our analysis, we will consider mainly the Gemini spectra, along with one object from the MMT (SDSS-J235144).

We used two techniques to measure the equivalent width after normalizing the shape of the spectrum near the \ion{Na}{1}~D lines. Both summing the values in the spectra or fitting two Gaussians (with fixed means to match the line separation) yielded essentially identical results. We will use the Gaussian fit equivalent widths. There are 10 objects with detectable lines. For the spectra without obvious lines, we calculated an upper limit for the equivalent width. Based on a prescription from \citet{hobbs84}, \citet{leonard01} derive this formula for the 3$\sigma$ upper limit of the equivalent width of a feature (in \AA):

\begin{equation}
 \mathrm{EW}_{\lambda}(3\sigma) = 3 \cdot \Delta\lambda \cdot \Delta I \sqrt{\frac{\strut W_{\rm line}}{\Delta\lambda \cdot B}}
\label{eqn:ew_upperlimit}
\end{equation}

\begin{figure*}[htpb]
    \centering
    \includegraphics[width=\textwidth]{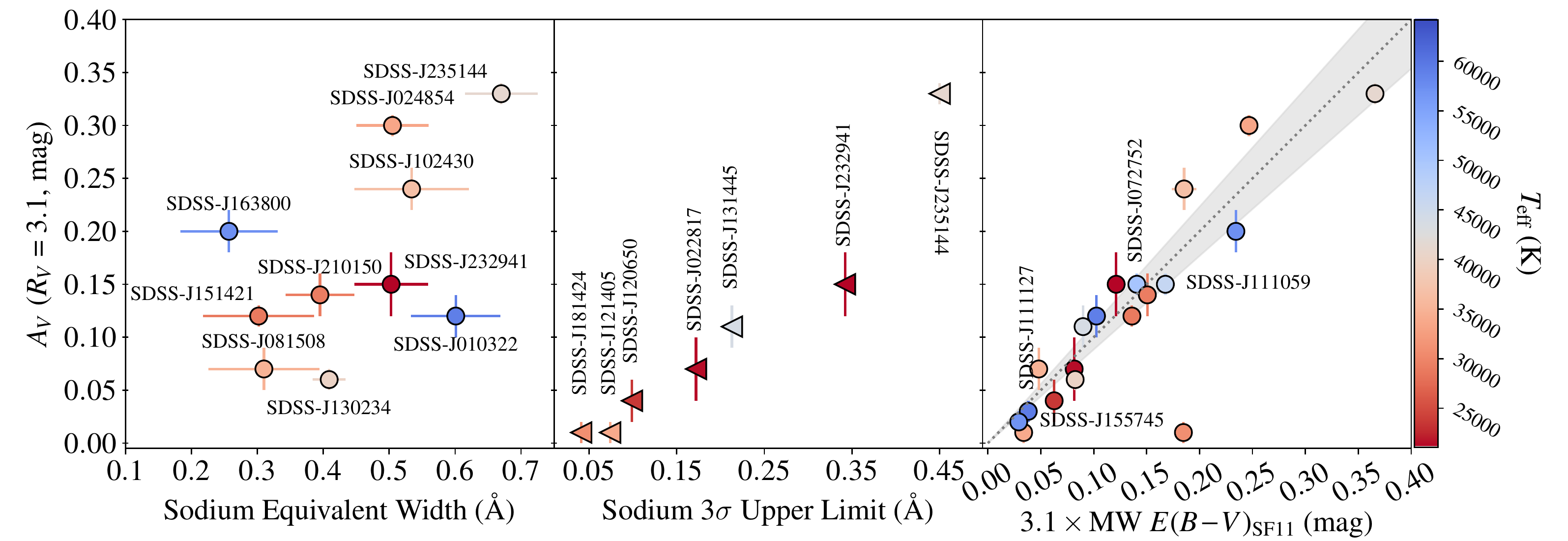}
    \caption{Extinction $A_V$ inferred from our model fits compared to independent measurements: (Left) vs equivalent width of the \ion{Na}{1}~D measured from the spectra. The sodium line equivalent widths correlate with $A_V$ ($r~=~0.51$). (Center) vs 3$\sigma$ upper limits that also correlate with $A_V$ ($r~=~0.96$). Note that two stars, SDSS-J232941 and SDSS-J235144, appears in both left and center panels as these objects have two spectra taken at different epochs with different spectrographs at different resolutions. In both cases, one of the spectra had insufficient S/N to measure a equivalent width. (Right) vs average extinction from the Galactic dust map of \citet[][denoted as SF11]{Schlafly11}. This average extinction is determined over a region with radius 5\arcmin\ radius centered on the position of each DA white dwarf, assuming $R_V = 3.1$. A 1:1 relationship is indicated by the dotted grey line. Our distant sources suffer extinction, but are not behind the full dust column, and our inferred $A_V$ will generally be lower than the estimate derived from the dust maps, which is appropriate for extragalactic sources. \citet{Schlafly16} finds the dispersion in $R_V$ along different lines of sight to be 0.18. The 2$\sigma$ region of the extinction $A_V$ given the color excess $E(B-V)$ is indicated by the shaded grey region. Points are colored by effective temperature indicated by the color bar at far right.}
    \label{fig:sodium-ew}
\end{figure*}

Here, $\Delta\lambda$ is the width of a resolution element in \AA, $B$ is the width of a resolution element in pixels at the native dispersion, $\Delta I$ is the 1$\sigma$ RMS fluctuation of the flux around a normalized continuum level, and $W_{\rm line}$ is the width of the line feature in \AA\ (taken to be 10\AA\ for interstellar \ion{Na}{1}~D). $\Delta\lambda$ and $B$ are determined from each spectrum using the FWHM from a line in the comparison lamp near the wavelength of \ion{Na}{1}~D and are likely to be \emph{overestimates} given the observing conditions. $\Delta I$ is measured from the normalized spectrum. The limit is essentially an estimate of the noise in a resolution element. There are 7 spectra with reasonable upper limits.

In Figure \ref{fig:sodium-ew}, we compare the measured sodium equivalent widths and upper limits with the estimates of extinction $A_V$ inferred from our model fits to the same spectra. For both measured values and upper limits, the sodium equivalent widths do correlate with the values of A$_V$, providing an independent validation that the model fitting procedure is deriving reasonable estimates of interstellar extinction.

\subsection{Comparing Inferred Extinction to Extinction from Galactic Dust Maps}
We can compare our inferred extinction to each DA white dwarf against an independent estimate of the extinction derived from the Galactic dust maps of \citet{Schlafly11}. Assuming $R_V = 3.1$, appropriate for diffuse interstellar dust in the Milky Way, we can convert the color excess estimate MW~$E(B-V)$ from the dust map to an extinction. This quantity is a measure of the average extinction of \emph{extragalactic} sources determined from an ensemble of main sequence stars in a region of radius 5\arcmin\ centered on the DA white dwarf position. Our inferred $A_V$ are line-of-sight estimates from our observations of each DA white dwarf. These two quantities can be substantially different as is the case for the nearby CALSPEC primary standards, two of which have MW~$E(B-V) > 0.3$~mag despite \citetalias{Bohlin2014} constraining them to $A_V < 0.005$~mag. Nevertheless, the extinction inferred for our more distant white dwarfs should be correlated with the integrated extinction determined from the dust map. This is evident in Fig.~\ref{fig:sodium-ew} where are our inferred $A_V$ are generally smaller than the estimate from the dust map. The dispersion about a 1:1 relationship is consistent with the variation of $R_V$ along different lines of sight found by \citet{Schlafly16}.

If $R_V$ was significantly different from 3.1 for any specific program star, we would see correlated residuals as a function of wavelength in all our passbands that grow with the estimate of the extinction $A_V$. We do not observe any such trend for any object. Moreover the magnitude of the effect is small for any star as it is on the order of the difference between the true value of $R_V$ and 3.1 times the color excess $E(B-V)$ i.e. $\sim \sigma_{R_V} \times E(B-V) =$~3--4~mmag for any object. As our stars are spread across the sky, the difference between the latent $R_V$ and 3.1 is not correlated across all objects. Any uncertainty that arises from the difference between the true reddening law and the canonical \citetalias{Fitzpatrick99} model is a random effect, i.e. fixing $R_V$ to 3.1 may add dispersion, but does not cause a systematic bias.

\subsection{Expected Magnitudes on Common Photometric Systems}
\begin{figure}
    \centering
    \includegraphics[width=0.47\textwidth]{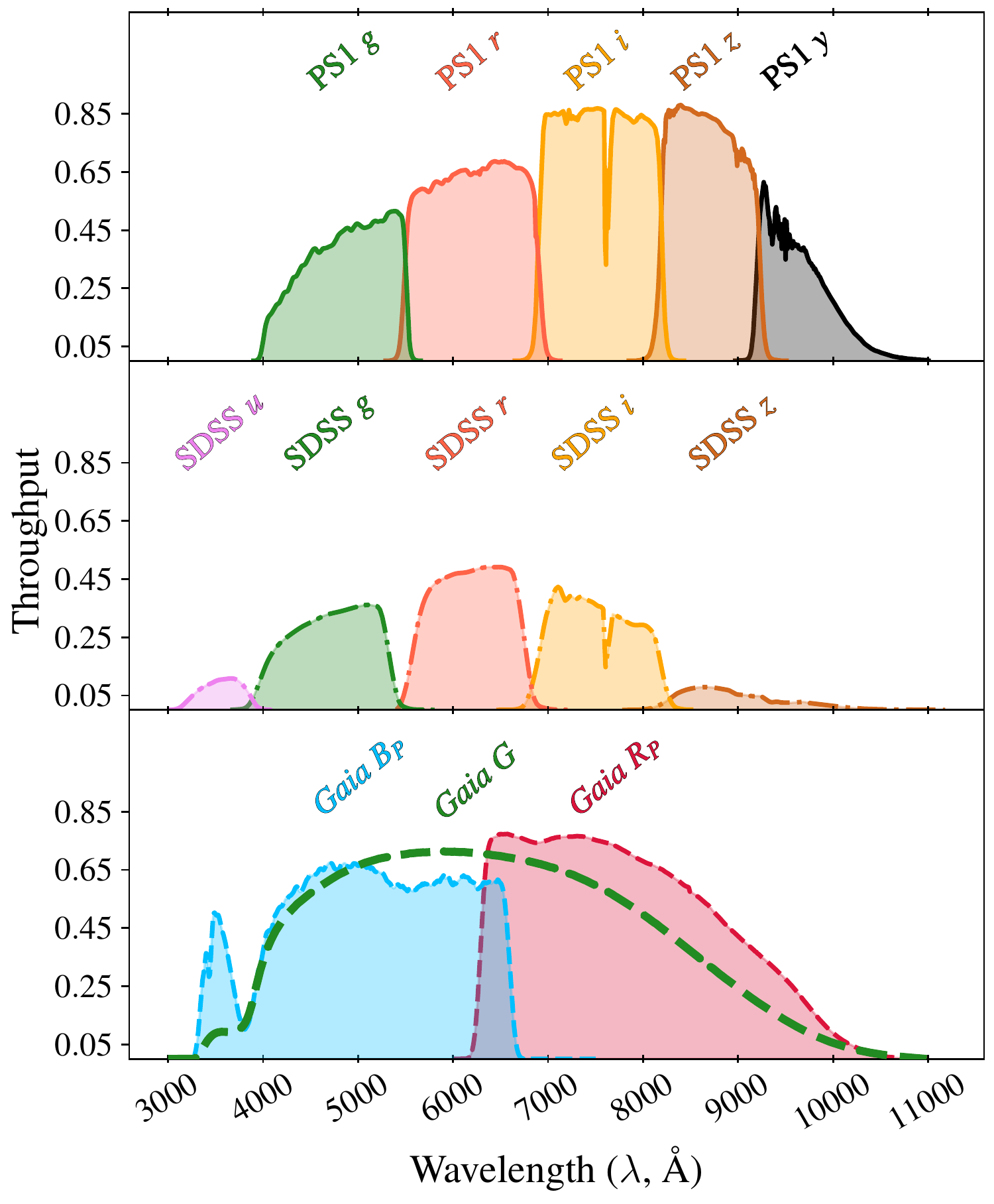}
    \caption{Passband transmissions for surveys/facilities that provide an independent comparison of observed DA white dwarf photometry (from top to bottom: Pan-STARRS PS1, SDSS, and \textit{Gaia}). The passband names are indicated in the labels above each passband, centered on the pivot wavelength. Throughput curves are illustrated as reported, and we have not normalized them.}
    \label{fig:comp_passbands}
\end{figure}

In this section, we compare our synthetic photometry to independently observed photometry from PS1, SDSS, and \textit{Gaia}. While comparing the synthetic photometry of our DA white dwarfs against catalogs from other surveys can be informative, it is also challenging with ground-based surveys owing to the systematic effects listed in \S\ref{sec:HSTphot}. Indeed, avoiding these systematic effects was the motivation for us to obtain above-atmosphere \emph{HST} photometry for our program. Nevertheless, these data are a valuable test of consistency. Historically, optical surveys have used such comparisons synthetic and observed photometry from spectrophotometric references such as BD+17\degree4708 to quantify the \emph{inconsistency} of the survey flux scale with the AB flux scale, and derive offsets to their natural system magnitudes. We account for these offsets where they are available in the literature. For surveys that do not report calibrated AB magnitudes, we compare synthetic and observed natural system photometry up to an overall constant. The passband responses of the different surveys are shown in Fig.~\ref{fig:comp_passbands} and machine-readable tables are included within our \texttt{WDmodel} package together with routines to generate synthetic photometry from our SED models. The results of the comparison of each survey are discussed below.

\subsubsection{Comparison with PS1 DR1}
\citet[][hereafter \citetalias{Supercal}]{Supercal} have derived offsets between photometry from the first 1.5 years of PS1~\citep{Schlafly12} to the AB flux scale using comparisons of the synthetic photometry of 7 CALSPEC standards\footnote{The \citetalias{Supercal} analysis used the \texttt{stisnic\_005} SEDs of blue CALSPEC standards \href{ftp://ftp.stsci.edu/cdbs/calspec/snap1\_stisnic\_005.fits}{Snap-1}, \href{ftp://ftp.stsci.edu/cdbs/calspec/wd1657\_343\_stisnic\_005.fits}{WD1657+343}, \href{ftp://ftp.stsci.edu/cdbs/calspec/lds749b\_stisnic\_005.fits}{LDS749B}, and red CALSPEC standards \href{ftp://ftp.stsci.edu/cdbs/calspec/sf1615\_001a\_stisnic\_005.fits}{SF1615+001A}, \href{ftp://ftp.stsci.edu/cdbs/calspec/snap2\_stisnic\_005.fits}{Snap-2}, \href{ftp://ftp.stsci.edu/cdbs/calspec/c26202\_stisnic\_005.fits}{C26202}, as well as the \texttt{stisnic\_004} SED of red standard \href{ftp://ftp.stsci.edu/cdbs/calspec/kf06t2\_stisnic\_004.fits}{KF06T2}. \href{ftp://ftp.stsci.edu/cdbs/calspec/gd153\_stisnic\_005.fits}{GD153} is included with the other standards to determine the AB offsets for \textit{z}$_{\text{PS1}}$ and \textit{y}$_{\text{PS1}}$ as the \textit{gri}$_{\text{PS1}}$ measurements are saturated.}. This initial release of PS1 photometry has been superseded by the Pan-STARRS PS1 Data Release\footnote{\url{https://panstarrs.stsci.edu/}} (DR1) photometry. The internal PS1 calibration ladder is described in \citet{ps1cal} and the DR1 calibration is described in \citet{ps1recal}, which apply the results of the \citetalias{Supercal} analysis. There have been numerous changes in the PS1 Image Processing Pipeline between the catalogs used in \citetalias{Supercal} and the DR1, and we therefore recompute the AB offsets using the observed DR1 photometry and the same CALSPEC standards used in the \citetalias{Supercal} analysis. 

\begin{table*}[htb]
\scriptsize
\begin{centering}
\begin{tabular}{c|c|ll|ll|ll|ll}
\hline
\hline
Object & PS1 ObjID & \multicolumn{2}{c}{PS1 \textit{g}} &  \multicolumn{2}{c}{PS1 \textit{r}} & \multicolumn{2}{c}{PS1 \textit{i}}  & \multicolumn{2}{c}{PS1 \textit{z}} \\
  & & \multicolumn{8}{c}{mag (mmag)} \\
\hline
Snap-1 & 171512473989538506 & 15.506 (6) & 15.498 & 15.892 (4) & 15.894 & 16.207 (2) & 16.202 & 16.425 (2) & 16.425 \\
WD1657+343 & 149172547130238286 & 16.230 (4) & 16.228 & 16.700 (2) & 16.693 & 17.074 (2) & 17.074 & 17.375 (7) & 17.360 \\
SF1615+001A & 108002445593133309 & 16.988 (4) & 16.991 & 16.560 (2) & 16.563 & 16.381 (2) & 16.384 & 16.314 (3) & 16.317 \\
Snap-2 & 174682449420946620 & 16.443 (5) & 16.443 & 16.053 (3) & 16.045 & 15.912 (3) & 15.905 & 15.873 (2) & 15.874 \\
C2602 & 74560531369524156 & 16.669 (4) & 16.673 & 16.365 (5) & 16.368 & 16.258 (4) & 16.264 & 16.250 (3) & 16.243 \\
KF06T2 & 188132696582938100 & 14.406 (2) & 14.418 & 13.613 (4) & 13.607 & 13.272 (NaN) & 13.260 & 13.093 (1) & 13.087 \\
GD153 & 134431942595767273 & 13.134 (7) & 13.128 & 13.598 (1) & 13.591 & 13.990 (11) & 13.978 & 14.261 (2) & 14.263 \\
\hline
CALSPEC $\mu, \sigma$ &  & $-$0.006 (1) & 0.006 & +0.003 (1) & 0.005 & +0.001 (1) & 0.005 & +0.003 (1) & 0.006 \\
\hline
SDSS-J010322 & 107580158424764461 & 19.093 (10) & 19.120 (5) & 19.570 (19) & 19.563 (5) & 19.979 (17) & 19.933 (5) & 20.130 (64) & 20.209 (7) \\
SDSS-J022817 & 97850370715284759 & 19.837 (14) & 19.827 (11) & 20.188 (53) & 20.163 (6) & 20.523 (36) & 20.477 (7) & 20.803 (117) & 20.728 (10) \\
SDSS-J024854 & 148510422289556735 & 18.351 (7) & 18.392 (8) & 18.699 (6) & 18.740 (5) & 18.972 (12) & 19.052 (3) & 19.198 (31) & 19.296 (5) \\
SDSS-J072752 & 146681119698436001 & 18.018 (10) & 18.026 (3) & 18.475 (11) & 18.450 (2) & 18.806 (12) & 18.809 (2) & 19.127 (24) & 19.079 (3) \\
SDSS-J081508 & 117031237865415713 & 19.781 (40) & 19.747 (5) & 20.328 (37) & 20.180 (6) & 20.625 (73) & 20.547 (7) & 20.710 (165) & 20.823 (7) \\
SDSS-J102430 & 107351561288148089 & 18.885 (9) & 18.936 (9) & 19.292 (23) & 19.311 (7) & 19.440 (98) & 19.641 (12) & 19.758 (31) & 19.896 (18) \\
SDSS-J111059 & 87401677476402284 & 17.895 (5) & 17.889 (2) & 18.302 (9) & 18.307 (2) & 18.607 (15) & 18.664 (2) & 18.957 (26) & 18.934 (3) \\
SDSS-J111127 & 155931678637970024 & 18.412 (15) & 18.456 (3) & 18.886 (11) & 18.929 (3) & 19.260 (11) & 19.319 (3) & 19.586 (16) & 19.607 (3) \\
SDSS-J120650 & 110431817100904541 & 18.693 (10) & 18.691 (4) & 19.096 (29) & 19.056 (5) & 19.388 (24) & 19.388 (6) & 19.645 (34) & 19.648 (7) \\
SDSS-J121405 & 162761835213366810 & 17.779 (5) & 17.787 (4) & 18.236 (7) & 18.229 (3) & 18.570 (10) & 18.605 (2) & 18.849 (17) & 18.890 (3) \\
SDSS-J130234 & 120251956434903476 & 17.052 (3) & 17.066 (3) & 17.494 (3) & 17.505 (2) & 17.858 (6) & 17.877 (2) & 18.114 (9) & 18.157 (3) \\
SDSS-J131445 & 104111986877165205 & 19.078 (14) & 19.130 (7) & 19.556 (21) & 19.562 (6) & 19.887 (40) & 19.928 (4) & 20.240 (69) & 20.203 (6) \\
SDSS-J151421 & 108952285886717975 & 15.720 (2) & 15.729 (4) & 16.101 (4) & 16.110 (2) & 16.434 (2) & 16.448 (2) & 16.715 (5) & 16.712 (2) \\
SDSS-J155745 & 174922394391413855 & 17.487 (5) & 17.505 (2) & 17.958 (7) & 17.978 (2) & 18.356 (5) & 18.367 (2) & 18.647 (11) & 18.656 (3) \\
SDSS-J163800 & 108942495015106395 & 18.860 (13) & 18.864 (6) & 19.314 (22) & 19.277 (4) & 19.611 (13) & 19.627 (4) & 19.816 (53) & 19.891 (5) \\
SDSS-J181424 & 202682736002931825 & 16.573 (5) & 16.568 (2) & 17.007 (3) & 16.999 (1) & 17.358 (4) & 17.369 (2) & 17.651 (9) & 17.650 (2) \\
SDSS-J210150 & 101083154611083390 & 18.652 (9) & 18.677 (6) & 19.052 (8) & 19.056 (3) & 19.410 (18) & 19.393 (4) & 19.703 (33) & 19.655 (6) \\
SDSS-J232941 & 108223524222323007 & 18.134 (6) & 18.163 (6) & 18.452 (5) & 18.468 (5) & 18.772 (8) & 18.762 (4) & 19.003 (17) & 19.000 (5) \\
SDSS-J235144 & 153513579345744806 & 18.085 (4) & 18.099 (6) & 18.447 (13) & 18.450 (4) & 18.776 (10) & 18.764 (3) & 19.100 (36) & 19.010 (4) \\
\hline
DA WD $\mu, \sigma$ &  & $-$0.012 (2) & 0.019 & $-$0.004 (2) & 0.017 & $-$0.013 (2) & 0.021 & $-$0.007 (4) & 0.030 \\
\hline
\end{tabular}
\footnotesize{\tablecomments{Measurements in each passband are presented in the corresponding column with both observed (left) and synthetic (right) magnitudes. Uncertainties are reported parenthetically in millimags. The weighted mean offset $\mu$ and standard deviation $\sigma$ of the residuals are reported for both the CALSPEC standards used in \citetalias{Supercal} and our DA WD. All catalog magnitudes are reported, however we impose selection cuts detailed in the text to determine which stars are used to determine the mean offset. All quantities are rounded to a millimag.}}
\caption{Comparison of PS1 DR1 aperture magnitudes and synthetic magnitudes derived from our DA white dwarf SEDs.\label{table:ps1comp}}
\end{centering}
\end{table*}

We use aperture photometry from the PS1 DR1 ``MeanObject'' table, which we test against the bitmask \texttt{0x1C138} --- see \citet{ps1db} for a detailed description of the PS1 DR1 schema, and Table 15 for an explanation of the meaning of each bit. We also require that the reported photometry be the average of at least 5 detections in each passband, and have S/N$\geq 10$. Finally, we impose a restriction that the PS1 DR1 and aperture magnitudes must agree to within 0.05~mag to exclude contaminated sources. All DR1 photometry of CALSPEC standard LDS749B are removed because of these selection criteria, and we exclude this standard from the comparison. KF06T2 has a valid \textit{i}$_{\text{PS1}}$ magnitude but the uncertainty is reported as NaN. We report this measurement as presented in the PS1 DR1 catalogs, but do not include it in determining the mean AB offset. Additionally, while we can determine an AB offset in \textit{y}$_{\text{PS1}}$ from the bright CALSPEC standards, there is no DR1 photometry of our DA white dwarfs that match our selection criteria. 

The result of our analysis to determine the AB offsets from the CALSPEC standard used in \citetalias{Supercal} and our DA white dwarfs is presented in Table~\ref{table:ps1comp}. These indicate that DR1 is consistent with the analysis in \citetalias{Supercal} to within a few millimag in \textit{griz}$_{\text{PS1}}$. These offsets remain consistent when considering the median difference or the 3$\sigma$-clipped difference.

We find the following weighted mean offsets between the PS1 DR1 magnitudes and the synthetic magnitudes of our DA white dwarfs in the PS1 passbands: $\{$\textit{g,r,i,z}$\}_{\text{PS1}} = \{-12,-4,-13,-7\}$~mmag. We propagate the uncertainties on observed and synthetic magnitudes to uncertainties in the weighted mean, which are 2--4~mmag. The sign and scale of the AB offsets are consistent across passbands, indicating a real, albeit small difference between the PS1 and CALSPEC flux scale.

\begin{figure*}[htb]
    \centering
    \includegraphics[width=\textwidth]{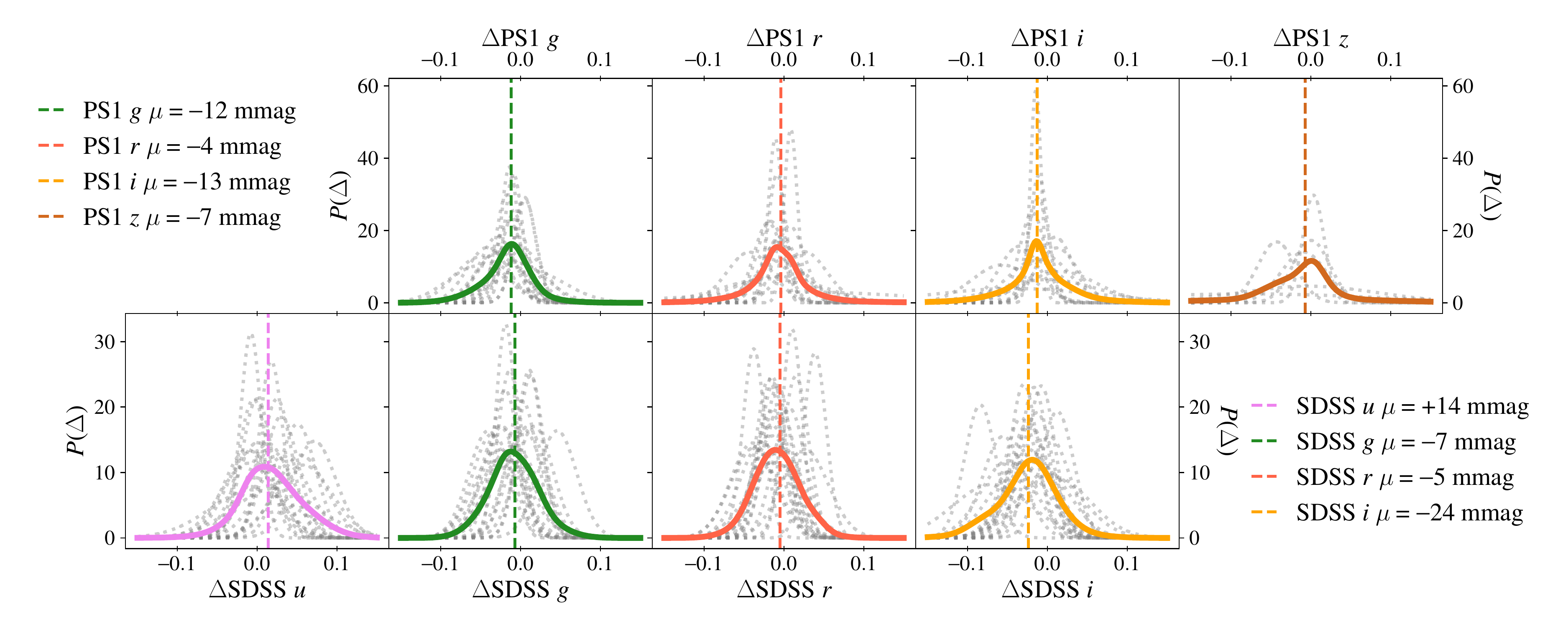}
    \caption{Distribution of residuals of our DA white dwarfs in PS1 (top row) and SDSS (bottom) row. Residuals and associated uncertainties for each star are indicated by the dotted grey Gaussian distributions. A kernel density estimate of the distribution of the residuals is indicated by the solid colored line in each panel. The dashed colored line indicates the weighted mean offset from zero in each band. The values of these offsets are presented in Table~\ref{table:ps1comp} and \ref{table:sdsscomp}, as well as the legend on each row. The reported uncertainties are scaled by a factor of 2.5 and 1.2, for PS1 and SDSS respectively, to obtain a reduced $\chi^{2}$ statistic of unity.}
    \label{fig:residdistrib}
\end{figure*}

The offsets in \textit{gr}$_{\text{PS1}}$ determined from our DA white dwarfs are comparable to the offsets determined from the CALSPEC standards. However, the reported uncertainties on the observations of our DA white dwarfs do not describe the dispersion of the residuals. The reduced $\chi^{2}$ statistic for our faint DA white dwarfs is much higher than for the bright CALSPEC standards. This is likely why the offsets in \textit{iz}$_{\text{PS1}}$ are larger than those determined directly from the bright CALSPEC standards, which are additionally much redder on the mean than our DA white dwarfs. We find that the PS1 errors must be scaled by a factor of 2.5 for a reduced $\chi^{2}$ of unity. The distribution of residuals for PS1 and SDSS (discussed in the following subsection) are shown in Fig.~\ref{fig:residdistrib}.

We obtained additional imaging of our DA white dwarfs from the Foundation survey, operated on the PS1 telescope, but reduced with the \texttt{photpipe} image processing pipeline~\citep[originally described in ][but significantly updated thereafter]{photpipe}. The Foundation PS1 images indicate that the photometric repeatability is $\sim1.5$\%. The AB offsets determined from the Foundation PS1 images are correlated with those determined from our comparison to the PS1 DR1 catalog, as the latter is used to determine the zeropoints of the former. We will use these data to constrain the temporal variability of our standards together with our LCO observations, and establish secondary standards around our DA white dwarfs in future work.

\subsubsection{Comparison with SDSS DR12}
\begin{table*}[htb]
\scriptsize
\begin{centering}
\begin{tabular}{c|c|ll|ll|ll|ll}
\hline
\hline
Object & SDSS ID & \multicolumn{2}{c}{SDSS \textit{u}} & \multicolumn{2}{c}{SDSS \textit{g}} & \multicolumn{2}{c}{SDSS \textit{r}}  & \multicolumn{2}{c}{SDSS \textit{i}} \\
  & & \multicolumn{8}{c}{mag (mmag)} \\
\hline
SDSS-J010322 & SDSS J010322.19-002047.7 & 18.626 (45) & 18.631 (5) & 19.056 (28) & 19.067 (6) & 19.546 (21) & 19.560 (5) & 19.912 (29) & 19.921 (5) \\
SDSS-J022817 & SDSS J022817.16-082716.4 & 19.798 (42) & 19.775 (11) & 19.785 (24) & 19.806 (11) & 20.153 (29) & 20.163 (6) & 20.398 (40) & 20.467 (7) \\
SDSS-J024854 & SDSS J024854.96+334548.2 & 18.116 (22) & 18.115 (7) & 18.361 (18) & 18.357 (8) & 18.721 (17) & 18.738 (5) & 18.957 (16) & 19.042 (3) \\
SDSS-J072752 & SDSS J072752.76+321416.0 & 17.569 (15) & 17.570 (2) & 17.957 (10) & 17.976 (3) & 18.457 (10) & 18.447 (2) & 18.768 (14) & 18.797 (2) \\
SDSS-J081508 & SDSS J081508.78+073145.8 & 19.416 (29) & 19.358 (8) & 19.676 (19) & 19.700 (5) & 20.203 (24) & 20.177 (6) & 20.538 (33) & 20.535 (7) \\
SDSS-J102430 & SDSS J102430.93-003207.0 & 18.577 (28) & 18.588 (9) & 18.889 (22) & 18.896 (10) & 19.305 (22) & 19.309 (7) & 19.596 (28) & 19.631 (12) \\
SDSS-J111059 & SDSS J111059.42-170954.2 & 17.477 (17) & 17.447 (4) & 17.858 (19) & 17.841 (3) & 18.312 (17) & 18.305 (2) & 18.620 (18) & 18.653 (2) \\
SDSS-J111127 & SDSS J111127.30+395628.0 & 17.984 (26) & 17.930 (4) & 18.407 (19) & 18.398 (3) & 18.918 (14) & 18.926 (3) & 19.282 (19) & 19.307 (3) \\
SDSS-J120650 & SDSS J120650.40+020142.4 & 18.541 (24) & 18.553 (4) & 18.650 (23) & 18.663 (4) & 19.028 (22) & 19.055 (5) & 19.328 (27) & 19.377 (6) \\
SDSS-J121405 & SDSS J121405.11+453818.5 & 17.370 (25) & 17.378 (3) & 17.711 (24) & 17.740 (4) & 18.211 (19) & 18.227 (3) & 18.538 (21) & 18.593 (2) \\
SDSS-J130234 & SDSS J130234.43+101238.9 & 16.614 (16) & 16.619 (2) & 16.976 (20) & 17.016 (3) & 17.463 (23) & 17.503 (2) & 17.840 (24) & 17.865 (2) \\
SDSS-J131445 & SDSS J131445.05-031415.5 & 18.716 (26) & 18.683 (5) & 19.045 (20) & 19.080 (7) & 19.522 (23) & 19.560 (6) & 19.927 (36) & 19.917 (4) \\
SDSS-J151421 & SDSS J151421.27+004752.8 & 15.482 (12) & 15.464 (2) & 15.681 (12) & 15.694 (4) & 16.089 (14) & 16.108 (2) & 16.421 (14) & 16.438 (2) \\
SDSS-J155745 & SDSS J155745.39+554609.7 & 16.975 (10) & 16.983 (2) & 17.493 (20) & 17.447 (2) & 17.990 (13) & 17.975 (2) & 18.334 (19) & 18.355 (2) \\
SDSS-J163800 & SDSS J163800.36+004717.7 & 18.461 (18) & 18.410 (5) & 18.826 (11) & 18.815 (6) & 19.269 (16) & 19.274 (4) & 19.597 (25) & 19.616 (4) \\
SDSS-J181424 & SDSS J181424.12+785402.9 & 16.236 (15) & 16.212 (2) & 16.498 (19) & 16.524 (2) & 16.959 (11) & 16.997 (2) & 17.344 (17) & 17.357 (2) \\
SDSS-J210150 & SDSS J210150.65-054550.9 & 18.483 (21) & 18.412 (9) & 18.655 (12) & 18.643 (6) & 19.042 (14) & 19.055 (3) & 19.385 (21) & 19.382 (4) \\
SDSS-J232941 & SDSS J232941.32+001107.8 & 18.173 (25) & 18.156 (3) & 18.149 (14) & 18.147 (7) & 18.453 (12) & 18.468 (5) & 18.745 (14) & 18.752 (4) \\
SDSS-J235144 & SDSS J235144.29+375542.6 & 17.760 (23) & 17.747 (4) & 18.042 (13) & 18.061 (6) & 18.487 (11) & 18.448 (4) & 18.766 (17) & 18.754 (3) \\
\hline
DA WD $\mu, \sigma$ &  & +0.014 (6) & 0.025 & $-$0.007 (4) & 0.021 & $-$0.005 (4) & 0.020 & $-$0.024 (5) & 0.025 \\
\hline
\end{tabular}
\footnotesize{\tablecomments{The format of this table matches Table~\ref{table:ps1comp}.}}
\caption{Comparison of SDSS DR12 PSF magnitudes and synthetic magnitudes derived from our DA white dwarf SEDs.\label{table:sdsscomp}}
\end{centering}
\end{table*}

\citet{Betoule13} derive AB offsets for SDSS using careful measurements of the three CALSPEC primary standards with the photometric telescope transferred to the primary survey telescope. Unfortunately, these determinations of the magnitudes of the primary standards differ significantly from the reported SDSS DR12~\citep{sdssdr12} photometry, which is saturated in most passbands. As we cannot independently recompute the AB offsets for DR12, we compare our synthetic magnitudes directly to the reported SDSS DR12 PSF magnitudes in Table~\ref{table:sdsscomp}. 

The uncertainties on the SDSS DR12 PSF magnitudes accurately describe the dispersion in \textit{ugr}. There is much larger dispersion in SDSS \textit{iz} where our DA white dwarfs are the most faint. Several of our targets do not have reliable photometry in the \textit{z} band, and we exclude it from this comparison. While we impose the same S/N $\geq 10$ threshold we used for the PS1 comparison, all reported SDSS DR12 photometry satisfy this criterion. We find the following weighted mean offsets between the SDSS DR12 magnitudes and the synthetic magnitudes of our standards in the SDSS passbands: $\{$\textit{u, g, r, i}$\}_{\text{PS1}} = \{+14,-7,-5,-24\}$~mmag. The reported uncertainties from SDSS are reasonable, with the reduced $\chi^{2}$ statistic indicating they must be scaled by only a factor of 1.2 to fully describe the dispersion. The standard deviation of the SDSS DR12 residuals is larger than we find for PS1, indicating the latter does have more internally consistent photometry, albeit with significantly underestimated uncertainties.

The offset in \textit{gr} are consistent with zero and the offset in \textit{u} is only significant at the 2$\sigma$ level. The residuals in the \textit{i} band are not significant at the 1--2$\sigma$ level for any individual object, but are consistent across all of our 19 DA white dwarfs. This likely reflects a real different between the CALSPEC flux scale and the SDSS DR12 flux scale. The $-24$~mmag AB offset in \textit{i} agrees with the corresponding value of $-$27~mmag determined by \citet{Betoule13}.

\subsubsection{Comparison with \emph{Gaia} DR2}
\begin{table*}[htpb]
\scriptsize
\begin{centering}
\begin{tabular}{c|c|ll|ll|ll}
\hline
\hline
Object & \textit{Gaia} DR2 ID & \multicolumn{2}{c}{\textit{Gaia} {B$_P$}} & \multicolumn{2}{c}{\textit{Gaia} {R$_P$}}  & \multicolumn{2}{c}{\textit{Gaia} {G}} \\
  & & \multicolumn{6}{c}{mag (mmag)} \\
 \hline
G191-B2B & 266077145295627520 & 11.487 (15) & 11.458 (2) & 12.067 (2) & 12.057 (1) & 11.738 (1) & 11.722 (2) \\
GD153 & 3944400490365194368 & 13.081 (5) & 13.064 (2) & 13.629 (1) & 13.614 (2) & 13.322 (1) & 13.307 (2) \\
GD71 & 3348071631670500736 & 12.770 (12) & 12.774 (2) & 13.299 (2) & 13.289 (2) & 13.026 (2) & 13.001 (2) \\
\hline
CALSPEC $\mu, \sigma$ &  & +0.015 (5) & 0.030 & +0.012 (1) & 0.037 & +0.018 (1) & 0.017 \\
\hline
SDSS-J010322 & 2536159496590552704 & 19.154 (30) & 19.046 (5) & 19.577 (72) & 19.571 (6) & 19.356 (4) & 19.285 (5) \\
SDSS-J022817 & 5176546064064586624 & 19.869 (139) & 19.823 (10) & 20.192 (141) & 20.130 (8) & 20.046 (10) & 19.964 (9) \\
SDSS-J024854 & 139724391470489472 & 18.333 (47) & 18.347 (7) & 18.704 (31) & 18.695 (4) & 18.561 (3) & 18.520 (6) \\
SDSS-J072752 & 892231562565363072 & 17.944 (7) & 17.956 (3) & 18.458 (36) & 18.448 (3) & 18.232 (3) & 18.184 (3) \\
SDSS-J081508 & 3097940536009636992 & 19.695 (44) & 19.694 (6) & 20.278 (166) & 20.187 (8) & 19.996 (5) & 19.915 (6) \\
SDSS-J102430 & 3830980604624181376 & 18.940 (59) & 18.882 (8) & 19.297 (105) & 19.284 (12) & 19.120 (5) & 19.075 (9) \\
SDSS-J111059 & 3559181712491390208 & 17.852 (11) & 17.822 (3) & 18.347 (20) & 18.304 (2) & 18.089 (2) & 18.045 (2) \\
SDSS-J111127 & 765355922242992000 & 18.365 (22) & 18.378 (3) & 18.955 (75) & 18.956 (3) & 18.690 (3) & 18.634 (3) \\
SDSS-J120650 & 3891742709551744640 & 18.651 (17) & 18.677 (5) & 18.957 (30) & 19.038 (6) & 18.885 (2) & 18.840 (5) \\
SDSS-J121405 & 1539041748872771968 & 17.757 (11) & 17.732 (4) & 18.154 (38) & 18.248 (2) & 18.002 (1) & 17.960 (3) \\
SDSS-J130234 & 3734528631432609920 & 17.044 (6) & 17.001 (3) & 17.527 (12) & 17.518 (2) & 17.268 (1) & 17.234 (3) \\
SDSS-J131445 & 3684543213630134784 & 19.082 (42) & 19.064 (6) & 19.631 (82) & 19.568 (5) & 19.354 (4) & 19.294 (6) \\
SDSS-J151421 & 4419865155422033280 & 15.743 (9) & 15.694 (3) & 16.119 (5) & 16.096 (2) & 15.905 (1) & 15.879 (3) \\
SDSS-J155745 & 1621657158502507520 & 17.452 (14) & 17.427 (2) & 18.019 (18) & 18.005 (2) & 17.721 (2) & 17.683 (2) \\
SDSS-J163800 & 4383979187540364288 & 18.853 (19) & 18.793 (5) & 19.313 (40) & 19.266 (5) & 19.065 (2) & 19.015 (5) \\
SDSS-J181424 & 2293913930823813888 & 16.570 (9) & 16.522 (2) & 17.031 (7) & 17.013 (2) & 16.773 (2) & 16.739 (2) \\
SDSS-J210150 & 6910475935427725824 & 18.654 (21) & 18.642 (6) & 19.095 (44) & 19.040 (5) & 18.867 (2) & 18.826 (5) \\
SDSS-J232941 & 2644572064644349952 & 18.187 (21) & 18.161 (6) & 18.394 (28) & 18.416 (4) & 18.323 (2) & 18.284 (5) \\
SDSS-J235144 & 2881271732415859072 & 18.056 (16) & 18.040 (5) & 18.417 (14) & 18.408 (4) & 18.272 (2) & 18.224 (5) \\
\hline
DA WD $\mu, \sigma$ &  & +0.027 (3) & 0.032 & +0.017 (4) & 0.040 & +0.041 (1) & 0.015 \\
\hline
\end{tabular}
\footnotesize{\tablecomments{The format of this table matches Table~\ref{table:ps1comp}.}}
\caption{Comparison of \textit{Gaia} DR2 magnitudes and synthetic magnitudes derived from our DA white dwarf SEDs.\label{table:gaiacomp}}
\end{centering}
\end{table*}

\emph{Gaia} provides photometry and parallax measurements~\citep{gaiadr2p1,gaiadr2p2} for most of our stars (these parallax measurements are reported in Table 1 of \citetalias{Calamida18}). For our faint DA white dwarfs, the \emph{Gaia} parallax errors have a mean precision of $25\%$. Additionally, the parallax errors increase as a function of magnitude, with three sources having relative errors of $>50\%$. Two sources in our sample do not have any reported parallax. In contrast to the heterogeneous parallax measurements, we increased exposure times for our spectroscopic and \emph{HST} observations to ensure that all our DA white dwarfs have comparable S/N. 

Despite their heterogeneity, there are potential gains to incorporating parallax measurements into our inference. Intrinsic DA white dwarf parameters derived from \emph{Gaia} observations with S/N~$> 20$ have been compared against measurements inferred from Pan-STARRS and SDSS~\citep{GentileFusillo19}, and these are in good agreement. The \emph{Gaia} parallax measurements can be very useful for distinguishing double-degenerate systems that masquerade as a single star. However, there are potential systematic issues with modeling the \emph{Gaia} measurements together with our spectroscopic and photometric observations. \citet{Tremblay2018} compare DA white dwarfs with intrinsic parameters that are well measured from spectroscopy to stellar parameters derived from \emph{Gaia} measurements and find that photometric and spectroscopic temperature scales differ systematically by a few percent. Additionally, there is a residual systematic in the inferred values of $\log g$ for DA white dwarfs with 11,000~$K < T_{\text{eff}} < $13,000~K. As \emph{Gaia} parallaxes are not presently available for all our stars and incorporating the extant measurements into the likelihood function in \S\ref{sec:wdmodel} could potentially introduce systematics, we do not use them directly at this time. 

Comparison with the \emph{Gaia} measurements is still informative, and the mission also reports photometry. These photometric observations are reported with respect to Vega by default, but the data release also provides AB-based zeropoints. For consistency with the remainder of this work, we use these AB zeropoints. We measure the offsets by comparing the \emph{Gaia} DR2 photometry with synthetic photometry of our SEDs of the CALSPEC primary standards. We then determined the offsets between the DR2 and synthetic photometry of our DA white dwarfs. The results of the comparison are presented in Table~\ref{table:gaiacomp}. As with the comparison against SDSS, we require that the S/N be $\geq 10$. This rejects 1 measurement in \textit{B}$_P$ and two in \textit{R}$_P$. These rejected measurements are consistent with our synthetic magnitudes to within 1$\sigma$ of the reported photometric uncertainties.

\begin{figure}[h!tb]
    \centering
    \includegraphics[width=0.47\textwidth]{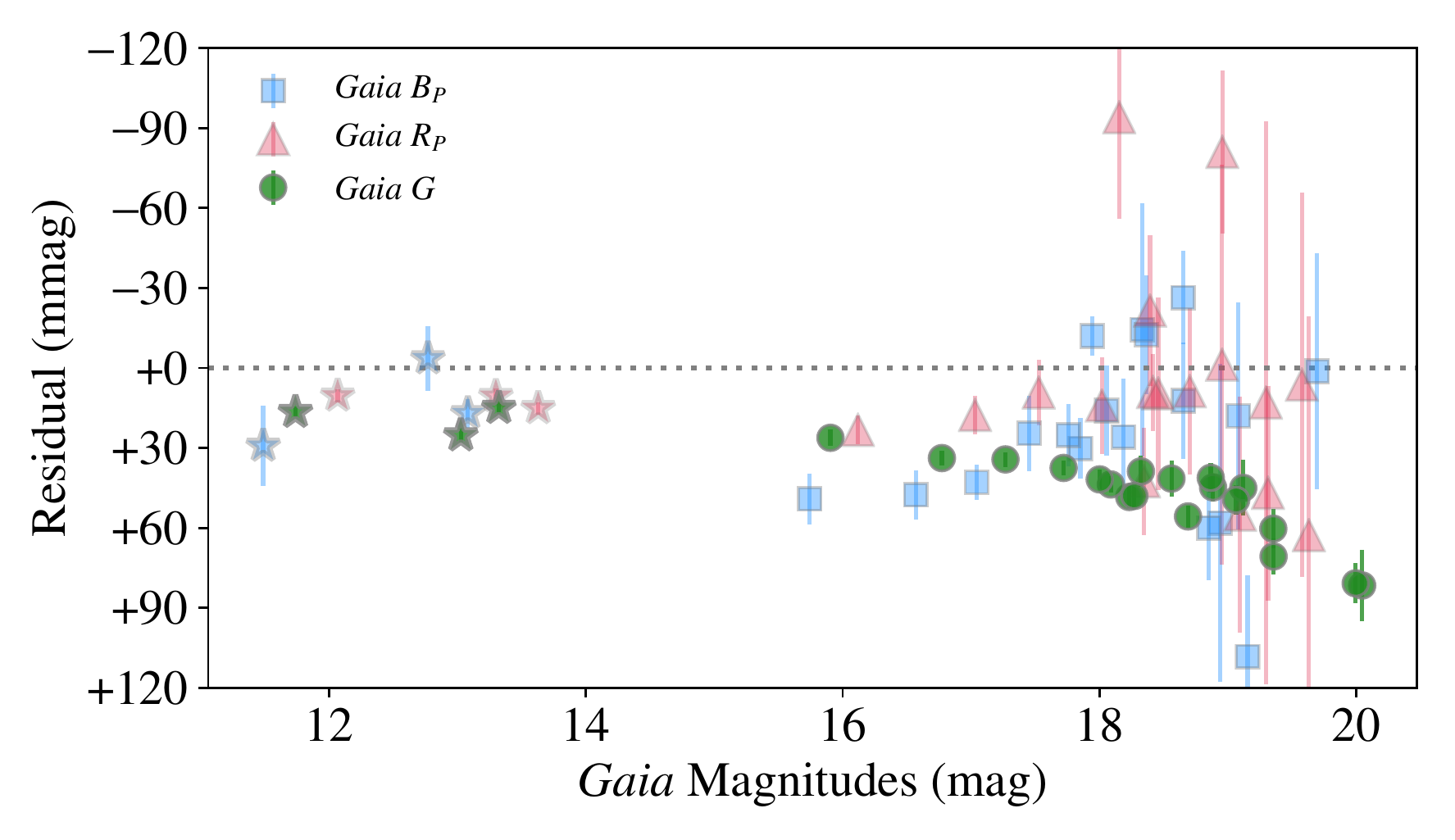}
    \caption{Residuals of CALSPEC primary standards (stars) and our DA white dwarfs (markers indicated in the legend for each passband) vs magnitude. The \emph{Gaia} \textit{G} residuals exhibit bias and a pronounced non-linearity with magnitude, not seen with other surveys. \citet{gaianonlin} also report a trend with respect to CALSPEC for 6~mag $\leq G \leq 16.5$~mag.}
    \label{fig:gaiagnonlin}
\end{figure}

We find offsets of $\{$\textit{B}$_P$, \textit{R}$_P$, \textit{G} $\} = \{+27, +17, +41\}$~mmag from our DA white dwarfs. These offsets are consistent with those determined from the CALSPEC primary standards, where we find offsets of $\{+15, +13, +18\}$~mmag. The residuals in \emph{Gaia} \textit{G} show an increasing trend with faint amplitudes. Weak trends can be seen in the other passbands, but are less significant because of the much larger uncertainties near the faint limit. \citet{gaianonlin} report on this bias using an independent comparison to CALSPEC, and provide a linear correction valid from 6~mag $ \leq G \leq 16.5$~mag. Their linear correction removes the linear-component of the residual, but a non-linear component remains. While it is possible to define an empirical relation and model the non-linearity, without understanding the physical origin of this effect it cannot be properly mitigated. No similar trends are seen in the comparisons with PS1 and SDSS. The reported uncertainties on the observations of our DA white dwarfs do not describe the dispersion of the residuals. The reduced $\chi^{2}$ statistic is much greater than unity for all the \emph{Gaia} passbands, indicating the uncertainties are underestimated, or that there are additional sources of dispersion with the photometers that have not been accounted for. 


\section{Summary and Discussion}\label{sec:futurework}

In \citetalias{Narayan16}, we identified various potential improvements of our proof-of-concept analysis. This work implements every one of those improvements, as well as refining other elements of the \citetalias{Narayan16} analysis. The calibrated SEDs provided in this work can be used to derive synthetic photometry in any passband from the ultraviolet to the infrared, which in turn can be used to calibrate observations on to our photometric system tied directly to the three CALSPEC primary standards. The internal precision of our network is better than 5~mmag in the optical, and the accuracy to which our photometric system is tied to CALSPEC is conservatively 4~mmag. This meets or exceeds the needs of most planned facilities. 

In the course of our analysis, we identified and quantified the few systematic effects that may affect our program, the largest of which are how well our network is tied to the CALSPEC flux scale, and a count rate non-linearity exhibited by the IR channel of \emph{HST/WFC3}. The biases in the inferred SEDs caused by these effects is small, and in a future analysis we expect to be able to mitigate these sources of systematic error further. Our ability to measure systematic effects at the few mmag level reflect the precision of our program --- these subtle signals would be swamped by other statistical systematic errors with a purely ground-based analysis. Our network of DA white dwarfs are the most internally accurate with CALSPEC and precise spectrophotometric references with 16.5~mag $ < V < 19$~mag presently available. This is a strong claim, but one that we feel is validated by the suite of diagnostic tests performed in \S\ref{sec:validation}.

Our network of standards can be of immediate benefit to several ongoing surveys. The SDSS and the Supernova Legacy Survey (SNLS) AB offsets~\citep{Betoule13} were determined from the three primary CALSPEC standards, observed with a different facility and transformed to the survey telescope using comparisons of observed photometry and synthetic stellar spectral libraries. \citetalias{Supercal} determined additional corrections to place these magnitudes on the same system as the Pan-STARRS PS1 magnitudes, with its own AB offsets determined using 7 CALSPEC standards. Each of these steps introduces potential systematic errors, and are only necessary because the existing CALSPEC standards are too bright and the primary standards are inaccessible from the south. Our network of faint northern and equatorial DA white dwarfs already addresses both these limitations, and simplifies the calibration procedure to i) observe DA white dwarf stars ii) determine synthetic magnitudes of DA white dwarfs from the SEDs published with this work iii) determine the difference.

\subsection{Future Work}
The simplest expansion on this work is to expand the dataset. Our cycle 25 \emph{HST/WFC3} observations of southern DA white dwarfs have been executed, and we are refining the data reduction and analysis drawing on lessons learned from \citetalias{Calamida18} and this work. Our temporal monitoring of these southern standards is almost complete. Our next analysis will use this expanded dataset. With the addition of any southern DA white dwarfs that meet the stringent criteria in this work, we will have established our all-sky network. While \emph{HST}'s lifetime is limited, our network will extend its legacy of precise calibration well into the future. Moreover, the methodology developed in this work can be used to expand this network further in the future. In particular, there is no conceptual difficulty in establishing faint DA white dwarfs as spectrophotometric standards tied to CALSPEC located within LSST Deep-Drilling Fields (DDF) or WFIRST SN Survey fieleds.

Mitigating the sources of systematic error that affect our SEDs beyond the level accomplished in this work requires more complex changes to our methodology. Our present analysis treats each white dwarf separately, fitting a single spectrum and the photometry for each. A minimal extension would be to infer results incorporating multiple spectra for each object where available. This would allow us to infer the SEDs of each star, including the primary standards, coherently from all available high S/N data. 

As in our previous analysis, this work establishes DA white dwarf models as good \emph{differential} predictors of measured flux ratios. The flux scale is itself set by the CALSPEC SEDs of the three primary standards, and there is no way to determine if this flux scale is accurately tied to the AB system within our framework. The only way to avoid tying our network to the CALSPEC flux scale is to establish a single common flux scale for the CALSPEC standards and our faint DA white dwarfs. Our analysis in this work treats the DA white dwarfs hierarchically when inferring their apparent magnitudes, but individually in order to infer the parameters that describe their SED. The advantage to this two step approach is that the model in \S\ref{sec:photmodel} is completely independent of the intrinsic and extrinsic DA white dwarf parameters. While this simple separable framework is conceptually appealing, the apparent magnitudes are directly tied to the CALSPEC SEDs of the three primary standards, which may have their own systematic errors. 

It is possible to construct a single hierarchical model to describe the instrumental measurements and spectroscopy of all the DA white dwarfs directly and CALSPEC standards, without the intermediate hierarchical model to infer the apparent magnitudes. This model could be simultaneously conditioned on the three primary standards and our DA white dwarfs to establish a single photometric system from $V \sim$ 9--19~mag, incorporating measurements of laboratory or satellite-born flux standards, with the model atmosphere grid to calibrate the absolute flux. This model would be significantly more complex than the analysis presented in this work --- instead of a 10 dimensional posterior distribution, we would be constraining $N_{\text{s}} \times (5 + 5\cdot N^{\lambda}_{s}) + 3 \cdot N_{\text{PB}}$ parameters at once, where $N_{\text{s}}$ is the number of objects, $N^{\lambda}_{s}$ is the number of spectra per object and $N_{\text{PB}}$ is the number of independent passbands. Even with conservative assumptions about the number of DA white dwarfs from our cycle 25 program that make good spectrophotometric standards, this would be $\sim 350$ dimensional problem. Inference of the parameters of this model would require the development of bespoke sampling algorithms and significant computational resources. 

There are few astrophysical sources that are as simple to model as DA white dwarfs, and capable of delivering the level of photometric accuracy achieved in our analysis. While our focus is on the accuracy of the fluxes of our SED models, various studies have focused on testing the absolute accuracy of inferred white dwarf intrinsic parameters such as temperature, surface gravity and mass using a variety of techniques including determinations from eclipsing binaries~\citep{Parsons17}, gravitational lensing~\citep{Sahu17}, dynamical studies~\citep{Bond17} and gravitational redshifts~\citep{Joyce18b}, and comparisons to other space-based missions including \emph{FUSE}~\citep{Joyce18a} and \emph{Gaia}~\citep{Tremblay2018}. These efforts may lead to refinements in the existing DA white dwarf model atmosphere grid, which in turn can be propagated to our inferred SEDs, yielding higher photometric accuracy.

The colors of our DA white dwarf stars are much bluer than the main-sequence, and we encourage their use to determine relative zeropoints, rather than determine color-transformations --- the latter would require red standards within the range covered by the main-sequence. We feel that this is largely a drawback of how transformation equations are parameterized, as they conflate establishing zeropoints with determining color-terms between different surveys photometric systems. The choice of which color to use to parameterize these transformations, and indeed the choice to restrict the transformation equations to a single color is entirely arbitrary. There are more sophisticated statistical methods of establishing the latter that model the non-linear shape of the stellar locus~\citep[e.g.][]{slr,bigmacs}, rather than simple linear equations. We are examining the feasibility of combining such stellar locus regression techniques with the methodology in this work. It is valuable to have calibrated red stars within a few arcminutes of our DA white dwarfs to verify the transformation equations between surveys, irrespective of how the transformations are derived. For this reason, we obtained \emph{HST/ACS} parallel observations in \textit{F475W} and \textit{F775W} to provide field stars that can be tied to our white dwarfs with color information. The properties of these stars will be presented in a future work. 

The most complex extension of our analysis would be incorporating photometry from major surveys to constrain the shape of the entire stellar locus, solving for relative offsets within each survey from overlapping images and establishing the absolute zeropoints using our network of DA white dwarfs. Such a model would combine the {\"U}bercal~\citep{Ubercal} method to establish \emph{uniform} internal photometry, with stellar locus regression to determine \emph{relative} offsets between surveys as a function of color, and use our DA white dwarf stars and laboratory references to set the \emph{absolute} flux scale. This would be an invaluable all-sky photometric catalog and enable numerous new studies.  


\acknowledgments

GN is supported by the Lasker Fellowship at the Space Telescope Science Institute. E. Olszewski was also partially supported by the NSF through grants AST-1313006 and AST-1815767.

Based on observations made with the NASA/ESA Hubble Space Telescope, obtained from the Mikulski Archive for Space Telescopes (MAST) at the Space Telescope Science Institute. STScI is operated by the Association of Universities for Research in Astronomy, Inc. under NASA contract NAS 5-26555. These observations are associated with program \href{http://www.stsci.edu/cgi-bin/get-proposal-info?id=12967&observatory=HST}{GO-12967} and \href{http://www.stsci.edu/cgi-bin/get-proposal-info?id=13711&observatory=HST}{GO-13711}. We thank the staff at STScI, and in particular our program coordinators, Tricia Royle and Miranda Link, for their assistance with implementing our program.  

Observations reported here were obtained at the MMT Observatory, a joint facility of the University of Arizona and the Smithsonian Institution. Based on observations obtained at the Gemini Observatory, which is operated by the Association of Universities for Research in Astronomy, Inc., under a cooperative agreement with the NSF on behalf of the Gemini partnership: the National Science Foundation (United States), the National Research Council (Canada), CONICYT (Chile), Ministerio de Ciencia, Tecnolog\'{i}a e Innovaci\'{o}n Productiva (Argentina), and Minist\'{e}rio da Ci\^{e}ncia, Tecnologia e Inova\c{c}\~{a}o (Brazil). Based on observations obtained at the Southern Astrophysical Research (SOAR) telescope, which is a joint project of the Minist\'{e}rio da Ci\^{e}ncia, Tecnologia, Inova\c{c}\~{o}es e Comunica\c{c}\~{o}es (MCTIC) do Brasil, the U.S. National Optical Astronomy Observatory (NOAO), the University of North Carolina at Chapel Hill (UNC), and Michigan State University (MSU). 
This work makes use of observations from the LCOGT network, the Pan-STARRS PS1 Survey and the Sloan Digital Sky Survey (SDSS). 

The Pan-STARRS1 Surveys (PS1) have been made possible through contributions of the Institute for Astronomy, the University of Hawaii, the Pan-STARRS Project Office, the Max-Planck Society and its participating institutes, the Max Planck Institute for Astronomy, Heidelberg and the Max Planck Institute for Extraterrestrial Physics, Garching, The Johns Hopkins University, Durham University, the University of Edinburgh, Queen's University Belfast, the Harvard-Smithsonian Center for Astrophysics, the Las Cumbres Observatory Global Telescope Network Incorporated, the National Central University of Taiwan, the Space Telescope Science Institute, the National Aeronautics and Space Administration under Grant No. NNX08AR22G issued through the Planetary Science Division of the NASA Science Mission Directorate, the National Science Foundation under Grant No. AST-1238877, the University of Maryland, and Eotvos Lorand University (ELTE).

Funding for SDSS-III has been provided by the Alfred P. Sloan Foundation, the Participating Institutions, the National Science Foundation, and the U.S. Department of Energy Office of Science. The SDSS-III web site is \url{http://www.sdss3.org/}.

SDSS-III is managed by the Astrophysical Research Consortium for the Participating Institutions of the SDSS-III Collaboration including the University of Arizona, the Brazilian Participation Group, Brookhaven National Laboratory, Carnegie Mellon University, University of Florida, the French Participation Group, the German Participation Group, Harvard University, the Instituto de Astrofisica de Canarias, the Michigan State/Notre Dame/JINA Participation Group, Johns Hopkins University, Lawrence Berkeley National Laboratory, Max Planck Institute for Astrophysics, Max Planck Institute for Extraterrestrial Physics, New Mexico State University, New York University, Ohio State University, Pennsylvania State University, University of Portsmouth, Princeton University, the Spanish Participation Group, University of Tokyo, University of Utah, Vanderbilt University, University of Virginia, University of Washington, and Yale University. 

This work has made use of data from the European Space Agency (ESA) mission {\it Gaia} (\url{https://www.cosmos.esa.int/gaia}), processed by the {\it Gaia} Data Processing and Analysis Consortium (DPAC, \url{https://www.cosmos.esa.int/web/gaia/dpac/consortium}). Funding for the DPAC has been provided by national institutions, in particular the institutions participating in the {\it Gaia} Multilateral Agreement.

The analyses in this paper were run on the Odyssey cluster supported by the FAS Division of Science, Research Computing Group at Harvard University.

We are grateful to Adam Riess for insightful discussions about binarity in white dwarf systems and the CRNL, as well as serving as contact scientist for our cycle 22 program. Our analysis benefitted from observations obtained by the Foundation Survey (P.I.s: Armin Rest, Daniel Scolnic, Ryan Foley) with the Pan-STARRS PS1 telescope. We thank the Foundation team for obtaining these images, and David Jones in particular for reducing the data and providing us with catalogs. We thank Daniel Foreman-Mackay for his thoughts on Gaussian process kernel design and producing valuable open source tools, Eddie Schlafly for providing background on the dust maps, Daniel Scolnic for discussions on Supercal and for providing PS1 passband response functions, and members of the Dark Energy Survey calibration team, in particular Deborah Guellidge and Douglas Tucker for testing our \texttt{WDmodel} code and providing feedback. This work would not be possible without the diligent efforts of IT staff at STScI, NOAO, and Harvard University and we express our thanks to Michael Peralta (NOAO) and Eric Winter (STScI) for their tireless assistance. GN is grateful to Helmut Abt for several useful and entertaining discussions on the history of photometric calibration and providing him with M\&Ms during long remote observing runs.

This work made extensive use of several open source packages including:
\software{\href{http://www.astropy.org}{\texttt{astropy}}~\citep{astropy:2013, astropy:2018}, \href{https://celerite.readthedocs.io/en/stable/}{\texttt{celerite}}~\citep{celerite}, \href{http://dfm.io/emcee/current/}{\texttt{emcee}}~\citep{dfm2013}, \href{https://extinction.readthedocs.io/en/latest/}{\texttt{extinction}}, \href{https://iminuit.readthedocs.io/en/latest/}{\texttt{iminuit}}~\citep{iminuit}, \href{https://matplotlib.org/index.html}{\texttt{matplotlib}}~\citep{Hunter:2007}, \href{http://www.numpy.org/}{\texttt{numpy}}~\citep{Oliphant:2015}, \href{https://docs.pymc.io/}{\texttt{pymc3}}~\citep{Salvatier2016}, \href{https://pysynphot.readthedocs.io/en/latest/}{\texttt{pysynphot}}~\citep{pysynphot}, \href{https://www.scipy.org/}{\texttt{scipy}}~\citep{scipy}, and the \href{https://conda.io/docs/}{\texttt{conda}} package manager.}
\facilities{\emph{HST} (\emph{WFC3}), Gemini:North (GMOS), Gemini:South (GMOS), MMT (Blue Channel), SOAR, Magellan:Baade (IMACS), LCOGT}.

\bibliographystyle{yahapj}
\bibliography{references}



\end{document}